\documentclass[aps,pra,onecolumn,superscriptaddress,showkeys]{revtex4}
\usepackage{graphicx}

\begin{document}

\title{High-resolution spectroscopy on trapped molecular ions in rotating electric fields:\\ A new approach for measuring the electron electric dipole moment}

\affiliation{JILA, National Institute of Standards and Technology
and University of Colorado, and Department of Physics, University
of Colorado, Boulder, Colorado 80309-0440, USA}

\affiliation{Department of Physics, University of Michigan, Ann
Arbor, Michigan 48109-1040, USA}

\affiliation{Physics Department, Harvard University, Cambridge,
Massachusetts, 02138, USA}

\author{A.E. Leanhardt}
\affiliation{Department of Physics, University of Michigan, Ann
Arbor, Michigan 48109-1040, USA}
\author{J.L. Bohn}
\affiliation{JILA, National Institute of Standards and Technology
and University of Colorado, and Department of Physics, University
of Colorado, Boulder, Colorado 80309-0440, USA}
\author{H. Loh}
\affiliation{JILA, National Institute of Standards and Technology
and University of Colorado, and Department of Physics, University
of Colorado, Boulder, Colorado 80309-0440, USA}
\author{P. Maletinsky}
\affiliation{Physics Department, Harvard University, Cambridge,
Massachusetts, 02138, USA}
\author{E.R. Meyer}
\affiliation{JILA, National Institute of Standards and Technology
and University of Colorado, and Department of Physics, University
of Colorado, Boulder, Colorado 80309-0440, USA}
\author{L.C. Sinclair}
\affiliation{JILA, National Institute of Standards and Technology
and University of Colorado, and Department of Physics, University
of Colorado, Boulder, Colorado 80309-0440, USA}
\author{R.P. Stutz}
\affiliation{JILA, National Institute of Standards and Technology
and University of Colorado, and Department of Physics, University
of Colorado, Boulder, Colorado 80309-0440, USA}
\author{E.A. Cornell}
\email{ecornell@jilau1.colorado.edu}
\homepage[URL: ]{http://jilawww.colorado.edu/bec/CornellGroup/}
\affiliation{JILA, National Institute of Standards and Technology
and University of Colorado, and Department of Physics, University
of Colorado, Boulder, Colorado 80309-0440, USA}

\date{\today}

\begin{abstract}

High-resolution molecular spectroscopy is a sensitive probe for violations of fundamental symmetries.  Symmetry violation searches often require, or are enhanced by, the application of an electric field to the system under investigation.  This typically precludes the study of molecular ions due to their inherent acceleration under these conditions.  Circumventing this problem would be of great benefit to the high-resolution molecular spectroscopy community since ions allow for simple trapping and long interrogation times, two desirable qualities for precision measurements.  Our proposed solution is to apply an electric field that rotates at radio frequencies.  We discuss considerations for experimental design as well as challenges in performing precision spectroscopic measurements in rapidly time-varying electric fields.  Ongoing molecular spectroscopy work that could benefit from our approach is summarized.  In particular, we detail how spectroscopy on a trapped diatomic molecular ion with a ground or metastable ${^3}\Delta_1$ level could prove to be a sensitive probe for a permanent electron electric dipole moment (eEDM).

\end{abstract}

\keywords{High-resolution spectroscopy, radio-frequency, fundamental symmetries, Stark and Zeeman interactions, molecular ions}

\maketitle

\section{Introduction}
\label{s:intro}

\subsection{High-Resolution Molecular Spectroscopy as a Probe of Fundamental Physics}

The quest to verify the most basic laws of nature, and then to search for deviations from them, is an ongoing challenge at the frontier of precision metrology.  To this end, high resolution spectroscopy experiments have made significant contributions over the years.  For example, the coupling strengths and transition energies between atomic and molecular levels are predominantly determined by the electromagnetic interaction.  However, the Standard Model does include fundamental processes, e.g.\ the weak interaction~\cite{Commins1983}, which have spectroscopic signatures that are both theoretically calculable and experimentally detectable.  Parity-violating transition amplitudes, forbidden by the electromagnetic interaction but allowed in the presence of the weak interaction, have been calculated and measured in atomic cesium~\cite{Wood1997,Guena2005} and ytterbium~\cite{Tsigutkin2009} with sufficient precision to test electroweak theory at the $\sim1\%$ level.  In addition, high-resolution molecular spectroscopy experiments are underway to probe parity violation in chiral polyatomic molecules~\cite{Crassous2005,Quack2008,Hegstrom1980,Harris2002} and to probe nuclear spin-dependent parity violation in diatomic molecules~\cite{DeMille2008a,Isaev2010}.  Looking outside of the Standard Model, precision molecular spectroscopy experiments have been designed to search for time-variation of fundamental constants, such as the electron-to-proton mass ratio~\cite{Flambaum2007,Zelevinsky2008,DeMille2008b,Kajita2009,Bakalov2010} and the fine structure constant~\cite{Flambaum2007,Hudson2006}, as well as to search for simultaneous parity and time-reversal symmetry violation in the form of permanent electric dipole moments (EDMs)~\cite{DBB00,HMU02,KBB04,KOD02,PTI05,HST02,SAH06,SHA06,KFD87,DKK92,VCG10,MB08,LEE09,MBD06}.

In most cases, atoms and molecules that are either neutral or ionic can be studied in an effort to observe the same underlying physics; however, typically there are technical advantages to selecting one system over the other.  Systems of neutral, as opposed to ionic, particles are attractive for precision spectroscopic studies due to the relative ease of constructing high-flux neutral particle beams, the relatively weak interactions between neutral particles, and the lack of coupling between the translational motion of neutral particles and external electromagnetic fields.  Conversely, charged particles are favored due to the relative ease of constructing ion traps and the long interrogation times that come with studying trapped particles.  Indeed, some of the most stringent tests of the Standard Model have been performed using trapped ions~\cite{Rainville2005,Gabrielse2006,OHD06,GHK06}, and spectroscopy on trapped molecular ions is of fundamental interest for studying interstellar chemistry~\cite{Schwalm2007,Roth2006,Roueff2009}.  Looking to combine the techniques of ion trapping and high-resolution molecular spectroscopy, several research groups are working to develop experimental platforms for studying ensembles of trapped molecular ions~\cite{Willitsch2008,Staanum2010,Schneider2010,Tong2010,Svendsen2010,Chen2011}.

The additional degrees-of-freedom afforded to molecular systems, in comparison with simple atomic systems, provide additional interaction mechanisms and correspondingly more routes for experimental investigation.  For example, molecular levels are inherently more sensitive to applied electric fields due to the presence of nearby states of opposite parity, e.g.\ rotational levels and/or $\Lambda$-doublet levels.  On the surface, this means that the Stark shifts observed in molecular spectra will be significantly larger than the corresponding shifts to atomic transitions.  More fundamentally, this means that in relative weak electric fields the quantum eigenstates of an atomic system are still dominated by a \textit{single} parity eigenstate, while the quantum eigenstates of molecular systems asymptotically approach an \textit{equal admixture} of even and odd parity eigenstates.  There are several classes of atomic and molecular symmetry violation experiments where larger Stark mixing amplitudes give rise to larger signals.  For example, the parity violation signals already attained in atomic systems~\cite{Wood1997,Guena2005,Tsigutkin2009} are expected to be exceeded by the next-generation of experiments using polarized diatomic molecules~\cite{DeMille2008a,Isaev2010}.  Similarly, in experiments designed to search for permanent electric dipole moments, the expected signal size scales with the ability to thoroughly mix parity eigenstates and increases dramatically when going from atoms to diatomic molecules~\cite{Sandars1967,SAN75,CSH89}.

Herein lies the conundrum for symmetry violation searches using trapped molecular ions: the electric field required to fully polarize the molecules will interfere with the electromagnetic fields necessary for trapping the ions with the likely result of accelerating the ions out of the trap.  Our solution to this problem is to apply an electric field that rotates at radio frequencies.  Under these conditions, the ions will still accelerate, however they will undergo circular motion similar to charged particles in a Penning trap~\cite{Rainville2005,Gabrielse2006,OHD06} or storage ring~\cite{KHR98,FJM04,OMS06,OSH10,AKO10,BBB09}.  The nuances of performing high-resolution electron spin resonance spectroscopy in this environment will be the main focus of this work, with the ultimate goal of demonstrating that such an experiment on the valence electrons in a ground or metastable ${^3}\Delta_1$ level could prove to be a sensitive probe for a permanent electron electric dipole moment (eEDM).

\subsection{Motivation for Electric Dipole Moment Searches}

The powerful techniques of spin resonance spectroscopy, as applied
to electrons, muons, nuclei, and atoms, have made possible
exquisitely precise measurements of electric and magnetic dipole
moments.  These measurements in turn represent some of the most
stringent tests of existing theory, as well as some of the most
sensitive probes for new particle physics.  As an example, the
recent improved measurement of the electron's magnetic
moment~\cite{OHD06} agrees with the predictions~\cite{GHK06} of
quantum electrodynamics out to four-loop corrections.  Compared to
the electron work, muonic g-2 measurements~\cite{BBB02,BBB04} are
less accurate but are nonetheless more sensitive (due to the
muon's greater mass) to physics beyond the Standard Model. Digging
a new-physics signal out of the muon g-2 measurement is made
difficult by uncertainty in the hadronic contributions to the
Standard Model prediction~\cite{HUK99}. One of the primary
motivations for experimental searches for electric dipole moments
(EDM) is the absence of such Standard Model backgrounds to
complicate the interpretation of these studies.  In the case of
the electron, for example, the Standard Model predicts an electric
dipole moment less than $10^{-38}$~e cm~\cite{PK91}. The natural
scale of the electron electric dipole moment (eEDM) predicted by
supersymmetric models is $10^{-29}$ to $10^{-26}$~e
cm~\cite{BES91,BAR93,POR05} (Table~\ref{t:predictions}). The
current experimental limit is $|d_e| < 1.6 \times 10^{-27}$~e
cm~\cite{RCS02}. With predictions of new physics separated by nine
orders of magnitude from those of ``old'' physics, and with the
current experimental situation such that a factor-of-ten
improvement in sensitivity would carve deeply into the predictions
of supersymmetry, an improved measurement of the eEDM is a
tempting experimental goal. In this paper we will describe an
ongoing experiment that we believe will be able to improve on the
existing experimental upper limit for an eEDM by a factor of
thirty in a day of integration time.

\begin{table}
\begin{center}
\caption{Theoretical predictions of the electron electric dipole moment, $d_e$.  Current listings
are taken from Ref.~\cite{CRD94}, which extracted the numbers from
Refs.~\cite{BES91,BAR93}.\label{t:predictions}}
\begin{tabular}{|c|c|}
\hline CP Violating Model & $|d_e|$ [e cm] \\
\hline Standard Model & $|d_e| < 10^{-38}$ \\
\hline Supersymmetric models & $|d_e| < 10^{-27}$ \\
\hline Left-right symmetric models & $10^{-28} < |d_e| < 10^{-26}$ \\
\hline Higgs models& $10^{-28} < |d_e| < 3 \times 10^{-27}$ \\
\hline Lepton flavor changing models & $10^{-29} < |d_e| < 10^{-26}$ \\
\hline
\end{tabular}
\end{center}
\end{table}

\subsection{A Brief Overview of the JILA Experiment}

Our JILA eEDM experiment will be based on electron spin resonance
(ESR) spectroscopy in a sample of trapped diatomic molecular ions.
We will use an $\Lambda$-doubled molecular state that can be
polarized in the lab frame with a lab frame electric field of only
a few volts/cm.  The very large internal electric field of the
molecule, coupled with relativistic effects near the nucleus of a
heavy atom, will lead to a large effective electric field,
$\mathcal{E}_\mathrm{eff}$, on the electron spin.  Confining the
molecules in a trap leads to the possibility of very long
coherence times and therefore high sensitivity.  Trapping of
neutral molecules has been experimentally realized recently, but
it remains an extremely difficult undertaking. Conversely,
trapping of molecular ions is straight forward to implement with
long-established technology.

On the face of it, measuring the electric dipole moment of a
charged object is problematic.  Even for a relatively polarizable
object like a molecule, one must apply sufficient electric field
to mix energy eigenstates of opposite parity.  This field will
cause the ion to accelerate in the lab-frame and limit trapping
time.  We will circumvent this problem via the application of a
rotating electric bias field, which will drive the ion in a
circular orbit.  The rotation rate will be slow enough that the
molecule's polarization can adiabatically follow the electric
field, but rapid enough that the orbit diameter is small compared
to the trap size.  The ESR spectroscopy will be performed in the
rotating frame.  We note that this approach is conceptually
related to efforts measuring electric dipole moments of charged
particles in storage rings~\cite{KHR98,FJM04,OMS06,OSH10,AKO10,BBB09}, but in our
case the radius of the circular trajectory will be measured in
millimeters, not meters. Precision spectroscopy in time-varying
fields can be afflicted with novel sources of decoherence and
systematic error, which will be discussed in
Secs.~\ref{s:spectroscopy}, \ref{s:collisions}, and
\ref{s:conclusion}.

\subsection{A Comparative Survey of Ongoing Experimental Work}

The primary purpose of this section will be to review
experimental searches for eEDM.  We will make no attempt to
survey the rapidly increasing diversity of low-energy \cite{RAS08} and
astrophysical searches for physics beyond the Standard Model.  A
subset of that broad area of endeavor is the search for permanent
electric dipole moments (EDMs), and a subset within that focuses on
electrons (eEDMs). For comparative surveys of the discovery
potential of various EDM studies see \cite{CHU09,JUN07,RIT09,PAU09,WHJ10}, we
summarize here by saying that from the point of view of new
physics, experiments on leptons provide physics constraints
complementary to those on diatomic atoms and to those directly on
bare nucleons and nuclei. As for the lepton experiments, there is
work on the tau lepton \cite{AAA04}, on muons \cite{AKO10,BBB09} and of course
on electrons as discussed in some detail below. The current best
neutron EDM measurement was done at ILL \cite{BDG06}; there
are ongoing neutron EDM searches \cite{BDG06,BBD06,SNS}. Beam-line measurements on bare nucleons are
envisioned \cite{OND06}. The current best atomic dipole
measurement is an experiment is in the diamagnetic species, Hg,
by the Washington group \cite{GSL09}. Many other groups are looking
for EDMs in diamagnetic (that is, net electron spin $S=0$),
ground-state electronic levels in Hg \cite{GSL09}, Xe \cite{YIU08,ROC01,LSR05,Fierlinger}, Rn \cite{TBC08}, Yb \cite{PRP10} and
Ra \cite{SGV09,GSA07,HAB10,WHJ10}. Experiments on diamagnetic atoms (with net
electron spin $S=0$) are sensitive to new physics predominantly via
the nucleonic contribution to the Schiff moment of the
corresponding atomic nucleus. Higher-order contributions from
eEDM contribute to the atomic EDM of $S=0$ atoms \cite{DFP09}, but
these are probably too small to provide a competitive eEDM limit.

For 20 years the most stringent limits on the eEDM have been the
atomic-beam experiments of Commins'
group at Berkeley~\cite{RCS02,CRD94,ACC90}. That work set a standard against
which one can compare ongoing and proposed experiments to improve
the limit.  Here is a brief survey of ongoing experiments of which
we are aware.

For evaluating the sensitivity of an eEDM experiment the key
figure-of-merit is $\mathcal{E}_\mathrm{eff} \tau \sqrt{N}$, where
$\mathcal{E}_\mathrm{eff}$ is the effective electric field on the
unpaired electron, $\tau$ is the coherence time of the resonance,
and $N$ is the number of spin-flips that can be counted in some
reasonable experimental integration time, for instance one week. The statistics-limited sensitivity to the eEDM is just the inverse of our figure-of-merit. We will discuss the three terms in order.

The conceptually simplest version of an eEDM experiment would
simply be to measure the spin-flip frequency of a free electron in
an electric field $\mathcal{E}_\mathrm{lab}$, $\omega_d = d_e
\mathcal{E}_\mathrm{lab}$, where $d_e$ is the electric dipole
moment of the electron~\cite{units}. Alas, a free electron in a
large electric field would not stay still long enough for one to
make a careful measurement of its spin-flip frequency; in practice
all eEDM experiments involve heavy atoms with unpaired electron
spins. An applied laboratory electric field distorts the atomic
wavefunction, and the eEDM contribution to the atomic spin-flip
frequency $\omega_d$ is enhanced by relativistic effects occurring
near the high-Z nucleus~\cite{SAN65,SAN66}, so that $\omega_d =
d_e \mathcal{E}_\mathrm{eff}$, where the effective electric field
$\mathcal{E}_\mathrm{eff}$ can be many times larger than the
laboratory electric field $\mathcal{E}_\mathrm{lab}$.  The
enhancement factor is roughly proportional to $Z^3$ although
details of the atomic structure come into play such that the
enhancement factors for thallium ($Z=81$) and cesium ($Z=55$) are
$-585$~\cite{LIK92} and $+114$~\cite{HLM90}, respectively.
Practical DC electric fields in a laboratory vacuum are limited by
electric breakdown to about $10^5$~V/cm. The Commins experiment
used a very high-Z atom, thallium, and achieved an
$\mathcal{E}_\mathrm{eff}$ of about $7 \times
10^7$~V/cm~\cite{RCS02}.  There have been proposed a number of
experiments in cesium~\cite{AMG07,FFW09,Heinzen} that expect to
achieve $\mathcal{E}_\mathrm{eff}$ of about $10^7$~V/cm. A
completed experiment at Amherst~\cite{MKL89} achieved
$\mathcal{E}_\mathrm{eff}=4.6 \times 10^5$~V/cm in Cs by using
$\mathcal{E}_\mathrm{lab}=4$~kV/cm.

It was pointed out by Sandars~\cite{Sandars1967,SAN75,CSH89} that much larger
$\mathcal{E}_\mathrm{eff}$ can be achieved in polar diatomic
molecules. In these experiments, the atomic wavefunctions of the
high-Z atom are distorted by the effects of a molecular bond,
typically to a much lighter partner atom, rather than by a
laboratory electric field. One still applies a laboratory electric
field, but it need be only large enough to align the polar
molecule in the lab frame.  The Imperial College group~\cite{HST02} is
working with YbF, for which the asymptotic value of
$\mathcal{E}_\mathrm{eff}$ is
26~GV/cm~\cite{HST02,KOE94,TME96,KOZ97,QSG98,PAR98,MKT98}.
The Yale group~\cite{DBB00,HMU02,KBB04} uses PbO, with an
asymptotic value of $\mathcal{E}_\mathrm{eff}$ $\simeq$
25~GV/cm~\cite{KOD02,IPM04,PTI05}.  The Oklahoma
group~\cite{SHA06} has proposed to work with PbF, which has a
limiting value of $\mathcal{E}_\mathrm{eff}$ $\simeq$
29~GV/cm~\cite{KFD87,DKK92}.  The ACME collaboration ~\cite{VCG10}
will use ThO, with $\mathcal{E}_\mathrm{eff}$ $\simeq$
100~GV/cm~\cite{MB08}. The Michigan group is working with WC,
with $\mathcal{E}_\mathrm{eff}$ $\simeq$ 54~GV/cm~\cite{LEE09}. We
will discuss candidate molecules for our experiment in
Sec.~\ref{ss:molecule}; we anticipate having an
$\mathcal{E}_\mathrm{eff}$ of around 25~to
90~GV/cm~\cite{MBD06,MB08,PMI07}.

After $\mathcal{E}_\mathrm{eff}$, the next most important quantity
for comparison is the coherence time $\tau$, which determines the
linewidth in the spectroscopic measurement of $\omega_d$.  In
Commins' beams experiment, $\tau$ was limited by transit time to
2.4~ms. Future beams experiments may do better with a longer beam
line ~\cite{SHA06}, or with a decelerated beam~\cite{TBH04}.
Groups working in laser-cooled cesium anticipate coherence times
of around 1~s, using either a fountain~\cite{AMG07} or an optical
trap ~\cite{FFW09,Heinzen}. The PbO experiment has $\tau$ limited
to $80~\mu$s by spontaneous decay of the metastable electronic
level in which they perform their ESR.  Coherence in ThO
experiment will be limited by the excited-state lifetime to
2~ms~\cite{VCG10}. A now discontinued experiment at Amherst~\cite{MKL89} achieved $\tau = 15$~ms in a vapor cell with
coated walls and a buffer gas. The JILA experiment will work with
trapped ions. The mechanisms that will limit the coherence time in
our trapped ions are discussed in Secs.~\ref{s:spectroscopy} and
\ref{s:collisions}. We anticipate a value in the vicinity of
300~ms.

The quantity $\mathcal{E}_\mathrm{eff}$ converts a hypothetical
value of $d_e$ into a frequency $\omega_d$, and $\tau$ sets the
experimental linewidth of $\omega_d$.  The final component of the overall figure-of-merit is $\sqrt{N}$,
which, assuming good initial polarization, good final-state
sensitivity, and low background counts, determines the fractional
precision by which we can split the resonance line. Since we
have defined $N$ as the number of spin flips counted, detection
efficiency is already folded into the quantity.
Vapor-cell experiments such as those at Amherst or Yale can
achieve very high values of effective $N$, atomic beams machines
are usually somewhat lower, and molecular beams usually lower yet
(due to greater multiplicity of thermally occupied states.) Atomic
fountains and atomic traps have still lower count rates, but the
worst performers in this category are ion traps.  The JILA
experiment may trap as few as 100 ions at a time, and observe only
4 transitions in a second.

The discussion above is summarized in Table~\ref{t:figureofmerit}.
To improve on the experiment of Commins, it is necessary to do
significantly better in at least one of the three main components
of the figure-of-merit.  The various ongoing or proposed eEDM
experiments can be sorted into categories according to the
component or components in which they represent a potential
improvement over the Commins' benchmark.  The prospects of large
improvements in both $\tau$ and $\mathcal{E}_\mathrm{eff}$ put
JILA's experiment in its own category.  This combination means
that our resonance linewidth, expressed in units of a potential
eEDM shift, will be $10^5$ times narrower than was Commins'.
Splitting our resonance line by even a factor of 100 could lead to
an improved limit on the eEDM. This is an advantage we absolutely
must have, because by choosing to work with trapped, charged
molecules, we have guaranteed that our count rate, $\dot{N}$, will
be far smaller than those of any of the competing experiments.

We note that there are in addition ongoing experiments attempting
to measure the eEDM in solid-state
systems~\cite{LAM02,HEC05,BLS06,LIL04}. These experiments may also
realize very high sensitivity, but because they are not strictly
speaking spectroscopic measurements, it is not easy to compare
them to the other proposals by means of the same figure-of-merit.

Finally, atoms with diamagnetic ground states may have  $S\neq0$
metastable states amenable to an eEDM search \cite{PLS70}. Closely spaced opposite parity states in Ra can give
rise to an $\mathcal{E}_\mathrm{eff}$ on the electron spin larger \cite{DFG00} than in Tl or Cs, but very short coherence times \cite{DFG00} may make complicate efforts \cite{WHJ10} to measure the eEDM in Ra.

\begin{table*}
\begin{center}
\caption{Figure-of-merit comparison between several recently
completed and ongoing eEDM experiments.  For ongoing experiments
these numbers are subject to change and are often
order-of-magnitude estimates.  For the JILA entry M is Hf, Th, or
Pt and x is H or F.\label{t:figureofmerit}}
\begin{tabular}{|c|c|c|c|c|c|c|}
\hline Group & Refs. & Species & $\mathcal{E}_\mathrm{lab}$ [V/cm] & $\mathcal{E}_\mathrm{eff}$ [V/cm] & $\tau$ [s] & $\dot{N}$ [s$^{-1}$] \\
\hline Berkeley & \cite{RCS02} & Tl & $1.23 \times 10^5$ & $7 \times 10^7$ & $2.4 \times 10^{-3}$ & $10^9$ \\
\hline Amherst & \cite{MKL89} & Cs & $4 \times 10^3$ & $4.6 \times 10^5$ & $1.5 \times 10^{-2}$ &   \\
\hline LBNL & \cite{AMG07} & Cs & $10^5$ & $10^7$ & $1$ & $10^9$ \\
\hline Texas & \cite{Heinzen} & Cs & $10^5$ & $10^7$ & $1$ &  \\
\hline Penn State & \cite{FFW09} & Cs & $10^5$ & $10^7$ & $1$ &  \\
\hline Yale & \cite{DBB00,HMU02,KBB04,KOD02,PTI05} & PbO & $10$ & $2.5 \times 10^{10}$ & $8 \times 10^{-5}$ &  \\
\hline Imperial & \cite{HST02,SAH06} & YbF & $8.3 \times 10^3$ & $1.3 \times 10^{10}$ & $10^{-3}$ &  \\
\hline Oklahoma & \cite{SHA06,KFD87,DKK92} & PbF & $7 \times 10^4$ & $2.9 \times 10^{10}$ &  &  \\
\hline ACME & \cite{VCG10,MB08} & ThO & $10^2$ & $10^{11}$ & $2 \times 10^{-3}$ & $10^5$ \\
\hline Michigan & \cite{LEE09} & WC & $$ & $5.4\times 10^{10}$ & $10^{-3}$ & $$ \\
\hline JILA & This work & Mx$^+$ & 5 & $3-9\times 10^{10}$ & $0.2-1$ & $\sim$ 10 \\
\hline
\end{tabular}
\end{center}
\end{table*}

\subsection{Outline}

A brief overview on the molecular level structure where the eEDM
will be measured and on how the measurement will be performed is
given below in Sec.~\ref{s:idea}.  Some aspects of the
experimental design, including production of molecular ions and
ion trapping will be covered in Sec.~\ref{s:apparatus}.
 Difficulties in performing precision spectroscopy in time-varying
and inhomogeneous electric and magnetic fields will be discussed
in Sec.~\ref{s:spectroscopy}. This will include discussions of
trap imperfections, stray magnetic fields, and effects of rotating
bias fields. Experimental chops used to minimize systematic errors
will also be explained.  In Sec.~\ref{s:collisions}, the effects
on spin coherence time and systematic errors of ion-ion collisions
will be investigated.  An estimate for experimental sensitivity to
the eEDM will be given in Sec.~\ref{s:conclusion}. The Appendix
gives a listing of variables used throughout the paper and a
sample set of experimental parameters.

\section{Molecular Structure and the Basic Spectroscopic Idea}
\label{s:idea}

\subsection{Molecular Notation}
\label{ss:notation}

As we prepare this paper, we have not made a final decision as to
which molecule we will use.  For reasons discussed below, the main
candidates are diatomic molecular ions Mx$^+$, where M = Hf, Pt,
or Th and x = H or F.  In the case of molecules such as HfF$^{\rm
+}$, {\it ab initio} methods~\cite{MBD06,PMI07} enable us to
determine that the ${}^3\Delta$ state is well described by a set
of Hund's case (a) quantum numbers:
$J,S,\Sigma,\Lambda,\Omega,M_J,e/f$.  Here $J$ is the sum of
electronic plus rotational angular momentum, $S$ the total
electronic spin angular momentum, $\Sigma$ the projection of $S$
onto the molecular axis, $\Lambda$ the projection of $L$, the
electronic orbital angular momentum, onto the molecular axis, and
$\Omega$ the projection of $J$ onto the molecular axis. In a case
(a) ${}^3\Delta$ molecule $|\Omega|$ can take the values one, two
or three.  $M_J$ is the projection of $J$ along the quantization
axis and the labels $e/f$ specify the parity of the molecular
state.

In addition to these quantum numbers, the experiment will be
concerned with the nuclear spin quantum number $I$, the total
angular momentum quantum number $F$, given by the vector sum of
$J$ and $I$, and $m_F$ the projection of $F$ along the
quantization axis. Throughout this paper we shall assume a total
nuclear spin of $I=1/2$, the nuclear spin of fluorine or hydrogen.
This leads to the values $F=3/2$ and $F=1/2$ for the states of
experimental interest.

\subsection{Choosing a Molecule}
\label{ss:molecule}

In selecting a molecular ion for this experiment we have several
criteria.  First, we want a simple spectrum.  Ideally, we would
like the supersonic expansion to be able to cool the molecules
into a single internal quantum state so that every trapped
molecule could contribute to the contrast of the spectroscopic
transition. Failing that, we want to minimize the partition
function by using a molecule with a large rotational constant,
most likely a diatomic molecule with one of its atoms being
relatively light. Small or vanishing nuclear spin is to be
preferred, as are atoms with only one abundant isotope. Second, we
need to be able to make the molecule.  This requirement favors
more deeply bound molecules and is the main reason we anticipate
working with fluorides rather than hydrides. Third, the molecule
should be polarizable with a small applied electric field, i.e.\
it should have a relatively small $\Lambda$-doublet splitting,
$\omega_\mathrm{ef}$. Fourth, and most important, the molecule
should have unpaired electron spin that experiences a large value
of $\mathcal{E}_\mathrm{eff}$.

These latter two requirements would appear to be mutually
exclusive: a small $\Lambda$-doublet splitting requires a large
electronic orbital angular momentum, which prohibits good overlap
with the nucleus required for a large $\mathcal{E}_\mathrm{eff}$.
Fortunately, working with two valance electrons in a triplet state
allows us to satisfy our needs.  One valance electron can carry a
large orbital angular momentum making the molecule easily
polarizable, while the other can carry zero orbital angular
momentum giving it good overlap with the nucleus and generating a
large $\mathcal{E}_\mathrm{eff}$.  This concept was detailed by
some of us in Ref.~\cite{MBD06} and for the $^3\Delta_1$ state of
interest here, the two valance electrons occupy molecular $\sigma$
and $\delta$ orbitals.  Our calculations, as well as those of
Ref.~\cite{PMI07}, indicate that in the $^3\Delta_1$ state of
ThF$^+$ and HfF$^+$ we should expect $\omega_\mathrm{ef} \lesssim
2\pi \times 40$~kHz with $\mathcal{E}_\mathrm{eff} \approx
90$~GV/cm for ThF$^+$ and $\mathcal{E}_\mathrm{eff} \approx
30$~GV/cm for HfF$^+$~\cite{MB08,PMI07}.

\subsection{$|\Omega|=1$ vs.\ $|\Omega|=3$}

We mention one final valuable feature we look for in a candidate
molecule: a small magnetic g-factor, so as to reduce the
vulnerability to decoherence and systematic errors arising from
magnetic fields.  To the extent that spin-orbit mixing does not
mix other $|\Omega|=1$ states into a nominally $^3\Delta_1$
molecular level, it will have a very small magnetic moment, a
feature shared by PbF in the $^2\Pi_{1/2}$ state~\cite{SHA06}.
This is because $\Sigma = -\Lambda/2$, and because the spin
g-factor is $\sim$2 times the orbital g-factor. Under these
conditions, the contributions of the electronic spin and orbital
angular momentum to the net molecular magnetic dipole moment
nominally cancel.  In HfF$^+$, the magnetic moment of a stretched
magnetic sublevel level of the $^3\Delta_1,\ J=1$ rotational
ground state is about $0.05~ \mu_B$. This is a factor of 20 less
than the magnetic moment of ground state atomic cesium. In the
$^3\Delta_3$ level, on the other hand, the magnetic moment in the
stretched zeeman level is $4.0 ~\mu_B$. The $|\Omega| = 3$ state
may nonetheless be of scientific interest. The $^3\Delta_1$ and
$^3\Delta_3$ levels have $\mathcal{E}_\mathrm{eff}$ equal in
magnitude but opposite in sign. If one could accurately measure
the science signal, $\omega_d$, in the $^3\Delta_3$ level despite
its larger sensitivity to magnetic field background (and despite
its shorter spontaneous-decay lifetime), the comparison with the
$^3\Delta_1$ result would allow one to reject many systematic
errors.

\subsection{$|\Omega|=1,\ J=1\ \Lambda$-doublet}

Since we have not made a final decision as to which molecule we
will use, and also because we have yet to measure the hyperfine
constants of our candidate molecules, the discussion of level
schemes in this section will be qualitative in nature, usually
emphasizing general properties shared by all the molecules we are
investigating.  To simplify the discussion, we will specialize to
discussing spectroscopy within the $J=1$ rotational manifold of a
molecular $^3\Delta_1$ level.

\begin{figure*}
\begin{center}
\includegraphics{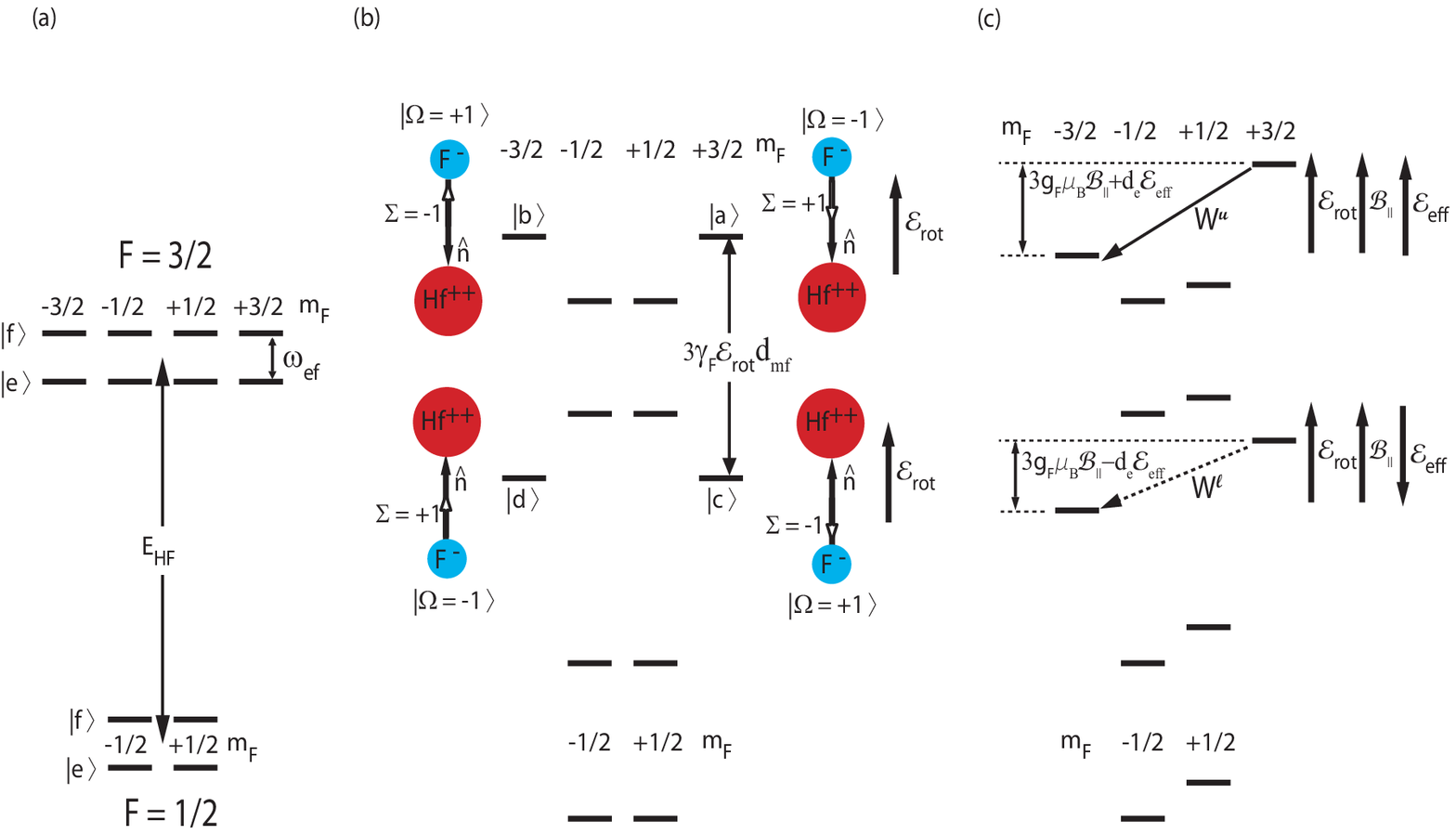}
\caption{Energy levels of HfF$^+$ in the $^3\Delta_1,\ J=1$ state
including hyperfine structure associated with the fluorine $I=1/2$
nucleus.  $\Lambda$ and $\Sigma$ are defined as the projection
along the molecular axis of the electronic orbital angular
momentum, and spin, respectively. $\Omega=\Lambda+\Sigma$.  (a) In
zero electric field, the eigenstates of the system are states of
good parity, $|e\rangle = (|\Omega=+1\rangle -
|\Omega=-1\rangle)/\sqrt{2}$ and $|f\rangle = (|\Omega=+1\rangle +
|\Omega=-1\rangle)/\sqrt{2}$, separated by a small
$\Lambda$-doublet splitting.  (b)  An electric field,
$\mathcal{E}_{\rm rot}$, mixes the parity eigenstates yielding
states with well defined $\Omega$. (c) A small magnetic field
lifts the degeneracy between states with the same value of
$m_F\Omega$.  A permanent electron electric dipole moment further
breaks this degeneracy, but with opposite sign for the upper
(solid arrow) and lower (dotted arrow) transition. Energy
splittings not to scale.\label{f:hyperfine_stutz}}
\end{center}
\end{figure*}

For Hunds' case (a) molecular levels with $|\Lambda| \geq 1$, each
rotational level is a $\Lambda$-doublet, that is, it consists of
two closely spaced levels of opposite parity.  We can think of the
even (odd) parity level as the symmetric (antisymmetric)
superposition of the electronic angular momentum lying
predominantly parallel and antiparallel to the molecular axis
[Fig.~\ref{f:hyperfine_stutz}(a)].  The parity doublet is split by
the $\Lambda$-doubling energy $\omega_\mathrm{ef}$.  A polar
diatomic molecule will have a permanent electric dipole moment,
$\vec{d}_\mathrm{mf}$, aligned along the internuclear axis
$\hat{n}$, but in states of good parity, there will be vanishing
expectation value $\langle \hat{n} \rangle$ in the lab frame.  An
applied laboratory electric field, $\mathcal{E}_\mathrm{rot}$,
will act on $d_\mathrm{mf}$ to mix the states of good parity.  In
the limit of $d_\mathrm{mf} \mathcal{E}_\mathrm{rot} \gg
\omega_\mathrm{ef}$, energy eigenstates will have nonvanishing
$\langle \hat{n} \rangle$ in the lab frame. More to the point,
$\Omega$, a signed quantity given by the projection of the
electron angular momentum on the molecular axis, $(\vec{L} +
\vec{S}) \cdot \hat{n}$, can also have a nonzero expectation value
[Fig.~\ref{f:hyperfine_stutz}(b)]. Heuristically, it is the large
electric fields developed internal to the molecule, along
$\hat{n}$, that gives rise to the large value of
$\mathcal{E}_\mathrm{eff}$ that the electron spin can experience
in polar molecules.  In the absence of the $\Lambda$-doublet
mechanism for polarizing the molecule, a much larger field would
be necessary, $d_\mathrm{mf} \mathcal{E}_\mathrm{lab} \gg 2B_e$,
to mix rotational states with splitting typically twice the
rotational constant $B_e$.  For HfF$^+$, we estimate
$\omega_\mathrm{ef}$ will be $2\pi \times 10$~kHz, whereas $B_e$
will be about $2\pi \times 10$~GHz.  For a dipole moment
$d_\mathrm{mf} = 4.3$~D, mixing the $\Lambda$-doublet levels will
take a field well under 1~V/cm, whereas ``brute force'' mixing of
rotational levels would require around 10~kV/cm. For an experiment
on trapped ions, the smaller electric fields are essential.

In the context of their eEDM experiment on the $a^3\Sigma_1$ level
in PbO, DeMille and his colleagues have explored in some
detail~\cite{DBB00,HMU02,KBB04} the convenient features of an
$|\Omega|=1,\ J=1$ state, especially with respect to the
suppression of systematic error. Our proposal liberally borrows
from those ideas.  In a molecule with at least one high-Z atom,
$^3\Delta_1$ states will be very similar to the $a{^3\Sigma}_1$
state of PbO, but with typically smaller values of
$\omega_\mathrm{ef}$ and much smaller values of magnetic g-factor.
Singly charged molecules with spin triplet states will necessarily
have an odd-Z atom, and thus the unavoidable complication of
hyperfine structure, not present in PbO.

In Fig.~\ref{f:hyperfine_stutz} we present the $^3\Delta_1$, J = 1
state with hyperfine splitting due to the fluorine I=1/2 nucleus.
 A key feature is the existence of two near-identical pairs of
$m_F$-levels with opposite parity.  As seen in
Fig.~\ref{f:hyperfine_stutz}(b), an external electric field,
$\mathcal{E}_{\rm rot}$, mixes these opposite parity states to
yield pairs of $m_F$-levels with opposite sign of
$\mathcal{E}_\mathrm{eff}$~\cite{KBB04} relative to the external
field.  Fig.~\ref{f:hyperfine_stutz}(c) shows the effect of a
rotating magnetic bias field, parallel with the electric field,
applied to break a degeneracy as described in
Sec.~\ref{ss:bigangle} below. Note that any two levels connected
by arrows in Fig.~\ref{f:hyperfine_stutz}(c) transform into each
other under time reversal.  Time reversal takes $m_F \rightarrow
-m_F$, $\Omega \rightarrow -\Omega$, and $\mathcal{B} \rightarrow
\mathcal{-B}$, where $\mathcal{B}$ is the magnetic field.  If we
measure the resonant frequency for the transition indicated by the
solid (or dashed) line once before and once after inverting the
direction of the magnetic field, time reversal invariance tells us
the difference between the two measurements should be zero. In the
presence of an eEDM, which violates time-reversal invariance, this
energy difference $W^u(\mathcal{B})-W^u(\mathcal{-B})$ will give
$2 d_e \mathcal{E}_\mathrm{eff}$. As well, under the same magnetic
field the transitions indicated by the solid and dashed lines
should be degenerate, if the magnetic g-factors are identical for
the states involved~\cite{gfactors}. With non-zero eEDM the energy
difference $W^u-W^l$ also gives $2 d_e \mathcal{E}_\mathrm{eff}$.

Potential additional shifts, due predominantly to Berry's
phase~\cite{BER84}, are discussed in Sec.~\ref{s:spectroscopy} but
for now we note only that in the absence of new physics (such as a
nonzero eEDM) the energy levels of a molecule in time-varying
electromagnetic fields obey time-reversal symmetry.  Reversing the
direction of the electric field rotation while chopping the sign
of the magnetic field amounts to cleanly reversing the direction
of time, and will leave certain transition energies rigorously
unchanged if $d_e=0$.  These are our ``science transitions'',
which we will measure with our highest precision.

\subsection{Electronic Levels, Spin Preparation, and Spin Readout}

\begin{figure}
\begin{center}
\includegraphics[width=8cm]{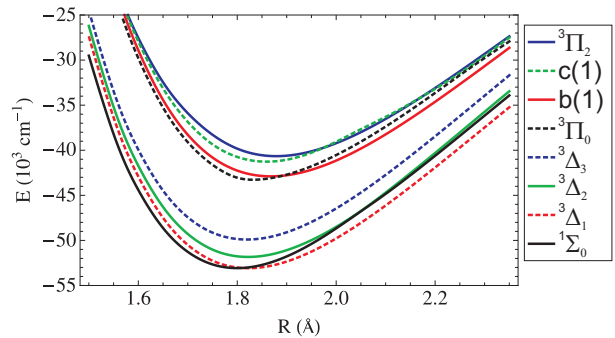}
\caption{Potential energy curves for select states of
HfF$^+$~\cite{MBD06}.  The b(1) and c(1) states are well-mixed
combinations of $^1\Pi_1, ^3\Pi_1$, and $^3\Sigma^-_1$
states.\label{f:HfH}}
\end{center}
\end{figure}

The density of trapped molecular ions will be too low to permit
direct detection of the radio frequency or microwave science
transitions.  (A possible exception could involve the use of a
superconducting microwave cavity, but this would add considerable
experimental complexity.)  We will of necessity rely on electronic
transitions to prepare the initial electron spin state, and on a
double resonance method to detect the spin flips. The details of
these steps will depend on the specific molecule we use.  For a
qualitative illustration, we present a schematic of the calculated
low-lying electronic potential curves of HfF$^+$
(Fig.~\ref{f:HfH}).  We note that HfH$^+$ and ThF$^+$ have similar
level structures~\cite{MBD06,PMI07}.

The molecules will be formed by laser ablation and cooled by
supersonic expansion such that a large portion of the molecular
population will be in $^1\Sigma_0$ ground state with a few
rotational levels occupied (Sec.~\ref{ss:beamline}). Spin-orbit
mixing between states of identical $|\Omega|$ are enhanced by
relativistic effects in the high-Z Hf atom.  The b(1) and c(1)
states are well-mixed combinations of $^1\Pi_1, ^3\Pi_1$, and
$^3\Sigma^-_1$ states, allowing for electric dipole transitions to
and from these states that do not respect spin selection rules.
The $^1\Sigma_0$ state, on the other hand, has no nearby
$|\Omega|=0$ state with which to mix, and thus $\Sigma$ and
$\Lambda$ are good quantum numbers. Similarly, the $^3\Delta_1$
state has so little contamination of $^1\Pi_1$ in it that a rough
calculation indicates that it is metastable against spontaneous
decay, with a lifetime of order 300~ms~\cite{MBD06,PMI07}.

The Ramsey resonance experiment will begin with a two-photon,
stimulated Raman pulse, off-resonant from the intermediate
$^{1,3}\Pi_1$ states, which will coherently transfer population
from the $^1\Sigma_0,\ J=0$ ground state to the two $|m_F|=3/2$
magnetic sublevels of the $^3\Delta_1,\ J=1$ level.  The relative
phase between the two magnetic levels evolves at a rate given by
the energy difference. After a variable dwell time, a second Raman
pulse is applied, which will coherently transfer a fraction of the
population back down to the $^1\Sigma_0$ state, with probability
determined by the accumulated relative phase.  By varying the
dwell time between Raman pulses, the population in the
$^1\Sigma_0$ state will oscillate at a frequency given by the
energy difference between the two spin states in the $^3\Delta_1$
manifold.

The final step in the resonance experiment is to measure the
number of molecules remaining in the $^3\Delta_1$ state.  This we
propose to do with state-selective photodissociation. Molecules in
the $^3\Delta_1$ state will be dissociated via a two-color pulse,
back up through the $^3\Pi_1$ state to a repulsive curve,
generating a Hf$^+$ atomic ion and a neutral fluorine atom.
Molecules in the $^1\Sigma_0$ state will not be affected by the
two-color laser pulse and will remain as HfF$^+$ molecular ions.
The Paul trap parameters will be adjusted to confine only ions
with the Hf$^+$ atomic mass, and not the HfF$^+$ molecular mass
with mass difference $\Delta M = 19$~amu. Finally, the potential
on an endcap electrode will be lowered, and the remaining ions in
the trap will be dumped onto a ion-counting device.

Details of this procedure will depend on the molecule ultimately
selected for this experiment.  We are also investigating
alternative modes of spin state readout, including
large-solid-angle collection of laser-induced fluorescence, and
high finesse optical cavities~\cite{YEH00}.

\section{Experimental Apparatus}
\label{s:apparatus}

\subsection{Molecular Beamline}
\label{ss:beamline}

\begin{figure*}
\begin{center}
\includegraphics[width=6in]{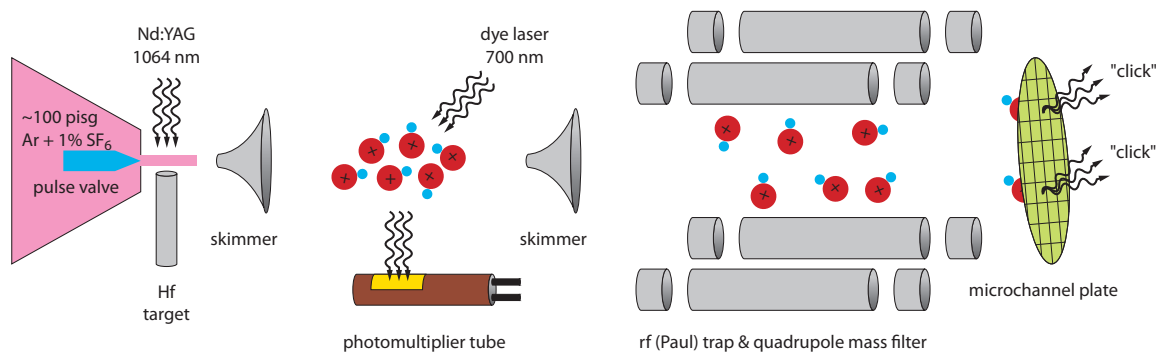}
\caption{Experimental setup.  Laser ablation of a metal Hf target
creates neutral Hf atoms and Hf$^+$ ions that react with SF$_6$ to
produce neutral HfF molecules and HfF$^+$ molecular ions,
respectively (Eqs.~\ref{e:reaction1} and \ref{e:reaction2}). The
molecules (both neutral and ionic) are cooled in a supersonic
expansion with a He buffer gas. The molecular beam is illuminated
with a pulse dye laser beam and the resulting fluorescence is
collected with a photomultiplier tube (PMT) yielding laser induced
fluorescence (LIF) spectra (Fig.~\ref{f:lif}).  At the end of the
beamline, the ions are loaded into an rf (Paul) trap where the
electron spin resonance experiment is performed.  The Paul trap
also acts as a quadrupole mass filter and ions of a particular
mass/charge ratio are detected with a microchannel plate (MCP)
(Fig.~\ref{f:massspec}). Additionally, the spatial resolution of
the MCP allows for the temperature of the ion cloud to be
determined from the detected cloud size.\label{f:beamline}}
\end{center}
\end{figure*}

We are interested in studying molecular radicals and therefore
must create the molecules \textit{in situ}.  As described in
Sec.~\ref{ss:molecule}, we have a small collection of molecules
that satisfy our selection criteria and our final choice of
molecule has not been made.  However, for clarity this section
will describe the production, detection, and characterization of a
beam containing neutral HfF molecules and HfF$^+$ molecular ions.

The molecules are made in a pulsed supersonic expansion
(Fig.~\ref{f:beamline}).  A pulse valve isolates $\sim$ 7
atmospheres of argon that is seeded with 1\% sulfur hexafluoride
(SF$_6$) gas from the vacuum chamber.  The pulse valve opens for
$\sim 200\ \mu$s allowing the Ar + 1\% SF$_6$ mixture to expand
into the vacuum chamber.  This creates a gas pulse moving at
550~m/s in the laboratory frame, but in the co-moving frame the
expansion cools the translational temperature of the Ar atoms to a
few Kelvin.

Immediately after entering the vacuum chamber, the gas pulse
passes over a Hf metal surface.  Neutral Hf atoms and Hf$^+$ ions
are ablated from this surface with a 50~mJ, 10~ns, 1064~nm Nd:YAG
laser pulse.  The ablation plume is entrained in the Ar + 1\%
SF$_6$ gas pulse and the following exothermic chemical reactions
occur:
\begin{eqnarray}
\label{e:reaction1}
    \textrm{Hf} + \textrm{SF}_6 & \longrightarrow & \textrm{HfF} + \textrm{SF}_5,\\
\label{e:reaction2}
    \textrm{Hf}^+ + \textrm{SF}_6 & \longrightarrow & \textrm{HfF}^+ + \textrm{SF}_5,
\end{eqnarray}
In the co-moving frame, the resulting neutral HfF molecules and
HfF$^+$ molecular ions are cooled through collisions with the Ar
gas to rotational, vibrational, and translational temperatures of
order a few Kelvin.  The molecular beam then passes through a
skimmer, first entering a region where laser induced fluorescence
(LIF) spectroscopy is performed and finally arriving at an rf
(Paul) trap where the ions are stopped and confined.

LIF spectroscopy is performed by transversely illuminating the
molecular beam with a $\sim 500\ \mu$J, 10~ns, $\sim 700$~nm dye
laser pulse. The linewidth of the dye laser is specified to be
less than 0.1~cm$^{-1}$.  Fluorescence photons are collected and
imaged onto a photomultiplier tube (PMT).

\begin{figure}
\begin{center}
\includegraphics{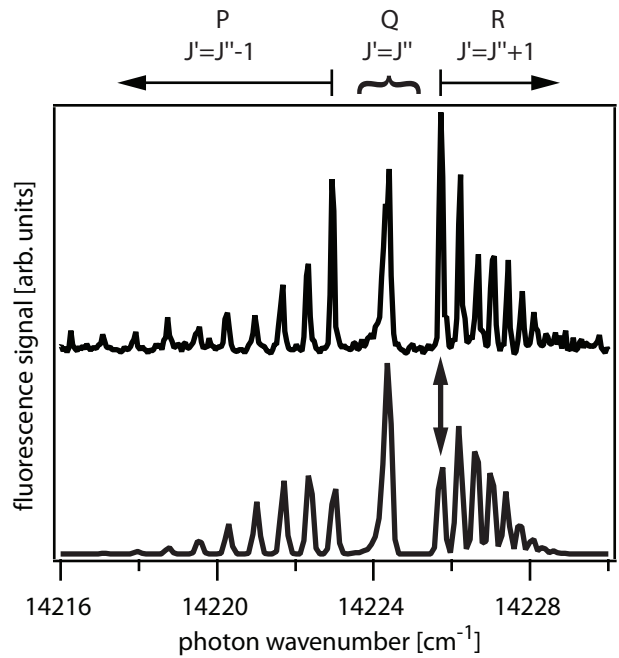}
\caption{Laser induced fluorescence (LIF) spectroscopy. The top trace is experimental data for a newly detected neutral
HfF transition: $[14.2]\ |\Omega|=3/2\ |v'=v'',J'\rangle
\leftarrow X{^2\Delta}_{3/2}\ |v'',J''\rangle$.  The transition
highlighted with a vertical arrow originates from the
rotational ground state. The bottom trace is a theoretical prediction
assuming a rotational temperature of $5$~K. The traces are offset
vertically for clarity.\label{f:lif}}
\end{center}
\end{figure}

Using this technique we have found previously unobserved neutral
HfF molecular transitions, one of which is shown in
Fig.~\ref{f:lif} (for previous neutral HfF spectroscopy see
Ref.~\cite{AHT04}). The data shows that entrained neutral HfF
molecules are cooled to rotational temperatures of order 5~K, with
a large fraction of the population in the rotational ground state.
We expect that entrained HfF$^+$ molecular ions should be
similarly cooled.

\begin{figure}
\begin{center}
\includegraphics{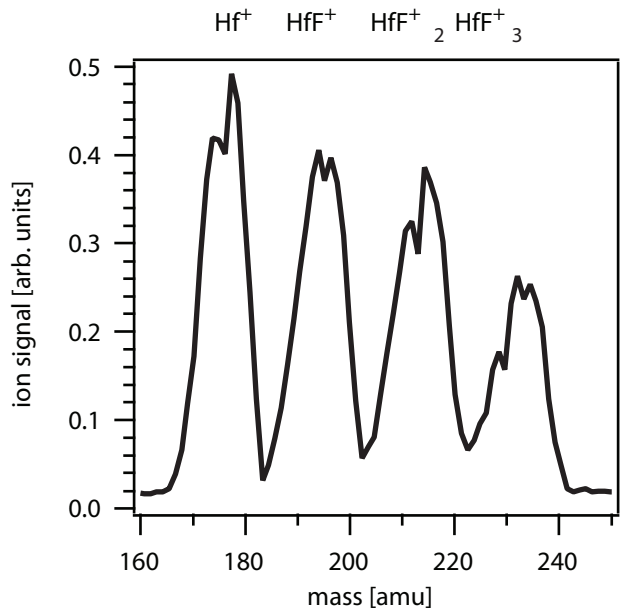}
\caption{Mass spectrometry.  Operating the rf (Paul) trap as a
quadrupole mass filter gives mass-dependent trapping potentials
such that Hf$^+$ ($M=180$~amu), HfF$^+$ ($M=199$~amu), HfF$^+_2$
($M=218$~amu), and HfF$^+_3$ ($M=237$~amu) can be separately
trapped and detected.  The ion detector signal is a non-linear
function of ion number, but a level of 0.4 corresponds to $\sim
100,000$ ions.\label{f:massspec}}
\end{center}
\end{figure}

To detect the presence of HfF$^+$ molecular ions in the beam the
rf (Paul) trap is operated as a quadrupole mass filter.  All of
the ions in the beam are stopped and loaded into the trap. The
voltages applied to the trap electrodes are then adjusted only to
confine ions of a particular mass/charge ratio.  Finally, the ions
remaining in the trap are released onto the ion detector and
counted.  A typical mass spectrum is shown in
Fig.~\ref{f:massspec}, which clearly resolves the HfF$^+$
molecular ions from the other atomic and molecular ions in the
trap.

Our experimental count rate will be limited by space charge
effects of the trapped ions.  Therefore, any ions trapped that are
not used in measuring the eEDM limit the statistical sensitivity
of our measurement.  In order to maximize our count rate, we wish
to create and trap only HfF$^+$ ions of a single Hf isotope and in
a single internal quantum state. One scheme is to filter out all
of the ions created from laser ablation and use photoionization
techniques to ionize neutral HfF in as state-selective a way as
possible. Using two color, two photon excitation, we excite to a
high lying Rydberg state, in an excited vibrational level, that
then undergoes vibrational autoionization~\cite{Field}. The ion
core of these Rydberg state molecules will occupy a single
rotational level and consist of a single Hf isotope.  The
autoionization process is seen, in our preliminary (unpublished)
data, to leave the ion core rotational level largely unperturbed.
It should be possible to excite a Rydberg level that corresponds
to an excited $^3\Delta_1$ ion core with $v$ = 1, $J$ = 1 (where v
is the vibrational quantum number).  The Rydberg state might then
vibrationally autoionize to the $v$ = 0, $J$ = 1 $^3\Delta_1$
level that will be used to measure the eEDM.

\subsection{Radio Frequency (Paul) Trap}
\label{ss:paultrap}

\begin{figure}
  \begin{center}
    \includegraphics{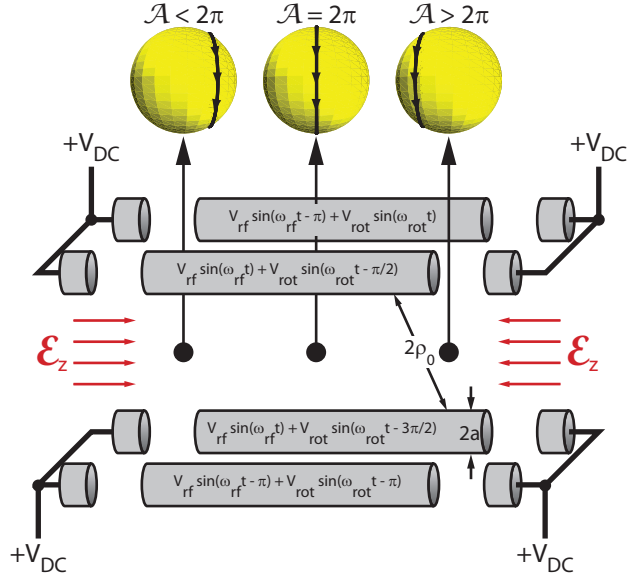}
    \caption{Linear rf (Paul) trap.  Neighboring cylindrical electrodes
      are driven with rf voltages $180^\circ$ out of phase. Axial
      confinement is provided by d.c.\ voltages applied to the end cap
      electrodes.  The cylindrical electrode rods have radius $a$ and
      the radial distance from the trap center to the nearest electrode
      surface is $\rho_0$.  See Ref.~\cite{PAU90} for further details of
      rf (Paul) trap operation. In addition to the voltages oscillating
      at $\omega_\mathrm{rf}$, there is also
      a component of the voltages oscillating at
      $\omega_\mathrm{rot}$. Over a period of time $2\pi/\omega_\mathrm{rot}$,
      the electric field at the axial
      center (z=0) of the trap will trace out a trajectory which
      subtends a solid angle $\mathcal{A}$ of exactly $2\pi$.  Ions
       to the left (right) of trap center will experience an electric
       field whose trajectory  subtends slightly less (greater than)
       $2\pi$. Consequences of this time variation are explored in
       discussed in Sec.~\ref{ss:rotatingfields}
       and~\ref{ss:bigangle}.  Not to scale.
      \label{f:axialosc2}}
  \end{center}
\end{figure}

For our preliminary studies of ion production, the ions are
confined by a linear rf (Paul) trap shown schematically in
Fig.~\ref{f:axialosc2}. The ideal hyperbolic electrodes are
replaced by cylinders of radius $a \approx 1.15\rho_0$, where
$\rho_0$ is the minimum radial separation between the trap center
and the surface of the electrodes.  This choice produces the best
approximation to a perfect radial two-dimensional electric
quadrupole field~\cite{DEN71}.

For HfF$^+$ ($M=199$~amu), an example set of operating parameters
for the ion trap would be $\rho_0=25$~mm, $V_\mathrm{rf}=550$~mV,
and $\omega_\mathrm{rf}=2\pi \times 15$~kHz.  This produces a
ponderomotive potential that is well within the harmonic
pseudo-potential approximation given by $U_\mathrm{rf}(\rho) =
M\omega_\mathrm{sec}^2\rho^2/2$, where the radial secular
frequency is approximately
$\omega_\mathrm{sec}=q\omega_\mathrm{rf}/\sqrt{8}$ with
$q=4eV_\mathrm{rf}/M\rho_0^2\omega_\mathrm{rf}^2$.  For the above
parameters, $q=0.2$, $\omega_\mathrm{sec}=2\pi \times 1$~kHz, and
$U_\mathrm{rf}(\rho_0) = 300$~K.  Under these conditions, an ion
cloud at a temperature of 15~K would have an rms radius of 5~mm.
The trap can also be operated in mass filter mode~\cite{DAW95}.

In addition to supplying the oscillating electric quadrupole field
for radial confinement, the cylindrical electrodes can also be
driven with voltages to produce the rotating electric bias field,
$\mathcal{E}_\mathrm{rot}$, needed to polarize the molecular ions
(Fig.~\ref{f:axialosc2}).  In order to generate
$\mathcal{E}_\mathrm{rot}$ neighboring electrodes will be driven
$90^\circ$ out of phase at a frequency $\omega_\mathrm{rot}$.  The
net voltage applied to each electrode is the sum of the voltages
$V_\mathrm{rf} + V_\mathrm{rot}$.

At present we are designing a second-generation ion trap with
geometry designed for optimal precision eEDM spectroscopy, rather
than for mass selection.  The perfect ion trap would have very
large optical access for collection of laser-induced fluorescence,
and idealized electric and magnetic fields as follows
\begin{equation}
\label{e:IdealEFields} \vec{\mathcal{E}} =
\mathcal{E}_\mathrm{rot} \hat{\rho}^{\prime} +
    \mathcal{E}^{\prime}_\mathrm{rf} (x \hat{x} - y \hat{y}) \cos
      (\omega_\mathrm{rf} t ) + \mathcal{E}^{\prime}_z (-z \hat{z} + y \hat{y}/2 + x
      \hat{x}/2)
\end{equation}
\begin{equation}
\label{e:IdealBFields}
 \vec {\mathcal B} = {\mathcal B}_\mathrm{rot} \hat{\rho}^{\prime}
\end{equation}
where $\hat{\rho}^{\prime} =  \cos (\omega_\mathrm{rot} t )
\hat{x} + \sin(\omega_\mathrm{rot} t ) \hat{y}$ and ${\mathcal
E}^\prime_{\rm rf}\approx-2V_{\rm rf}/\rho_0^2$.

If we assume $\omega_\mathrm{rot} \gg \omega_\mathrm{rf}$, that
$\omega_\mathrm{rot} / \omega_\mathrm{rf}$ is not a rational
fraction, and that $\omega_\mathrm{rf}^2 \gg e {\mathcal
E}^\prime_{\rm rf}/M$, then we can cleanly separate out the ion
motion into three components: rf micromotion, circular
micromotion, and secular motion.

rf micromotion involves a rapid oscillation at $\omega_{\rm rf}$
whose amplitude grows as the ion's secular trajectory takes it
away from trap center.   The kinetic energy of this motion,
averaged over an rf cycle, is given by
\begin{equation}
  \label{e:rfPonderomotive}
  {\rm E}_\mathrm{rf} = (x^2 + y^2)
  \frac{e^2\mathcal{E_{\mathrm{rf}}^{\prime}}^2}{4 M
  \omega_{\mathrm{rf}}^2}
\end{equation}
where x and y in this case refer to the displacement of the ion's
secular motion.

The displacement of the ion's circular micromotion is given by
\begin{equation}
  \label{e:rotDisplacement}
  \vec{r}_\mathrm{rot} = -\frac{e\vec{\mathcal{E}}_\mathrm{rot}}{M \omega_\mathrm{rot}^2}.
\end{equation}
The kinetic energy of the circular motion, averaged over a
rotation cycle, is given by
\begin{equation}
  \label{e:rotPonderomotive} {\rm E}_\mathrm{rot} =
  \frac{e^2 {\mathcal E}_\mathrm{rot}^2}{2 M \omega_\mathrm{rot}^2}.
\end{equation}
The time-averaged kinetic energies of the two micromotions act as
ponderomotive potentials that contribute to the potential that
determines the relatively slowly varying secular motion:
\begin{equation}
  \label{e:SecularPotential}
  U_\mathrm{sec} ={\rm E}_\mathrm{rot} (x,y,z) +
  {\rm E}_\mathrm{rf} (x,y,z) +  e \mathcal{E}^{\prime}_z (2z^2 - y^2 -
  x^2)/4.
\end{equation}
In the idealized case, the secular motion corresponds to 3-d
harmonic confinement with secular or ``confining" frequencies
\begin{equation}
  \label{e:SecularFrequencies}
  \omega_i = \left (\frac{1}{M} \frac{\partial^2 U_\mathrm{sec}}{\partial i^2}
  \right)^{1/2},
\end{equation}
for $i = x,y,z$.  In the idealized case, confinement is
cylindrically symmetric, $\omega_x = \omega_y$, and $\mathcal
{E}_\mathrm{rot}$ is spatially uniform, so the circular
micromotion does not contribute to the confining frequencies.

The density of ions will be low enough that there will be few
momentum-changing collisions during a single measurement.  Thus,
any given ion's trajectory will be well approximated by the simple
sum of three contributions:

  (i) a 3-d sinusoidal secular motion, specified by a magnitude and initial
phase for each of the $\hat{x}$, $\hat{y}$, and $\hat{z}$
directions. In a thermal ensemble of ions, the distribution of
initial phases will be random and the magnitudes,
Maxwell-Boltzmannian.  For typical experimental parameters (see
the Appendix) the secular frequencies $\omega_i$ will each be
about $2\pi \times 1$~kHz and the typical magnitude of motions, r,
will be about 0.5 cm.

  (ii) the more rapid, smaller amplitude rf micromotion, of characteristic
frequency about $2\pi \times 15$~kHz and radius perhaps 0.05 cm.
This rf micromotion, purely in the x-y plane, is strongly
modulated by the instantaneous displacement of the secular motion
in the x-y plane, and vanishes at secular displacement x=y=0.

  (iii) The still more rapid rotational micromotion, purely circular motion
in the x-y plane, at frequency $\omega_\mathrm{rot}$ about $2\pi
\times 100$~kHz and of radius comparable to the rf motion, around
0.05 cm. In the idealized case, the rotational micromotion (in
contrast to the rf micromotion) is not modulated by the secular
motion.

As described in Secs.~\ref{ss:axialosc} and~\ref{s:collisions}
below, for spectroscopic reasons we must operate with trapping
parameters such that ${\rm E}_\mathrm{rot} \gtrsim 30 k_B T$.
Under that condition, relatively small imperfections in
$\mathcal{E}_\mathrm{rot}$, say a spatial variation of $1.5\%$,
can give rise to contributions to $U_{\rm sec}$ of the same scale
as the ions' thermal energy, and thus significantly distort the
shape of the trapped ion cloud or even deconfine the ions.

For improved optical access we had to shrink the radius of the
linear electrodes $a$ with respect to their spacing $\rho_0$ The
spectroscopic requirement for highly uniform
$\mathcal{E}_\mathrm{rot}$ then forced the redesign of the
second-generation ion trap to be based on six near-linear elements
arranged on a hexagon, rather the four electrodes arranged on
square shown in Fig.~\ref{f:axialosc2}.  The trap will be
discussed in more detail in a future publication, but simulations
project spatial uniformity of $\mathcal{E}_\mathrm{rot}$ better
than $0.5\%$ with good optical access. The design led to
significant compromises in the spatial uniformity of ${\mathcal
E}_\mathrm{rf}$, so in future operation, mass selectivity in ion
detection will come not from a quadrupole mass filter, but rather
from pulsing ${\mathcal E}_{\rm rot}$ to a very high value for a
small fraction of a rotation cycle and then doing time-of-flight
mass discrimination on the ions thus ejected. ${\mathcal
B}_\mathrm{rot}$ will be imposed by means of time-varying currents
flowing lengthwise along the same electrodes that generate
$\mathcal{E}_\mathrm{rot}$.

\section{Spectroscopy in Rotating and Trapping Fields}
\label{s:spectroscopy}

On the face of it, an ion trap, with its inhomogeneous and rapidly
time-varying electric fields, is not necessarily a promising
environment in which to perform sub-Hertz spectroscopic
measurements on a polar molecule.  In this section we will explore
in more detail the effects of the various components of the
electric and magnetic fields on the transition energies relevant
to our science goals. The theoretical determination of the energy
levels of heavy diatomic molecules in the presence of time-varying
electric and magnetic fields is a tremendously involved problem in
relativistic few-body quantum mechanics. State-of-the-art {\it ab
initio} molecular structure calculations are limited to an energy
accuracy of perhaps $10^{13}$~Hz, a quantity which could be
compared with the size of a hypothetical ``science signal'', which
could be on the order of $10^{-3}$~Hz or smaller.

Fortunately, we can take advantage of the fact that at the energy
scales of molecular physics, time-reversal invariance is an exact
symmetry except to the extent that there is a time-violating
moment associated with the electron (or nuclear) spin. In this
section, except in those terms explicitly involving $d_e$, we will
assume that time-reversal invariance is a perfect symmetry in
order to analyze how various laboratory effects can cause
decoherence or systematic shifts in the relevant resonance
measurements. The results can be compared to the size of the line
shift that would arise from a given value of the electron EDM,
which is treated theoretically as a very small first-order
perturbation on the otherwise T-symmetric system.

In the subsections below, we bring in sequentially more realistic
features of the trapping fields.

\subsection{Basic Molecular Structure}

We begin by considering in detail the relevant molecular structure
in zero electric and magnetic fields, thus quantifying the
qualitative discussion of the experiment given in
Sec.~\ref{s:idea}.  Although the molecular structure cannot be
calculated in detail from {\it ab initio} structure calculations,
nevertheless its analytic structure is well known.  Because the
measurements will take place in nominally a single electronic,
vibrational, and rotational state, we will employ an effective
Hamiltonian within this state, as elaborated by Brown and
Carrington~\cite{BC03}. This approach will specify a few
undetermined numerical coefficients, whose values can be
approximated from perturbation theory, but which will ultimately
be measured.

Brown~\cite{BCM87,SNS90,NBE91} and co-workers have done thorough
work on deriving an effective Hamiltonian for ${}^3\Delta$
molecules. The complete Hamiltonian in the absence of $d_e$ is
given by
\begin{equation}\label{ed_effective}
  H_{\rm struct} = H_{\rm elec} + H_{\rm vib} + H_{\rm SO} + H_{\rm tum} +
  H_{\rm SS} + H_{\rm SR} + H_{\rm HFS} + H_{\rm LD} ,
\end{equation}
listed in rough order of decreasing magnitude. Since we are
concerned only with terms acting within the subspace of the
${}^3\Delta$ manifold, other electronic and vibrational states
will enter only as perturbations that help to determine the
effective Hamiltonian.  Thus we consider eigenstates of $H_{\rm
elec}$ and $H_{\rm vib}$.

The remaining terms in Eq.~(\ref{ed_effective}) are corrections to
the Born-Oppenheimer curves. They describe couplings between
various angular momenta ($H_{\rm SR}$, $H_{\rm HFS}$), parity
splittings ($H_{\rm LD}$, $H_{\rm HFS}$), and spin-dipolar
interactions ($H_{\rm SS}$, $H_{\rm HFS}$). In typical Hund's case
(a) molecules these interactions are small compared to the
rotational energy governed by $H_{\rm tum}$. The relevant
interactions that act within the $|\Omega|=1$ manifold of states
take the explicit form
\begin{eqnarray}
  H_{\rm SO} &=& A \Lambda \Sigma\\
  H_{\rm tum} &=& B_e ({\bf J}-{\bf S})^2-D({\bf J}-{\bf S})^4\\
  H_{\rm SS} &=& \frac{2}{3}\lambda(3\Sigma^2-{\bf S}^2)\\
  H_{\rm SR} &=& \gamma_{\rm SR}({\bf J}-{\bf S})\cdot{\bf S}\\
  H_{\rm HFS} &=& a I_z L_z + b_{\rm F}{\bf I}\cdot{\bf S}+\frac{c}{3}
  (3I_z S_z - {\bf I}\cdot{\bf S}) +
  \frac{1}{2}e_\Delta(J_+I_+S_+^2 + J_-I_-S_-^2) \\
  H_{\rm LD} &=& \frac{1}{2}(o_\Delta+3p_\Delta+6q_\Delta)(S_+^2J_+^2 +
  S_-^2J_-^2).
\end{eqnarray}
The constants in the first four terms are as follows: $A$ is the
molecular spin-orbit constant, $B_e$ the rotational constant for
the electronic level of interest, $D$ the effect of centrifugal
distortion on rotation (typically $D\sim B_e(m_e/m_{\rm mol})^2$,
with $m_e$ the electron mass and $m_{\rm mol}$ the reduced mass of
the molecule), $\lambda$ governs the strength of the spin-spin
dipolar interaction, and $\gamma_{\rm SR}$ determines the strength
of the interaction of the spin with the end-over-end rotation of
the molecule. These  four terms primarily describe an overall
shift of the $^3 \Delta_1$ $J$-level, and can be ignored in
evaluating energy differences in the states we care about. They
can, however, contribute small perturbations to these basic
levels, as we will describe below.

Within the $^3\Delta_1$, $J=1$ manifold of interest, the energy
levels are distinguished by the hyperfine and $\Lambda$-doubling
terms. The hyperfine Hamiltonian $H_{\rm HFS}$ includes the
familiar contact ($b_F$), nuclear-spin-orbit ($a$) and
spin-nuclear spin terms ($c$).  By estimating the parameters in
perturbation theory, it is expected that the resulting hyperfine
splitting is on the order of $2\pi \times 50$~MHz~\cite{PMI07}.
The hyperfine interaction also contains a previously unreported
term, with constant denoted $e_{\Delta}$, that is connected to the
$\Lambda$-doubling. This term is expected to be even smaller than
the already small $\Lambda$-doublet splitting itself~\cite{Meyer},
however, and will be ignored.

The $\Lambda$-doubling Hamiltonian arises from Coriolis-type
mixing of states with differing signs of $\Lambda$ due to
end-over-end rotation of the molecule.  For a $^3\Delta$ state
this interaction is characterized by three constants, of which the
parameter $o_{\Delta}$ is the dominant one. These terms describe
how the ${}^3\Delta$ state is perturbed by electronic states with
$^{\rm 2S+1}\Pi$ and $^{\rm 2S+1}\Sigma$ symmetry. Since we are
primarily concerned with terms in the Hamiltonian that affect the
ground rotational state of the ${}^3\Delta_1$ electronic level, we
only need to keep the term which connects $\Omega=1$ to
$\Omega=-1$. This term has the general form, with numerical
prefactors $C_{\Pi,\Sigma,\Pi'}$ that depend on Clebsch-Gordon
coefficients and wavefunction overlap,~\cite{BCM87}
\begin{equation}
  |o_\Delta+3p_\Delta+6q_\Delta| = {\tilde o}_\Delta \approx \sum_{\Pi,\Sigma,\Pi'}
  C_{\Pi,\Sigma,\Pi'}
  \frac{A^2\;B_e^2}{({\rm E}_\Delta-{\rm E}_\Pi)({\rm E}_\Delta-{\rm E}_\Sigma)({\rm E}_\Delta-{\rm E}_{\Pi'})},
\end{equation}
where the sum is over all intermediate $\Sigma$ and $\Pi$ states
of singlet and triplet spin symmetries. For HfF$^+$ this
perturbation leads to a $\Lambda$-doublet splitting on the order
of $2\pi \times 10$~kHz.  This estimate was carried out assuming a
$\sigma\delta$ molecular orbital configuration, where the $\delta$
orbital has total angular momentum $L=2$ in the pure precession
approximation. The ground $X{}^1\Sigma$ is a $\sigma^2$ molecular
orbital but has some admixture of atomic $d_0$ orbitals. We
therefore expand the molecular wavefunction into atomic orbitals
and reduce the amount of admixture by the factor $\epsilon_d$ that
describes the $d_0$ character. From here on, we shall express the
energy difference in parity levels for the $J=1$ as $\omega_{\rm
ef} = 4{\tilde o}_\Delta$, rather than ${\tilde o}_\Delta$ itself.

Thus the basic molecular structure of interest to the $^3
\Delta_1$, $J=1$ state is governed by two constants: the hyperfine
splitting ${\rm E}_{\rm hf}$ (given by $3 A_{||}/4$ for $J=1,
I=1/2$) and the $\Lambda$-doublet splitting $\omega_{\rm ef}$.
These constants give the structure depicted in
Fig.~\ref{f:hyperfine_stutz}(a). These basic levels may be
perturbed by couplings to other levels, especially rotational or
electronic excited states.  However, for the $J=1$ state of
interest, some simplifications are possible, namely: (1)
Off-diagonal couplings in $\Omega$ are zero since ${\bf
J}\cdot{\bf S}$ preserves the value of J (there is no level with
$J=1$ and $\Omega=2$); (2) Off-diagonal contributions that mix
$J=2$ into the $J=1$ manifold thus depend solely on the applied
fields and the hyperfine interactions. Since the value of the
spin-orbit constant is expected to be far larger than the
rotational constant and we are concerned with a $J=1$ state, the
operators that connect $\Omega$ to $\Omega\pm1$ will be ignored.
The contributions to the ground state characteristics by terms off
diagonal in $\Omega$ are smaller by a factor of the hyperfine
interaction energy to the spin-orbit separation energy, hence a
factor of $10^{\rm -6}$. This is the value which appears in front
of any term connecting $\Omega$ to $\Omega\pm1$ in the ground
$J=1$ state.

\subsection{Effect of Non-rotating Electric and Magnetic Fields}
\label{ss:staticfields}

The influence of external fields presents new terms in the
Hamiltonian of the form
\begin{eqnarray}
  H_{\rm Stark} &=& -\vec{\bf d}_{\rm mf}\cdot\vec{\bf \mathcal{E}}\\
  H_{\rm Zeeman} &=& -\vec{\bf \mu}\cdot\vec{\bf \mathcal{B}}.
\end{eqnarray}
Here $\vec{\bf \mathcal{E}}$ and $\vec{\bf \mathcal{B}}$ are the
electric and magnetic fields, assumed for the moment to be
collinear so that they define the axis along which $m_F$ is a good
quantum number;  while $\vec{\bf d}_{\rm mf}$ and $\vec{\bf \mu}$
are the electric and magnetic dipole moments of the molecule.

The electric dipole moment arises from the body-fixed molecular
dipole moment, at fields sufficiently small not to disturb the
electronic structure.  We assume that the field is sufficiently
large to completely polarize this dipole moment, i.e., $d_{\rm
mf}{\cal E} \gg \omega_{\rm ef}$, in which case the Stark
energies are given by
\begin{equation}\label{Stark_nr_energy}
{\rm E}_{\rm Stark} = - m_F  \Omega \gamma_F d_{\rm mf} {\cal E},
\end{equation}
where $\gamma_F$ is a geometric factor, analogous to a Land\'e
g-factor, which accounts for  the Stark effect in the total
angular momentum basis $F$. In the limit where the electric field
is weak compared to rotational splittings, it is given by
\begin{equation}
  \gamma_F=\frac{J(J+1)+F(F+1)-I(I+1)}{2F(F+1)J(J+1)}.
\end{equation}
Its numerical values in the $J=1$ state are therefore
$\gamma_{F=3/2}=1/3$ and $\gamma_{F=1/2} = 2/3$.  The electric
field therefore raises the energy of the states with $m_F \Omega
<0$ (denoted ``upper'' states with superscript $u$), and lowers
the energy of states with $m_F \Omega >0$ (``lower'' states with
superscript $\ell$). This shift in energy levels is shown in
Fig.~\ref{f:hyperfine_stutz}(b), where $|a\rangle$ and $|b\rangle$
are upper and $|c\rangle$ and $|d\rangle$ are lower states.

The form of the Zeeman interaction is somewhat more elaborate, as
the magnetic moment of the molecule can arise from any of the
angular momenta ${\bf L}$, ${\bf S}$, ${\bf J}$, and ${\bf I}$.
Quite generally, however, in the weak-field limit where $\mu_B
{\mathcal B} \ll {\rm E}_{\rm hf}$, the Zeeman energies are given
by $m_F g_F^{u / \ell}\mu_B {\mathcal B}$, where $\mu_B$ is the
Bohr magneton and $g_F^{u/\ell}$ are g-factors for the upper and
lower states.  In general, $g_F^u \ne g_F^\ell$, and this
difference can depend on electric field, a possible source of
systematic error. We will discuss this in Sec.~\ref{ss:deltagF}
below.

The leading order terms in the Zeeman energy are those that
preserve the signed value of $\Omega$.  They are given by
\begin{equation}
\label{e:zeeman}
  H_{\rm Zeeman} = \left(\gamma_F \left[\left((g_L+g_r)\Lambda +(g_S+g_r) \Sigma\right)
    \Omega -g_r J(J+1)\right]- g_I \kappa_F\right) m_F \mu_B {\mathcal B},
\end{equation}
where $\kappa_F=(F(F+1)+I(I+1)-J(J+1))/2F(F+1)$ is another
Land\'e-type g-factor, but for nuclear spin. The orbital and spin
g-factors are $g_L$ and $g_S$, while the rotation and nuclear spin
g-factors are $g_r$ and $g_I$.  Both $g_r$ and $g_I$ are small,
being on the order of the electron-to-molecular mass ratio $\sim
m_e/m_{\rm mol} \sim 10^{-3}$.  Thus for an idealized $^3\Delta_1$
molecule where $g_L=1$, $\Lambda = \pm 2$, $g_S = 2$, $\Sigma =
\mp 1$, we would expect molecular g-factors on the order of
$10^{-3}$. More realistically, $g_s$ differs from 2 by a number on
the order of $\alpha$, the fine structure constant, and a g-factor
$\sim 10^{-2}$ might be expected.  In heavy-atom molecules such as
ours for which spin-orbit effects mix $\Lambda$, we may expect
instead the difference $2g_L-g_S$ to be as large as $\sim 0.1$ in
magnitude. If we assume the dominant contribution comes from these
spin-orbit type effects, we can define the $g$-factor for the
$J=1$ state as
\begin{equation}
  g_{F=3/2} = \gamma_{F=3/2} (g_L\Lambda+g_S\Sigma)\Omega\lesssim 0.03,
\end{equation}
while
\begin{equation}
  g_{F=1/2} = 2g_{F=3/2}.
\end{equation}

Finally, the effect of the EDM itself introduces a small energy
shift
\begin{equation}
H_{\rm EDM} = -{\vec d}_e \cdot {\vec {\cal E}}_{\rm eff} = d_e
{\cal E}_{\rm eff} {\vec \sigma}_{1} \cdot {\hat n},
\end{equation}
where ${\vec \sigma}_1$ is the spin of the $s$-electron
contributing to the EDM signal; and ${\hat n}$ denotes the
intermolecular axis, with ${\hat n}$ pointing from the more
negative atom to the more positive one; in our case from the
fluorine or hydrogen to thorium, platinum, or hafnium.  Also in
this convention we take ${\mathcal E}_{\rm eff}$ as positive if it
is anti-parallel to ${\hat n}$. The energy shift arising from this
Hamiltonian depends only on the relative direction of the electron
spin and the internuclear axis, and is given by
\begin{equation}
{\rm E}_{\rm EDM} = -\frac {d_e {\mathcal E}_{\rm eff}} {2
|\Omega|} \Omega.
\end{equation}
Polarizing the molecule in the external field selects a definite
value of $\Omega$, hence a definite energy shift, positive or
negative, due to the EDM.  This additional shift is illustrated in
Fig.~\ref{f:hyperfine_stutz}(c).

For a range of field strengths and parameters, the energies of the
sublevels within the $J=1$ manifold are well approximated by a
linear expansion in the electric and magnetic fields. We define
\begin{equation}
  \vec{\mathcal B} = \mathcal{B}_{||}\frac{\vec{\mathcal E}}{|{\mathcal E}|}
    + \vec{\mathcal B}_\bot
\end{equation}
Taking $\omega_\mathrm{ef} \ll d_\mathrm{mf} \mathcal{E} \ll {\rm
E}_\mathrm{hf}$ and $d_\mathrm{mf} \mathcal{E} \gg g_F \mu_B
{\mathcal B}_{||}$, and setting ${\mathcal B}_\bot =0$,  we get
for the non-rotating energies,
\begin{equation}
    \label{e:linear}
    {\rm E}_\mathrm{nr}^{u/\ell}(F, m_F, \Omega;\mathcal{E}, \mathcal{B}) \approx
       \frac{1}{3}(F(F+1)-\frac{11}{4}){\rm E}_{\rm hf} - m_F \Omega \gamma_F d_\mathrm{mf} {\cal E} +
        m_F g^{u/\ell}_F \mu_B {\mathcal B} - (d_e {\cal E}_\mathrm{eff} / 2|\Omega|)\Omega,
\end{equation}
where $\Omega$ is either 1 or -1, and the prefactor in front of
${\rm E}_{\rm hf}$ is such that for the $J=1$ level, ${\rm
E}(F=3/2)$ - ${\rm E}(F=1/2)$ = $3A_{||}/4$ = ${\rm
E}_\mathrm{hf}$. $F$ and $\Omega$ are good quantum numbers only to
the extent that the electric field is neither too large nor too
small, but we will use $F$ and $\Omega$ as labels for levels even
as these approximations begin to break down.

For notational compactness, we introduce special labels for
particular states as follows (see Fig.
\ref{f:hyperfine_stutz}(b)):
\begin{eqnarray}
    \label{e:specialstates}
  |a\rangle &=& |F=3/2, m = 3/2,  \Omega =  -1\rangle \\
  |b\rangle &=& |F=3/2, m = -3/2, \Omega = 1\rangle \nonumber \\
  |c\rangle &=& |F=3/2, m = 3/2,  \Omega = 1\rangle \nonumber \\
  |d\rangle &=& |F=3/2, m = -3/2,  \Omega = -1\rangle \nonumber
\end{eqnarray}
with corresponding energies, ${\rm E}_a$, ${\rm E}_b$, ${\rm
E}_c$, and ${\rm E}_d$, and identify the energies of two
particularly interesting transitions, $W^u={\rm E}_a-{\rm E}_b$,
and $W^\ell = {\rm E}_c-{\rm E}_d$ such that
\begin{eqnarray}\nonumber
  W^u &=& 3 g_F^u \mu_B {\mathcal B} + d_e{\mathcal E}_{\rm eff}\\\label{e:W-trans}
  W^\ell &=& 3 g_F^\ell \mu_B {\mathcal B} - d_e{\mathcal E}_{\rm eff}.
\end{eqnarray}

Taking this analysis a step farther, it is possible that the
electric field energy $d_{\rm mf}{\cal E}$ is {\it not} small
compared to the hyperfine splitting ${\rm E}_{\rm hf}$.  In this
case the electric field mixes the different total-$F$ states and
perturbs the above energies.  Ignoring the magnetic field and EDM
energies, the energy levels take the form
\begin{widetext}
\begin{eqnarray}
    \label{e:enr-mo-pos}
    {\rm E}_\mathrm{nr}({\tilde F}\sim3/2, m_F\Omega=+1/2) & = & -\frac{1}{2}\left(\frac{
      d_{\rm mf}{\mathcal E}}{2}+\frac{{\rm E}_{\rm hf}}{3}\right)+\frac{1}{2}\sqrt{\left(
      \frac{d_{\rm mf}{\mathcal E}}{6}-{\rm E}_{\rm hf}\right)^2+2\left(\frac{
      d_{\rm mf}{\mathcal E}}{3}\right)^2}-\frac{3\omega_{\rm ef}^2}
    {4d_{\rm mf}{\mathcal E}}\\
    \nonumber
    {\rm E}_\mathrm{nr}({\tilde F}\sim1/2, m_F\Omega=+1/2) & = & -\frac{1}{2}\left(\frac{
      d_{\rm mf}{\mathcal E}}{2}+\frac{{\rm E}_{\rm hf}}{3}\right)-\frac{1}{2}\sqrt{\left(
      \frac{d_{\rm mf}{\mathcal E}}{6}-{\rm E}_{\rm hf}\right)^2+2\left(\frac{
      d_{\rm mf}{\mathcal E}}{3}\right)^2}-\frac{3\omega_{\rm ef}^2}
    {8d_{\rm mf}{\mathcal E}}\\
    \label{e:enr-mo-neg}
    {\rm E}_\mathrm{nr}({\tilde F}\sim3/2, m_F\Omega=-1/2) & = & \frac{1}{2}\left(\frac{
      d_{\rm mf}{\mathcal E}}{2}-\frac{{\rm E}_{\rm hf}}{3}\right)+\frac{1}{2}\sqrt{\left(
      \frac{d_{\rm mf}{\mathcal E}}{6}-{\rm E}_{\rm hf}\right)^2+2\left(\frac{
      d_{\rm mf}{\mathcal E}}{3}\right)^2}+\frac{3\omega_{\rm ef}^2}
    {4d_{\rm mf}{\mathcal E}}\\
    \nonumber
    {\rm E}_\mathrm{nr}({\tilde F}\sim1/2, m_F\Omega=-1/2) & = & \frac{1}{2}\left(\frac{
      d_{\rm mf}{\mathcal E}}{2}-\frac{{\rm E}_{\rm hf}}{3}\right)-\frac{1}{2}\sqrt{\left(
      \frac{d_{\rm mf}{\mathcal E}}{6}-{\rm E}_{\rm hf}\right)^2+2\left(\frac{
      d_{\rm mf}{\mathcal E}}{3}\right)^2}+\frac{3\omega_{\rm ef}^2}
    {8d_{\rm mf}{\mathcal E}},
\end{eqnarray}
\end{widetext}

The equations of this section have so far been to one degree or
another approximate results.  But in the absence of exotic
particle physics we can invoke time-reversal symmetry and write
$\textit{exact}$ relations:
\begin{equation}
    {\rm E}_\mathrm{nr} (F, m_F, \Omega;\mathcal{E}, {\mathcal B}) -
    {\rm E}_\mathrm{nr}(F, -m_F, -\Omega;\mathcal{E}, {\mathcal B})
    = {\rm E}_\mathrm{nr}(F, -m_F, -\Omega;\mathcal{E}, -{\mathcal B}) -
    {\rm E}_\mathrm{nr}(F, m_F, \Omega;\mathcal{E}, -{\mathcal B})
\end{equation}
which, for ${\mathcal B}=0$, becomes
\begin{equation}
    {\rm E}_\mathrm{nr}(F, m_F, \Omega;\mathcal{E}) =
    {\rm E}_\mathrm{nr}(F, -m_F, -\Omega;\mathcal{E}).
\end{equation}
This exact degeneracy is, in fact, an example of the Kramers
degeneracy that follows from time-reversal
invariance~\cite{SRG87}.  For our purposes, the key result here is
that, in the limit of non-rotating fields, zero applied magnetic
field, and an electron EDM, the energy of the science transitions
$|m_F, \Omega\rangle \leftrightarrow |-m_F, -\Omega\rangle$ (and
in particular, $W^u$ and $W^\ell$) are independent of the
magnitude of the electric field. This is an important property
because we are using spatially inhomogeneous electric fields to
confine the ions in the trap, and we want to minimize the
resulting decoherence.

This degeneracy in turn means that the energy differences $W^u$
and $W^l$ depend only on the magnetic field and, of course, the
EDM term as shown in Eq.~(\ref{e:W-trans}). The magnetic
contribution reverses sign upon reversing the direction of
${\mathcal B}$ with respect to the electric field direction (which
also sets the quantization axis, since $d_{\rm mf}{\cal E} \gg
\mu_B {\mathcal B}$).  Therefore the science measurement is given
by the combinations
\begin{eqnarray}
\label{e:WuWl}
 W^u( {\mathcal E}, {\mathcal B}) + W^u(
{\mathcal E},- {\mathcal B})
&=&  2  d_e {\mathcal E}_{\rm eff} \nonumber \\
W^l( {\mathcal E}, {\mathcal B}) + W^l( {\mathcal E},- {\mathcal
B}) &=& - 2  d_e {\mathcal E}_{\rm eff},
\end{eqnarray}
where a $+$ sign on ${\mathcal B}$ denotes that it points in the
same direction as ${\cal E}$.

\subsection{Rotating Fields, Small-Angle Limit}
\label{ss:rotatingfields}

Many EDM experiments over the years have been complicated by the
problem of ``Berry's phase", the term in this context used as a
catch-all to describe a variety of effects related to the motion
of the particles in inhomogeneous fields.

\begin{figure}
\begin{center}
\includegraphics{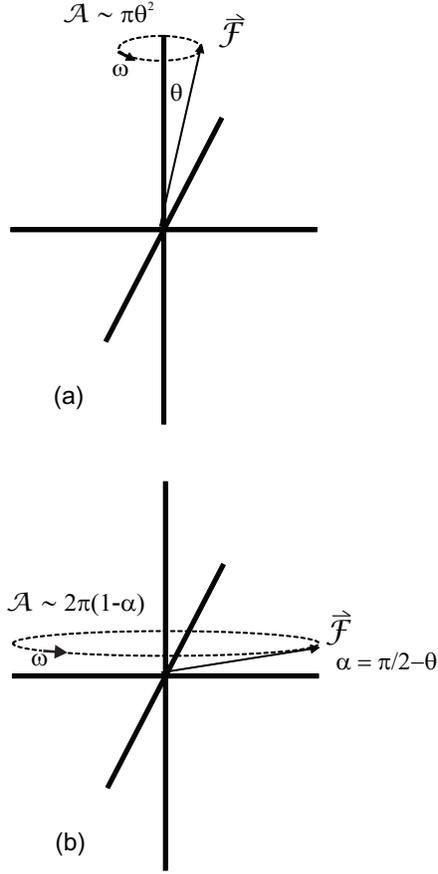}
\caption{(a) Small-angle limit.  When the quantization axis
$\mathcal{F}$ follows a slow periodic perturbation characterized
by tilt angle $\theta$, angular frequency $\omega$ and enclosed
solid angle $\mathcal{A}$, two states whose instantaneous
projection of angular momentum along $\mathcal{F}$ differs by
$\delta m$ will have their effective relative energy displaced by
a ``Berry's energ'' $\omega \mathcal{A}\delta m/2\pi$. (b)
Large-angle limit. When instead the quantization axis sweeps out a
full $2\pi$ steradians per cycle ($\alpha$=0), the differential
phase shift between the two levels is indistinguishable  from
zero, and in the most natural conceptual framework, the Berry's
energy vanishes.\label{f:berry}}
\end{center}
\end{figure}

The sketch in Fig.~\ref{f:berry}(a) illustrates the classic
Berry's phase result: if the field that defines the quantization
axis, as experienced locally by a particle (or atom, or molecule),
precesses about the laboratory axis at some angle, $\theta$, then,
in the limit of slow precession, with each cycle of the precession
the wave-function $\Psi$ picks up a phase given by
$m_F\mathcal{A}$, where $m_F$ is the instantaneous projection of
the particle's total angular momentum on the quantization axis,
and $\mathcal{A}$  is the solid angle subtended by the cone. If
the precession is periodic with period $\tau$, one can (with
provisos, as we will discuss) think of this phase-shift as being
associated with a frequency, or indeed energy,
$m_F\mathcal{A}/\tau$ .  In a spectroscopic measurement of the
energy difference between two states whose $m_F$ values differ by
$\delta m_F$, there will be a contribution to the transition
angular frequency $\mathcal{A}\delta m_F/\tau$.

In neutron EDM experiments, motional magnetic fields, in
combination with uncharacterized fixed gradients from magnetic
impurities, Berry's phase can be a dangerous systematic whose
dependence on applied fields can mimic an EDM signal~\cite{PHH04}.
In Sec.~\ref{ss:relativistic} we will see that the effects of
motional fields in our experiment are negligible.

Neutral atoms or molecules may be confined in traps consisting of
static configurations of electric or magnetic fields. These traps
are based on the interaction between the trapped species' magnetic
or electric dipoles and the inhomogeneous magnetic or electric
fields, respectively, of the trap.  Especially in cases where the
traps are axially symmetric, so that the single-particle
trajectory of an atom can orbit many times one way or the other
about the axis of the trap, the coherence time of an ensemble of
atoms with a thermal distribution of trajectories can be severely
restricted~\cite{RSR08}.  Our system is quite different, because
in an ion trap the forces arise from the interaction between the
trapping fields and the monopole moment of our trapped ion.
Assuming the temperature, size of bias field, and radius of
confinement are the same, the trapping fields for an ion are
spatially much more homogeneous than would be those for a neutral
molecule or atom.

That said, the fact that we can speak of a ``bias" electric field
at all in an ion trap comes at the cost of having the applied
electric field constantly rotating.

\subsection{Rotating Fields, Large-Angle Limit (Dressed States)}
\label{ss:bigangle}

The basic dressed-state idea is an extension of the more common
idea of an energy eigenstate: a system governed by a
time-invariant Hamiltonian $H$ will have solutions $\Psi$ such
that $\Psi(t+T) = e^{-i \omega T} \Psi (t)$ for all $T$ and $t$;
such a solution $\Psi$ is called an energy eigenstate, with
$\omega$ being then the corresponding energy. Similarly, a system
governed by a periodic Hamiltonian with period $\tau$ such that
$H(t + \tau) = H (t)$ for all values of $t$, will have so-called
``dressed-state" solutions $\Psi$ such that $\Psi(t+n \tau) =
e^{-i n \phi} \Psi(t)$ for all $t$ and all integer values of $n$.
It is tempting to call $\phi/\tau$ the ``energy" of the dressed
state, but there will be an ambiguity in that energy because we
can always replace $\phi$ with $\phi + 2\pi$.

Operationally, the dressed state energies are derived from the
eigenvalues of a formally time-independent Hamiltonian.  If $H_0$
denotes the Hamiltonian in the absence of the field, then the
appropriate rotation-dressed Hamiltonian is given by
\begin{equation}\label{geo_Hamiltonian}
  H_{\rm dressed} = H_0 - {\vec d}_{\rm mf} \cdot {\vec {\cal E}_{\rm rot}}
  +H_{\rm rot},
\end{equation}
$H_{\rm rot}$ is defined as~\cite{MLB09}
\begin{equation}
\label{e:Hrot}
  H_{\rm rot} = -\omega_{\rm rot} \left( \cos(\theta) F_z - \sin(\theta) F_x \right)
\end{equation}
where $F_z$and $F_x$ are the projections of the total angular
momentum ${\vec F}$ into a set of axes where $z$ coincides with
the instantaneous direction of the electric field.  We now make
explicit the rotating electric field with ${\vec {\cal E}_{\rm
rot}}$. The $\cos(\theta)$ term thus provides an energy which,
when multiplied by the rotational period $\tau = 2 \pi /
\omega_{\rm rot}$, gives the ordinary Berry phase,
\begin{equation}
- 2 \pi \cos (\theta) m_F \rightarrow 2 \pi  (1 - \cos (\theta))
m_F
\end{equation}
where we have taken the liberty of adding an arbitrary phase $2
\pi m_F$ to reveal explicitly  the solid angle $2 \pi (1- \cos
\theta)$.

\begin{figure}
\begin{center}
\includegraphics[width=8cm]{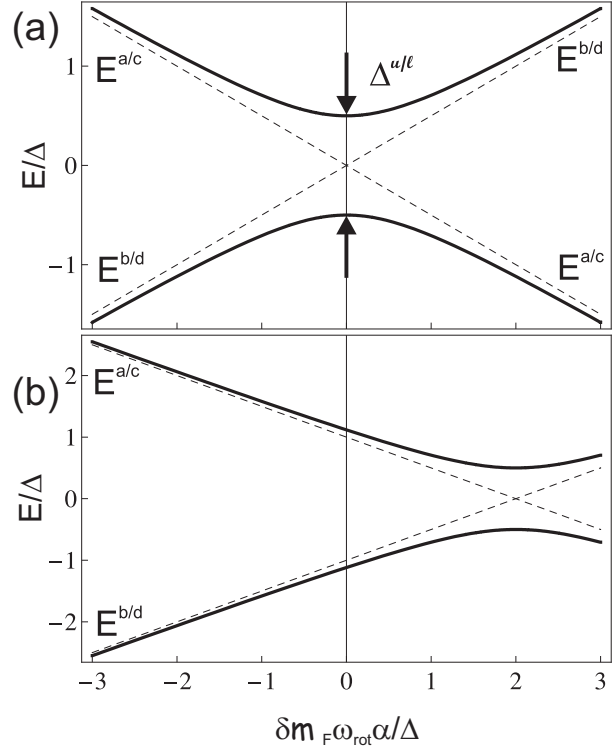}
\caption{The apparent energy shifts between $m_F = +3/2$ and $m_F
= -3/2$ states in upper (a,b) and lower (c,d) $\Lambda$-doublet
levels versus $\alpha$, the angle of the electric field to the
plane orthogonal the rotation axis of $\mathcal{E}_\mathrm{rot}$
($\alpha$ is shown in Fig.~\ref{f:berry}(b)). (a) At $\alpha = 0$,
there is an avoided crossing that mixes $m_F = \pm 3/2$ states,
with an energy splitting at the crossing of $\Delta^{u/\ell}$. (b)
Since $\alpha = 0$ at the axial trap center, and since we need
$m_F$ to be a signed quantity in order to measure $d_e$, we will
bias away from the avoided crossing using a magnetic field
$\mathcal{B}_\mathrm{rot}$. $\delta m_F g_F \mu_{B}
\mathcal{B}_\mathrm{rot}
> \Delta^{u/\ell}$ is required for $m_F$ to be a quantity of
definite sign.  This picture is intuitively correct in the limit
that $\Delta^{u/\ell} > \omega_\mathrm{max}$ (see
Sec.~\ref{ss:axialosc}). The experiment will be performed in the
opposite limit.  However, solving the time dependent
Schr\"{o}dinger equation (Eq.~\ref{e:timedependent}) gives the
same requirement of $\delta m_F g_F \mu_{B}
\mathcal{B}_\mathrm{rot}
> \Delta^{u/\ell}$ in both limits.\label{f:anticrossing}}
\end{center}
\end{figure}

In the experiment, the applied electric field should lie very
nearly in the plane orthogonal to the rotation axis, i.e., $\theta
\approx \pi/2$.  It is therefore useful to consider the small
angular deviation from this plane, $\alpha = \pi/2 - \theta$
(Fig.~\ref{f:berry}). Then the apparent energy shift arising from
the geometric phase is
\begin{equation}
  {\rm E}_{\rm geo} = - m_F \omega_{\rm rot} \sin (\alpha)
  \approx -m_F \omega_{\rm rot} \alpha.
\end{equation}
Now consider two states which are, in the absence of rotation,
degenerate, say the states $|a\rangle$, with $m=3/2,\Omega=-1$,
and the state $|b\rangle$, with $m = -3/2,\Omega=1$, indicated in
Fig~\ref{f:hyperfine_stutz}(b). Rotation breaks this degeneracy,
by adding the energies $\sim \pm (3/2) \omega_{\rm rot} \alpha$,
as shown by the dashed lines in Fig.~\ref{f:anticrossing}. These
levels cross at $\alpha = 0$, leading to their apparent degeneracy
when the electric field lies in the horizontal plane.

In addition, the rotation of the field also incurs coupling
between states with different $m_F$ values, arising from the $\sin
(\theta)$ term in Eq.~(\ref{e:Hrot}).  This perturbation, treated
at third-order in perturbation theory, connects the two levels and
turns the crossing into an avoided one, as shown by the solid
lines in Fig.~\ref{f:anticrossing}. Since the energy contribution
due to the rotating field is small compared to Stark energy
splittings, we can use ideas similar to the derivation of
$\Lambda$-doubling, i.e., we take a sum of the perturbing
components and take them to the appropriate power. We look for
terms in this expansion that can connect the state
$|a\rangle=|m_F\Omega\rangle$ to $|b\rangle=|-m_F-\Omega\rangle$.
Therefore, the power of perturbation theory needed is $2m_F+1$,
where the $2m_F$ takes $m_F\rightarrow-m_F$ and the extra power
takes $\Omega\rightarrow-\Omega$. The two terms in the Hamiltonian
that can do this are the $\Lambda$-doubling term and the
$m_F$-changing terms of the rotating electric field. Our expansion
is, schematically, the following
\begin{equation}
  H_{\rm coup} = \frac{(H_{\rm LD} + H_{\rm rot})^{\rm 2m_F+1}}
  {(\Delta {\rm E}_{\rm m_F})^{\rm 2m_F}}.
\end{equation}
The $(\Delta {\rm E}_{\rm m_F})^{\rm 2m_F}$ are the energy level
differences between states with different $m_F$ values, thus are
related to the Stark splittings. This tells us that
\begin{equation}
  \Delta \sim \omega_{\rm ef}\left(\frac{\omega_{\rm rot}}
    {d_{\rm mf}\mathcal{E}_{\rm rot}}\right)^{\rm 2m_F},
\end{equation}
where $\Delta$ is the energy splitting at the level crossing
between otherwise degenerate states with $m_F>0$ and $m_F<0$. The
numerical prefactor in this expression has a rather complicated
form within perturbation theory.  However, its value can be
computed by numerically diagonalizing the relevant
hyperfine-plus-rotation dressed Hamiltonian~\cite{Meyer}. The
result, for the $m_F = \pm 3/2$ states in
Fig.~\ref{f:hyperfine_stutz}(b), is
\begin{equation}
  \label{e:avoidedsplitting} \Delta^{u/\ell} \approx 170
  \omega_\mathrm{ef} \left( \frac {\omega_\mathrm{rot} } { d_\mathrm{mf}
    \mathcal{E}_\mathrm{rot}} \right)^3,
\end{equation}
where the superscript $u$ refers to mixing between the
$|a\rangle$ and $|b\rangle$ states, and the superscript $\ell$ to
mixing between $|c\rangle$ and $|d\rangle$ states. In the absence
of the hyperfine interaction, the average value of the numerical
prefactor is $170$ and the upper and lower states have the same
avoided crossing. However, small fractional differences between
$\Delta^u$ and $\Delta^{\ell}$ turn out to be significant, and
are discussed further below.

 The presence of the electric field causes the states with
$|F=1/2,m_F=\pm1/2\rangle$ and $|F=3/2,m_F=\pm1/2\rangle$ to mix.
 Including the hyperfine interaction into the numerical
diagonalization yields
\begin{eqnarray}
  \Delta = \frac{1}{2}(\Delta^u+\Delta^\ell) &\approx& 170  \omega_\mathrm{ef} \left( \frac {\omega_\mathrm{rot} } { d_\mathrm{mf}
    \mathcal{E}_\mathrm{rot}} \right)^3,\\\label{e:deltaDelta}
  \delta\Delta=\frac{1}{2}(\Delta^u-\Delta^\ell) &\approx& 127
  \omega_\mathrm{ef} \left( \frac {\omega_\mathrm{rot} } { d_\mathrm{mf}
    \mathcal{E}_\mathrm{rot}} \right)^3\left(\frac{d_{\rm mf}{\mathcal E}_{\rm rot}}
    {{\rm E}_{\rm hf}}\right)^2.
\end{eqnarray}
It is evident that the average shift is the same, but now the
upper and lower levels acquire a different splitting due to the
rotation-induced mixing within the sublevels. The difference is
suppressed relative to the average value of the splitting by a
factor of $(d_{\rm mf}{\mathcal E}/{\rm E}_{\rm hf})^2$,
reflecting the fact that higher orders of perturbation theory are
needed to include the effects of the hyperfine interaction.  For
$\omega_\mathrm{ef}$ = 2$\pi \times 10$~kHz, $\omega_\mathrm{rot}$
= 2$\pi \times 100$~kHz, $d_\mathrm{mf}\mathcal{E}_\mathrm{rot}$ =
$2\pi \times 10$~MHz, ${\rm E}_{\rm hf}$ = $2\pi \times 45$~MHz,
then $\Delta$ = $2\pi \times 2$~Hz and $\delta\Delta$ = $2\pi
\times 0.06$~Hz.

The magnitude of the rotation-induced mixing within any of the
four pairs of otherwise degenerate $m = \pm 1/2$ states is much
larger than the mixing within either pair of $m= \pm 3/2$ states,
$\Delta^u$ or $\Delta^\ell$. For this reason, the $m =  \pm 1/2$
levels are probably not great candidates for precision metrology
in rotating fields.

An ion in a trap will feel an axial force pushing it towards the
axial position where the axial electric field vanishes, that is,
the location at which $\alpha$ is identically zero.  This poses a
problem, because at $\alpha$ = 0, each dressed state is an equal
mixture of states with $\Omega =1$ and with $\Omega= -1$. In other
words, the dressed states right at the avoided crossing will have
vanishing eEDM signal.   The solution is to bias the avoided
crossing away from $\alpha = 0$ by adding to the trapping fields a
uniform, rotating magnetic field which is instantaneously always
parallel or anti-parallel to vector $\mathcal{E}_\mathrm{rot}$,
\begin{equation}
  \vec{ {\mathcal B}}_\mathrm{rot} = \frac{ {\mathcal B}_\mathrm{rot}}
      {\mathcal{E}_\mathrm{rot}}
      \vec{\mathcal{E}}_\mathrm{rot}.
\end{equation}
In our convention, ${\vec {\cal E}}_{\rm rot}$ defines the
quantization axis, so that the number ${\cal E}_{\rm rot}$ will
always be taken to be positive. The sign of ${\mathcal B}_{\rm
rot}$ then determines whether the co-rotating magnetic field is
parallel (${\mathcal B}_{\rm rot}>0$) or anti-parallel
(${\mathcal B}_{\rm rot}<0$).

The energy levels are now as shown in
Fig.~\ref{f:anticrossing}(b). As derived below in
Sec.~\ref{ss:axialosc}, in the limit ${\mathcal B}_\mathrm{rot}
g_F \mu_B \gg \Delta^{u/ \ell}$, the dressed states near $\alpha =
0$ are once again states of good $m_F$ and $\Omega $.  The energy
splitting between the two states, as altered by the rotation of
the field, are given approximately by
\begin{equation}
  \label{e:avoidedcurves}
  W^{u/\ell}({\cal E}_\mathrm{rot}, \mathcal{B}_\mathrm{rot})
  = {\rm E}_{a/c} - {\rm E}_{b/d} =
  -3 \alpha \omega_\mathrm{rot} + 3  g^{u/\ell}_F \mu_B {\mathcal B}_\mathrm{rot} +
  \frac { ( \Delta^{u/ \ell} )^2 }
    {6  g^{u/\ell}_F \mu_B{\mathcal B}_\mathrm{rot} - 6 \alpha
      \omega_\mathrm{rot} }
    \pm d_e \mathcal{E}_\mathrm{eff},
\end{equation}
where the $+$ sign corresponds to $u$ states, and the $-$ sign to
$\ell$ states.

Over the course of one axial oscillation of the ion in the trap,
$\alpha$ which is approximately proportional to the axial electric
field, will average to zero.  Unfortunately, the contributions to
$\delta W$ from $\mathcal{E}_\mathrm{rot}$ and from ${\mathcal
B}_\mathrm{rot}$ are larger than that from the scale of the
physics we most care about, $d_e \mathcal{E}_\mathrm{eff}$, and
the spatial and temporal variation in $\mathcal{E}_\mathrm{rot}$
and in $\mathcal{B}_\mathrm{rot}$ will reduce the coherence time
of the spectroscopy, as discussed in Sec.~\ref{ss:erot}
-~\ref{ss:strayB} below. But to the extent that one is able quite
precisely to chop ${\mathcal B}_\mathrm{rot}$ to $-{\mathcal
B}_\mathrm{rot}$ on alternate measurements, the science signal
still arises from the same combination as in Eq.~\ref{e:WuWl}:
\begin{equation}
  \label{e:bfieldchop} \langle W^{u/\ell}({\cal E}_\mathrm{rot},
  \mathcal{B}_\mathrm{rot})\rangle
  + \langle W^{u/\ell} ({\cal E}_\mathrm{rot},
  -\mathcal{B}_\mathrm{rot})\rangle \approx   \pm2 d_e \mathcal{E}_\mathrm{eff},
\end{equation}
where the +/- corresponds to the $u/\ell$ superscripts
respectively, and the brackets denote averaging over the
excursions of $\alpha$, which is assumed to vary symmetrically
about zero.

The equation above relies on several approximations. One needs in
particular that $d_\mathrm{mf} \mathcal{E}_\mathrm{rot} \gg
\omega_\mathrm{ef}$,  $3g_F\mu_B B_\mathrm{rot} \gg 170
\omega_\mathrm{ef}$ $(\omega_\mathrm{rot} / d_\mathrm{mf}
\mathcal{E}_\mathrm{rot})^3$,  and
$d_\mathrm{mf}\mathcal{E}_\mathrm{rot} \gg \omega_\mathrm{rot}$,
$\alpha \ll 1$, and $d_\mathrm{mf} \mathcal{E}_\mathrm{rot} < {\rm
E}_\mathrm{hf}$. These are all good approximations, but they are
not perfect. For example, using values from the Appendix,
$\omega_\mathrm{rot}/(d_\mathrm{mf} \mathcal{E}_\mathrm{rot})
\approx 0.01$, a small number, but not zero. To what extent will
imperfections in these approximations mimic an eEDM signal?

The driving principle of our experimental design is to measure
$d_e$ with as close to a null background as possible.   We are not
especially concerned if the right hand side of
Eq.~(\ref{e:bfieldchop}) is 1.9 $d_e {\mathcal E}_{\rm eff}$
rather than 2.0 $d_e {\mathcal E}_{\rm eff}$. More important to
us is that, if $d_e = 0$, the right-hand side of
Eq.~(\ref{e:bfieldchop}) be as close to zero as possible. As we
shall see, as long as we preserve certain symmetries of the
system we are guaranteed a very high quality null. A preliminary
remark is that the ``energy" of a dressed state, or more
precisely the phase shift per period $\tau$, is unaffected by an
offset in how the zero of time is defined.  A second observation
is that, in the absence of exotic particle physics (such as
nonzero eEDM), the energy levels of a diatomic molecule in
external electromagnetic fields are not affected by a global
parity inversion.

Under the action of this inversion, all the fields and
interactions in the Hamiltonian transform according to their
classical prescriptions, whereas quantum states are transformed
into their parity-related partners.  In a parity-invariant system,
parity thus changes quantum numbers, but leaves energies of the
eigenstates unchanged. This is true for the dressed states as
well, since their eigen-energies emerge formally from a
time-independent Hamiltonian.

To formulate the effect of inversion symmetry we write the
electric and magnetic fields as
\begin{eqnarray}
 \vec{\mathcal{E}} = \mathcal{E}_\mathrm{rot} \hat{\rho}' + \mathcal{E}_z
 \hat{z} \\
 \vec{ {\mathcal B}} = {\mathcal B}_\mathrm{rot} \hat{\rho}'
\end{eqnarray}
where $\hat{\rho}'
  = \cos (\omega_\mathrm{rot} t) \hat{x} + \sin(\omega_\mathrm{rot} t)
  \hat{y}$ and $\alpha = \tan^{-1} (\mathcal{E}_z/\mathcal{E}_\mathrm{rot})$.
The dressed states defined by the rotating field are characterized
by the projection $m_F$ of total angular momentum on the axis
defined by the rotating electric field, $\vec{\mathcal{E}}_{\rm
rot}/ \mathcal{E}_{\rm rot}$. Because the magnetic field is not
strictly collinear with the electric field, and because of the
field rotation, $m_F$ is only approximately a good quantum number.
Nevertheless, considering the effect of parity on all the $m_F$'s
simultaneously, we can still map each dressed eigenstate into its
parity-reversed partner.

Assuming the ions are ``nailed down" in their axial oscillation,
at a particular value of $\mathcal{E}_z$ and thus $\alpha$,  our
various spectroscopic measurements would give dressed energy
differences ${\rm E}(\mathcal{E}_\mathrm{rot}, {\mathcal
B}_\mathrm{rot}, \alpha, m_F, \Omega)$ $-$ ${\rm
E}(\mathcal{E}_\mathrm{rot}, {\mathcal B}_\mathrm{rot}, \alpha,
-m_F, -\Omega)$. Now we invoke the following symmetry argument: if
we take the entire system, electric fields, magnetic field, and
molecule, and apply a parity inversion, that will leave the energy
of the corresponding levels unchanged.  If further we then shift
the zero of time by $\pi/\omega_\mathrm{rot}$, in effect letting
the system advance through half a cycle of the field rotation,
that also will not change the corresponding energy levels of the
dressed state, which are after all defined over an entire period
of the rotation. This transformation effectively connects
measurements made for $\alpha >0$, above the mid-plane, to those
with $\alpha <0$, below the mid-plane.  The combined transform
acts as follows:
\begin{eqnarray}
  F &\rightarrow & F \\
  \vec{{\mathcal B}}_\mathrm{rot} &\rightarrow & - \vec{{\mathcal B}}_\mathrm{rot}
  \nonumber \\
  \vec{\mathcal{E}}_\mathrm{rot} & \rightarrow &
  \vec{\mathcal{E}}_\mathrm{rot} \nonumber \\
  \mathcal{E}_z        &\rightarrow & - \mathcal{E}_z \nonumber \\
  \alpha  &\rightarrow  & - \alpha \nonumber \\
  \vec{\omega}_\mathrm{rot} &\rightarrow & \vec{\omega}_\mathrm{rot}\nonumber \\
  m_F          &\rightarrow & -m_F \nonumber \\
   \vec{S} \cdot \vec{{\mathcal B}}  &\rightarrow  & \vec{S}  \cdot \vec{{\mathcal B}}
   \nonumber \\
   \hat{d} \cdot \vec{\mathcal{E}}  & \rightarrow & \hat{d} \cdot
   \vec{\mathcal{E}} \nonumber \\
  \Omega  &\rightarrow & - \Omega \nonumber
\end{eqnarray}
The last of these is equivalent to ${\hat n} \cdot {\vec
\sigma_1}$, i.e., our symmetry operation would change the sign of
the EDM energy shift.  However, in the {\it absence} of this
shift we can expect the following exact relations between the
dressed state energies:
\begin{eqnarray}
\\
{\rm E}(\mathcal{E}_\mathrm{rot}, {\mathcal B}_\mathrm{rot},
\alpha, m_F, \Omega)
   -  {\rm E}(\mathcal{E}_\mathrm{rot}, -{\mathcal B}_\mathrm{rot}, -\alpha, -m_F, -\Omega) =0 \nonumber \\
{\rm E}(\mathcal{E}_\mathrm{rot}, -{\mathcal B}_\mathrm{rot},
\alpha, m_F, \Omega)
   -  {\rm E}(\mathcal{E}_\mathrm{rot}, {\mathcal B}_\mathrm{rot}, -\alpha, -m_F, -\Omega) =0
   \nonumber \\
{\rm E}(\mathcal{E}_\mathrm{rot}, {\mathcal B}_\mathrm{rot},
-\alpha, m_F, \Omega)
   -  {\rm E}(\mathcal{E}_\mathrm{rot}, -{\mathcal B}_\mathrm{rot}, \alpha, -m_F, -\Omega) =0
   \nonumber \\
{\rm E}(\mathcal{E}_\mathrm{rot}, -{\mathcal B}_\mathrm{rot},
-\alpha, m_F, \Omega)
   -  {\rm E}(\mathcal{E}_\mathrm{rot}, {\mathcal B}_\mathrm{rot}, \alpha, -m_F, -\Omega)
   =0. \nonumber
\end{eqnarray}
Summing four equations and rearranging terms, we get that
\begin{equation}
W^{u/\ell}(\mathcal{E}_\mathrm{rot}, {\mathcal B}_\mathrm{rot},
\alpha) + W^{u/\ell}(\mathcal{E}_\mathrm{rot}, -{\mathcal
B}_\mathrm{rot}, \alpha) +W^{u/\ell}(\mathcal{E}_\mathrm{rot},
-{\mathcal B}_\mathrm{rot}, -\alpha) +
W^{u/\ell}(\mathcal{E}_\mathrm{rot}, {\mathcal B}_\mathrm{rot},
-\alpha)=0. \nonumber
\end{equation}

If we assume that the axial confinement is symmetric (not
necessarily harmonic), and that our spectroscopy averages over an
ensemble of ions oscillating in the axial motion with no preferred
initial phase of the axial motion (we will later explore the
consequences of relaxing this assumption) then the ions will spend
the same amount of time on average at any given positive value of
$\alpha$ as they do at the corresponding negative value of
$\alpha$, and thus the averaged results yield:
\begin{equation}
   <W^{u/\ell}(\mathcal{E}_\mathrm{rot}, {\mathcal B}_\mathrm{rot})>
+ <W^{u/\ell}(\mathcal{E}_\mathrm{rot}, -{\mathcal
B}_\mathrm{rot})> = 0
\end{equation}

The combined result, in the absence of exotic particle physics, is
zero by symmetry.  We did not need to invoke the various
approximations that went into Eq.~\ref{e:bfieldchop}. In
particular, this null result is, unlike the traditional Berry's
phase result, not based on the assumption of very small
$(\omega_\mathrm{rot}/d_\mathrm{mf} \mathcal{E}_\mathrm{rot})$.
Also, for conceptual simplicity we have discussed the result as
being based on an average over quasi-static values of $\alpha$,
but the symmetry argument does not hinge on the axial frequency
being infinitely slow compared to $\omega_\mathrm{rot}$.

\subsection{Frequency- or Phase-Modulation of Axial Oscillation}
\label{ss:axialosc}

The trapped ions will oscillate in the axial direction at a
frequency $\omega_z$, confined by an approximately harmonic axial
trapping potential $U_z = (1/2)M \omega_z^2 z^2$.  Upon moving
away from the mid-plane $z=0$, the ions will experience an
oscillating axial electric field ${\cal E}_{z}(t)  = -M \omega_z^2
z(t) /e$.  The geometric phase correction to the energy  is then $
-m_F \omega_{\rm rot} \alpha_{\rm max} \cos ( \omega_z t)$, where
$\alpha_{\rm max} = {\cal E}_{z, {\rm max}}/{\cal E}_{\rm rot}$ is
the maximum excursion of the tilt angle. Because the product
$\omega_{\rm rot} \alpha_{\rm max}$ is again an energy, it is
convenient to redefine the geometric energy contribution in terms
of a frequency $\omega_{\rm max}$,
\begin{equation}
{\rm E}_{\rm geo} = \omega_{\rm max} \cos (\omega_z t),
\end{equation}
with $\omega_{\rm max} = - \delta m_F \omega_{\rm rot}
\alpha_{\rm max} $.

For $\omega_z = 2\pi \times 1$~kHz, an ion cloud temperature of 15
K, an ion whose axial energy ${\rm E}_z$ is twice the thermal
value, for $\omega_\mathrm{rot}$ and $\mathcal{E}_{\rm rot}$ as
shown in the Appendix, then a transition such as $W^{u/\ell}$ with
$\delta m = 3$ will have a maximum frequency modulation
$\omega_{\rm max} = 2\pi \times 400$~Hz.

Thus the electric field at the ion's location undergoes two
motions, the comparatively fast radial rotation, and the
comparatively slow axial wobble. We exploit the different time
scales to create, for each instantaneous value of
 $\alpha$, the rotation-dressed states worked out in the previous section.
The effect of the axial wobble is then described by the time
variation of the amplitudes in these dressed states.  The
time-dependent Schr\"{o}dinger equation of motion for this is
\begin{widetext}
\begin{equation}
\label{e:timedependent}
  i\frac{\partial}{\partial t}\left(\begin{array}{c} a\\ b\end{array}\right) =
  \left(\begin{array}{cc} \frac{3}{2}g_F\mu_B {\mathcal B}_{\rm rot} +
    \frac{\omega_{\rm max}}{2}
    \cos(\omega_{\rm z} t) & \frac{\Delta}{2}\\ \frac{\Delta}{2} &
    -\frac{3}{2}g_F\mu_B {\mathcal B}_{\rm rot}-\frac{\omega_{\rm max}}{2}
    \cos(\omega_{\rm z} t)
    \end{array}\right)\left(\begin{array}{c} a \\ b\end{array}\right),
\end{equation}
\end{widetext}
where $a$ and $b$ are the probability amplitudes for being in the
$|a\rangle$ and $|b\rangle$ states, respectively.  For typical
experimental values, $\omega_{\rm z}$ is about $2\pi \times
1$~kHz, $\omega_{\rm max}$ will range as high as $2 \pi \times
1$~kHz, and $\Delta$ (given by Eq.~\ref{e:avoidedsplitting}) is
perhaps $2\pi \times 2$~Hz, and $3 \mu_B g_F {\mathcal B}_{\rm
rot}$ is about $2\pi \times 8$~Hz.

Eq.~\ref{e:timedependent} describes a system again governed by a
periodic Hamiltonian, and we will therefore follow a similar
course to Sec.~\ref{ss:bigangle} and search for dressed-state
solutions $\Psi$ such that $\Psi(t+n \tau) = e^{-i n \phi}
\Psi(t)$.  Of course, this will only be valid in the limit that
$\omega_\mathrm{rot} \gg \omega_\mathrm{z}$, a necessary condition
to write the time-dependent Hamiltonian in
Eq.~\ref{e:timedependent}.  First, we get rid of fast
time-dependence by guessing solutions:
\begin{eqnarray}
a(t) = A(t)\displaystyle\sum_{n=-\infty}^{\infty}J_{\rm
n}\left(\frac{\omega_{\rm max}}{2\omega_{\rm z}}\right)e^{\rm
-i\omega_{
\rm z}nt} \\
b(t) = B(t)\displaystyle\sum_{n=-\infty}^{\infty}J_{\rm
n}\left(\frac{-\omega_{\rm max}}{2\omega_{\rm z}}\right)e^{\rm
-i\omega_{ \rm z}nt}, \nonumber
\end{eqnarray}
where $J_{\rm n}$ are Bessel's functions of the first kind and
A(t) and B(t) are slowly varying functions. We then substitute our
trial solutions into Eq.~\ref{e:timedependent} and use the
recurrence relation $(2n/x) J_{\rm n}(x)=J_{\rm n-1}(x) + J_{\rm
n+1}(x)$.  We multiply through by
$\displaystyle\sum_{n'=-\infty}^{\infty} J_{\rm
n'}\left(\frac{\omega_{\rm max}}{2\omega_{\rm
z}}\right)e^{i\omega_{\rm z}n't}$ or
$\displaystyle\sum_{n'=-\infty}^{\infty} J_{\rm
n'}\left(\frac{-\omega_{\rm max}}{2\omega_{\rm
z}}\right)e^{i\omega_{\rm z}n't}$ as appropriate.  We then
integrate over an axial time period, $2\pi/\omega_{\rm z}$, and
make the approximation that A(t) and B(t) are unchanged over this
small time interval. This approximation should be good as long as
$\omega_{\rm z} \gg \Delta$ and $\omega_{\rm z} \gg g_F\mu_B
{\mathcal B}_{\rm rot}$. The integration then yields,
\begin{equation}
\label{e:doubled}
 i\frac{\partial}{\partial t}\left(\begin{array}{c} A\\ B\end{array}\right) =
  \left(\begin{array}{cc} \frac{3}{2}g_F\mu_B {\mathcal B}_{\rm
  rot}
  & \frac{\Delta_{\rm eff}}{2}\\ \frac{\Delta_{\rm eff}}{2} &
    -\frac{3}{2}g_F\mu_B {\mathcal B}_{\rm rot}
    \end{array}\right)\left(\begin{array}{c} A \\
    B\end{array}\right)\\
\end{equation}
with
\begin{equation}
\Delta_{\rm eff} = \displaystyle\sum_{n=-\infty}^{\infty} J_{\rm
n}\left(\frac{\omega_{\rm max}}{2\omega_{\rm z}}\right) J_{\rm
n}\left(\frac{-\omega_{\rm max}}{2\omega_{\rm z}}\right)\Delta =
J_{\rm 0}\left(\frac{\omega_{\rm max}}{\omega_{\rm
z}}\right)\Delta.
\end{equation}

\begin{figure}
\begin{center}
\includegraphics[width=8cm]{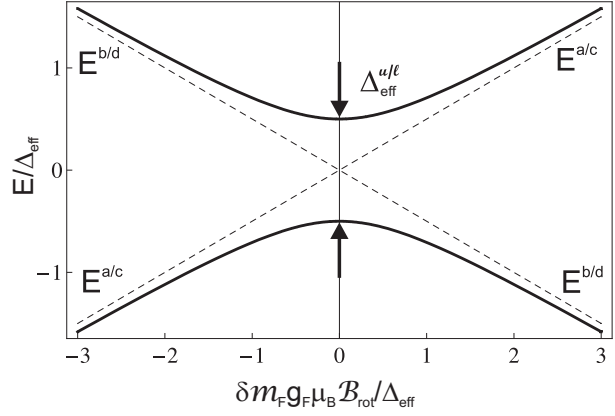}
\caption{The apparent energy shifts between $m_F = +3/2$ and $m_F
= -3/2$ states in upper (a,b) and lower (c,d) $\Lambda$-doublet
levels versus $\mathcal{B}_{\rm rot}$,``dressed'' first by the
electric field rotation ($\omega_{\rm rot}$) and then by the ion's
axial trap oscillation ($\omega_{z}$).  At $\mathcal{B}_{\rm rot}
= 0$, there is an avoided crossing that mixes $m_F = \pm 3/2$
states, with an energy splitting at the crossing of $\Delta_{\rm
eff}^{u/\ell}$. In the limit $\delta m_{F} g_F \mu_{B}
\mathcal{B}_{\rm rot} \gg \Delta_{\rm eff}$, the dressed states
are of good $m_F$ with an energy splitting slightly modified by
$\Delta_{\rm eff}$.\label{f:anticrossingBrot}}
\end{center}
\end{figure}

This results in dressed-state energies, now as a function of
${\mathcal B}_{\rm rot}$, and not $\alpha$, as seen in
Fig.~\ref{f:anticrossingBrot}. This clearly shows the requirement
of $3g_F\mu_B {\mathcal B}_{\rm rot}
> \Delta_{\rm eff}$ in order to keep $|a\rangle$ and $|b\rangle$ as the
dressed states. This is true despite the fact that an ion will
sample the avoided crossing in Fig.~\ref{f:anticrossing} during
its axial oscillation in the trap, as $\omega_{\rm max} \gg
3g_F\mu_B {\mathcal B}_{\rm rot}$ in our experiment.  $\Delta_{\rm
eff}$ will have a maximum value of $\Delta$ at $\omega_{\rm
max}/\omega_{\rm z}= 0$ and will oscillate about zero according to
$J_{\rm 0}(\omega_{\rm max}/\omega_{\rm z})$.

For finite $\omega_{\rm max}/\omega_{\rm z}$, the dressed states
from Eq.~\ref{e:doubled} only appear stationary if measured at
integer multiples of the axial trapping period, $2\pi/\omega_{\rm
z}$.
 Consider states $|+\rangle$ and $|-\rangle$, symmetric and antisymmetric combinations
of states $|a\rangle$ and $|b\rangle$, respectively. In the limit
that $\delta m_{F} g_F \mu_{B} \mathcal{B}_{\rm rot} \gg
\Delta_{\rm eff}$, an ion initially in state $|+\rangle$ will
oscillate between $|+\rangle$ and $|-\rangle$ at the precession
frequency $\omega_0 = ((3 {g^{u/\ell}_F \mu_B \mathcal
B}_\mathrm{rot} + d_e \mathcal{E}_{\rm eff})^2 + \Delta_{\rm
eff}^2)^{\frac{1}{2}}$, when measured at integer multiples of the
axial trapping frequency. However, if our EDM measurement is made
after a non-integer number of axial oscillations, or if the ions
have different axial frequencies in the trap, the $|+\rangle$ to
$|-\rangle$ oscillation will be frequency modulated at
$\omega_{\rm max}$. For the example parameters, the
frequency-modulation index $\omega_{\rm max}/\omega_z$ is less
than 1, and thus the spectral power of transition is
overwhelmingly at $\omega_0$, the quantity which symmetry
arguments above show is unaffected by Berry's phase. In an
ensemble of ions which have a random distribution of initial axial
motions, the sidebands on the transition average to zero, and
won't pull the frequency of the measured central transition. If
instead the process of loading ions into the trap has left the
ions with an initial nonzero axial velocity or axial offset from
trap center, the measured frequency can be systematically pulled
from $\omega_0$.

We note that increasing $\mathcal{E}_\mathrm{rot}$ or decreasing
$\omega_\mathrm{rot}$ reduces the value of $\omega_{\rm max}$ and
thus the frequency modulation index. On the other hand, these
changes also would have the effect of increasing the energy ${\rm
E}_{\rm rot}$ of the micromotion of the ions in the rotating
fields. For harmonic axial confinement, we find the frequency
modulation for a $\delta m = 3$ transition obeys the following
relation $\omega_{\rm max}/\omega_z =  3 ({\rm E}_z/{\rm
E}_\mathrm{rot})^{1/2}$. Thus to keep the modulation safely under
unity for a comfortable majority of an ensemble of ions with an
average ${\rm E}_z$ given by $T_z$, one needs to choose operating
parameters such that ${\rm E}_\mathrm{rot}
> 30 k_B T_z$. This inequality in turn places stringent requirements
on the spatial uniformity of $\mathcal{E}_\mathrm{rot}$.  On a
time-scale slow compared to $1/\omega_\mathrm{rot}$,
$\mathcal{E}_\mathrm{rot}$ acts like a sort of ponderomotive
potential analogous to the effective confining potential in a Paul
trap. If ${\rm E}_\mathrm{rot} = 30 k_B T_z$, then a spatial
inhomogeneity in $\mathcal{E}_\mathrm{rot}$ of only 1.5\% already
gives rise to structure in the ponderomotive potential comparable
to $T_z$.

To summarize the effect of axial motion: in the limit $3g_F \mu_B
{\mathcal B}_\mathrm{rot} > \Delta$, ions prepared, for instance
by optical pumping, in state $|a\rangle$ (or $|b\rangle$) will
remain in $|a\rangle$ (or $|b\rangle$). The energy difference
between dressed states which are predominantly either $|a\rangle$
or $|b\rangle$ will be slightly modified by the avoided crossing.
But the important combined measurement described by
Eq.~(\ref{e:bfieldchop}) will continue to yield zero for $d_e$ =
0, and the sensitivity of that combined measurement to a nonzero
EDM will not be much affected as long as $\omega_{\rm
max}/\omega_z \lesssim 1$.

\subsection{Structure of the Measurements. What Quantities Matter}
\label{ss:fourchop}

In the remainder of this section, we will look at the possible
effects of various experimental imperfections on our measurement.

The symmetry argument in Sec.~\ref{ss:staticfields} presupposes
the ability to impose a perfect ``${\mathcal B}$-chop", i.e., to
collect data with alternating measurements changing quite
precisely only the sign of ${\mathcal B}_\mathrm{rot}$. If not
only the sign but the magnitude of the rotating magnetic field
alternates, the situation is more complicated. There will likely
be contributions to the rotating magnetic field that are not
perfectly reversed in our ${\mathcal B}$-chop, including
displacement currents associated with sinusoidally charging the
electrodes that create the rotating electric field. These effects
can be quantified with a value ${\mathcal B}_\mathrm{rot}^{\rm
stray}$, and to lowest order they would appear as a frequency
offset in the chopped measurement:
\begin{equation}
  W^{u/\ell}({\vec {\mathcal E}},{\vec {\mathcal B}}_\mathrm{rot}
      + {\vec {\mathcal B}}_\mathrm{rot}^{\rm stray})
      + W^{u/\ell}({\vec {\mathcal E}},-{\vec {\mathcal B}}_\mathrm{rot}
      + {\vec {\mathcal B}}_\mathrm{rot}^{\rm stray})
      =  6 g_F^{u/\ell} \mu_B {\mathcal B}_\mathrm{rot}^{\rm stray} \pm 2
      d_e\mathcal{E}_\mathrm{eff}.
\end{equation}

This offset is very nearly the same for the upper and lower
states, to the extent that $g_F^u  \approx g_F^{\ell}$, i.e., to
the extent that
 $\delta g_F \equiv 1/2 (g_F^u - g_F^\ell) \ll g_F$. The effect of
the stray field is reduced by combining measurements from the
upper and lower states, in the form of  a ``four-way chop":
\begin{eqnarray}
  \label{e:fourwaychop} &&
  \left [ W^{u}({\vec {\cal E}}_{\rm rot},{\vec {\mathcal B}}_\mathrm{rot}
    + {\vec {\mathcal B}}_\mathrm{rot}^{\rm stray}) +
    W^{u}({\vec {\mathcal E}}_{\rm rot},-{\vec {\mathcal B}}_\mathrm{rot} +
    {\vec {\mathcal B}}_\mathrm{rot}^{\rm stray}) \right] \nonumber \\
  && - \left[ W^{\ell}({\vec {\mathcal E}}_{\rm rot},{\vec {\mathcal B}}_\mathrm{rot}
    + {\vec {\mathcal B}}_\mathrm{rot}^{\rm stray})
    + W^{\ell}({\vec {\mathcal E}}_{\rm rot},-{\vec {\mathcal B}}_\mathrm{rot} +
    {\vec {\mathcal B}}_\mathrm{rot}^{\rm stray}) \right] \nonumber \\
  && =  12 \delta g_F \mu_B  {\mathcal B}_\mathrm{rot}^{\rm stray} +4
  d_e\mathcal{E}_\mathrm{eff}.
\end{eqnarray}

It may prove to be advantageous to shim the ${\mathcal B}$-chop by
deliberately adding a non-chopped rotating magnetic field,
${\mathcal B}_\mathrm{rot}^{\rm shim}$, and adjusting its value
until experimentally we measure
\begin{equation}
  \label{e:bfieldshim}
  W^{u}({\mathcal B}_\mathrm{rot} + {\mathcal B}_\mathrm{rot}^{\rm stray}
  +{\mathcal B}_\mathrm{rot}^{\rm shim}) + W^{u}(-{\mathcal B}_\mathrm{rot} +
  {\mathcal B}_\mathrm{rot}^{\rm stray} +{\mathcal B}_\mathrm{rot}^{\rm shim}) =
  0.
\end{equation}
Then, a measurement in the lower $\Lambda$-doublet state gives
\begin{equation}
  W^{\ell}({\mathcal B}_\mathrm{rot} + {\mathcal B}_\mathrm{rot}^{\rm stray}
  +{\mathcal B}_\mathrm{rot}^{\rm shim}) + W^{\ell}(-{\mathcal B}_\mathrm{rot} +
  {\mathcal B}_\mathrm{rot}^{\rm stray} +{\mathcal B}_\mathrm{rot}^{\rm shim}) =
  -2(1+\frac{g^{\ell}_F}{g^u_F})d_e\mathcal{E}_\mathrm{eff},
\end{equation}
yielding a still more accurate value for $4
d_e\mathcal{E}_\mathrm{eff}$.

What we care about most then are: (1) Things that perturb $W^u$
and $W^\ell$ differently, in particular the quantity  $\delta
g_F$, but also the quantity $\delta \eta$, to be defined and
estimated in Sec.~\ref{ss:erot}, and (2) to a lesser extent, we
care about effects which affect $W^u({\mathcal
B}_\mathrm{rot})+W^u(-{\mathcal B}_\mathrm{rot})$ the same way as
they affect $W^\ell({\mathcal B}_\mathrm{rot})+W^\ell(-{\mathcal
B}_\mathrm{rot})$, because, to the extent that they lead to a
measurement
\begin{equation}
  W^{u}({\mathcal B}_\mathrm{rot}) + W^{u}(-{\mathcal B}_\mathrm{rot}) = + 2
  d_e\mathcal{E}_\mathrm{eff} + \delta_\mathrm{syst},
\end{equation}
we can mistake a nonzero value for $\delta_\mathrm{syst}$ as an
indicator for a nonzero value of ${\mathcal B}_\mathrm{rot}^{\rm
stray}$. In that case, the shimming procedure discussed above to
remove ${\mathcal B}_\mathrm{rot}^{\rm stray}$ would lead to a
combined result from the four-way chop of $4 d_e
\mathcal{E}_\mathrm{eff}  + (4 \delta g_F /
g_F)\delta_\mathrm{syst}$. This is down by a relative factor of
$(\delta g_F / g_F)$ compared to the effects that differentially
perturb $W^u$ versus $W^{\ell}$, but they could still be
troublesome. And (3) to a still lesser extent, we care about
imperfections that perturb individual measurements such as
$W^u({\mathcal B}_\mathrm{rot})$, even if they do not perturb the
${\mathcal B}$-chop measurement, $W^u({\mathcal
B}_\mathrm{rot})-W^u(-{\mathcal B}_\mathrm{rot})$, because, to the
extent that they vary over time, or depend on the trajectory of an
individual ion in the trap, they can reduce coherence times. This
leads not to systematic errors, but to a reduction in the overall
precision.

In addition to the ${\mathcal B}$ chop, state chop, and four-way
chop discussed above, we can perform a rotation chop, by changing
the sign of $\omega_\mathrm{rot}$.   Our hope is to keep
experimental imperfections to a level where the four-way chop is
by itself already good enough to suppress systematic error below
the desired level.  Then repeating the entire series of
measurements with the opposite sign of $\omega_\mathrm{rot}$
(rotating the field CW instead of CCW) will to the extent it
yields the same final value of $4 \mathcal{E}_\mathrm{eff} d_e$
provide a useful redundant check.

\subsection{An Estimate of $\delta g_{F=3/2}$}
\label{ss:deltagF}

There are two leading contributions to $\delta g_{F=3/2}$ in our
molecule. In the regime in which we will operate (a regime wherein
$\Omega$ is a signed quantity) they are to a good approximation
independent of each other. These two contributions are the
zero-field difference and the induced difference caused by the
applied electric field. In the zero-field limit, the former is
dominant. However, in the limit in which we are working, the
latter dominates.

The zero electric field contribution arises due to centrifugal
distortion effects in the molecular Hamiltonian. In
Sec.~\ref{ss:staticfields}, we wrote the Zeeman Hamiltonian in
Eq.~\ref{e:zeeman}. We omitted two terms which connect states of
$\Omega\rightarrow-\Omega$. Therefore, these states will give rise
to a parity dependent $g$-factor for each $J$-level. The
Hamiltonians which govern this interaction can be found with the
use of perturbation theory in a manner similar to the approach
used to find the $\Lambda$-doubling parameters. Brown \textit{et
al.}~\cite{BCM87} and Nelis \textit{et al.}~\cite{NBE91} have
written these terms as
\begin{eqnarray}\nonumber
  H_{\rm Zeeman Dist} &=& -\frac{1}{2}g_{rS}
  \mu_B({\mathcal B}_+J_-S_+S_- + {\mathcal B}_-J_+S_-S_+)\\
  H_{\rm Zeeman Doub} &=& \frac{1}{2}g_{rS}^\prime
  \mu_B({\mathcal B}_+J_+S_+^2 + {\mathcal B}_-J_-S_-^2),
\end{eqnarray}
where $H_{\rm Zeeman Dist}$ is the centrifugal distortion induced
by the magnetic field and $H_{\rm Zeeman Doub}$ is the Zeeman
induced $\Lambda$-doubling. $H_{\rm Zeeman Dist}$ is parity
independent while $H_{\rm Zeeman Doub}$ is parity dependent. Due
to the nature of the perturbation approach, we can estimate the
size of $g_{rS}^\prime$ in terms of the $\Lambda$-doubling $J=1$
energy splitting $\omega_{\rm ef}$
\begin{equation}\label{e:zero-field}
  |g_{rS}^\prime| \approx \left|\frac{\omega_{\rm ef}}{2B_e}\right|,
\end{equation}
In addition, if the ${}^3\Delta_1$ state of interest is composed
of a $(s)\sigma(d)\delta$ molecular orbital (where $(s)$ and $(d)$
refer to atomic orbitals with $l=0,2$), then
$g_{rS}=g_{rS}^\prime$ is expected. The difference in zero-field
$g$-factors is then given by twice the value in
Eq.~(\ref{e:zero-field}). It is evident that this effect is quite
small, of the order $10^{-6}$ for HfF$^+$.

The electric field dependent $g$-factor arises due to the mixing
of rotational levels $J$ in the molecule. The levels with $J=2$,
while far away in energy compared to the Stark energy $d_{\rm
mf}\mathcal{E}_{\rm rot}$, are perturbers. In the signed $\Omega$
basis, the $m_F$ sub-levels in the $J=2$ level have a smaller
$\gamma_F$ value than do the $m_F$ sub-levels in the $J=1$ level.
Therefore, the states which go up (down) in energy in the $J=1$
level ``gain'' (``run'') on (from) the $J=2$ level. When one
includes the effects of Hyperfine interactions, there are multiple
connections to each sub-level. In the $J=1$, $m_F=\pm3/2$ levels
that we are interested in, we can write an analytic expression for
the electric field dependent $\delta g_F$ factor
\begin{equation}\label{e:deltagF}
  \delta g_F (\mathcal{E}_{\rm rot}) = \sum_{J^\prime,F^\prime}
  \frac{d_{\rm mf}\mathcal{E}_{\rm rot}}{B_e(J+1)}\frac{g_F}{\gamma_{F}\Omega}
  [F,F^\prime,J,J^\prime]^2
  \left(\begin{array}{ccc}F & 1 & F^\prime \\ -m_F & 0 & m_F\end{array}\right)^2
    \left(\begin{array}{ccc}J & 1 & J^\prime \\ -\Omega & 0 & \Omega\end{array}\right)^2
      \left\{\begin{array}{ccc} F^\prime & J^\prime & I \\ J & F & 1\end{array}\right\}^2,
\end{equation}
where $[J,J^\prime,\dots]=\sqrt{(2J+1)(2J^\prime+1)\dots}$. The
terms in parentheses are $3J$-symbols while the term in curly
brackets is a $6J$-symbol. The sum runs on all states connected to
$|J,F\rangle$ by the electric field. In the case of HfF$^+$ with a
$J=1,F=3/2$ ground state, the sum contains the $J^\prime=2$ and
$F^\prime=3/2,5/2$ states. Since the rotation constant $B_e$ is
far larger than either $d_{\rm mf}\mathcal{E}_{\rm rot}$ or ${\rm
E}_{\rm hf}$, only $B_e$ is included in the perturbative
expression for $\delta g_F(\mathcal{E}_{\rm rot})$. For the
parameters here, this contribution is
\begin{equation}
  \frac{\delta g_{F=3/2}}{g_{F=3/2}} (\mathcal{E}_{\rm rot}) =
  \frac{9d_{\rm mf}\mathcal{E}_{\rm rot}}{40B_e},
\end{equation}
which means that the fractional shift $\delta
g_{F=3/2}/g_{F=3/2}$ is a few $10^{-4}$. The same approach gives
that the electric field ``g'' factor, $\gamma_F$, will shift in
the same manner such that $\delta\gamma_F/\gamma_F \approx
10^{-4}$.

For rotating fields, another contribution to $\delta g_F$ arises
from non-vanishing value of  $\omega_\mathrm{rot}/(d_{\rm
mf}\mathcal{E}_\mathrm{rot})$. The states with $\Omega=1$ and
$\Omega=-1$ are equally affected by the rotating field since they
have an equal Stark shift in the absence of hyperfine
interactions. However, because the levels with
$|F=3/2,m_F=\pm1/2\rangle$ are repelled by the lower
$|F=1/2,m_F=\pm1/2\rangle$ states, the effective Stark difference
between $m_F$ levels with $\Omega=-1$ (upper levels) is smaller
than the same $m_F$ levels with $\Omega=+1$ (lower levels). The
scale at which this difference will appear is then determined by
how much the lower hyperfine state pushes on the upper due to the
coupling induced by the electric field.
\begin{equation}
  \frac{\delta g_{F=3/2}}{g_{F=3/2}} = \frac{\sqrt{6}}{\gamma_{F=3/2}^2}
  \frac{\omega_{\rm rot}^2}{d_{\rm mf}\mathcal{E}_{\rm rot} {\rm E}_{\rm hf}}.
\end{equation}
This fractional shift is of the order a few $10^{-4}$ and is
therefore about the same magnitude as the electric field induced
mixing of higher rotational levels.

\subsection{Dependencies on $\mathcal{E}_\mathrm{rot}$}
\label{ss:erot}

Proximity to the avoided crossing shown in
Fig.~\ref{f:anticrossing}(b) means that the transitions $W^u$ and
$W^\ell$ will have residual dependencies on
$\mathcal{E}_\mathrm{rot}$, which in turn may lead to decoherence
or systematic errors.  We characterize the sensitivity of
$W^{u/\ell}$ to small changes in $\mathcal{E}_\mathrm{rot}$ with
the following expansion
\begin{equation} \label{e:etaexpansion}
  W^{u/\ell}({\cal E}_{\rm rot} ^0 + \delta  {\cal E}_{\rm rot},
  {\mathcal B}_\mathrm{rot}) =
  W^{u/\ell}( \mathcal{E}_\mathrm{rot}^0 , {\mathcal B}_\mathrm{rot})
  + \eta^{u/\ell} \delta \mathcal{E}_\mathrm{rot}
\end{equation}
with
\begin{equation}
  \eta^{u/\ell} \equiv
  \frac{ \partial W^{u/\ell} } {\partial \mathcal{E}_\mathrm{rot} }
  \vert_{\mathcal{E}_\mathrm{rot}^0,{\mathcal B}_{\rm rot} }
  = \frac{(\Delta^{u/\ell})^2}{g_F^{u/\ell}
    \mu_B{\mathcal B}_{\rm rot}{\mathcal E}_{\rm rot}},
\end{equation}
using the expressions in Eqs.~(\ref{e:avoidedsplitting})
and~(\ref{e:avoidedcurves}). Any spatial inhomogeneity in
$\mathcal{E}_\mathrm{rot}$ that does not average away with ion
motion will lead to a decoherence rate given approximately by
$\eta \delta \mathcal{E}_\mathrm{rot}$.

In terms of systematic errors, if chopping the sign of ${\mathcal
B}_{\rm rot}$ gives rise to an unintended systematic change in the
magnitude of ${\mathcal E}_{\rm rot}$ (call it $\delta{\mathcal
E}_{\rm chop}$), for instance due to motional fields discussed
later, or due to ohmic voltages generated by the eddy currents,
then there will be a frequency shift in a ${\mathcal B}$-chop
combination, $2\eta^{u/\ell}\delta{\mathcal E}_{\rm chop}$. To the
extent that $\delta\eta=\frac{1}{2}(\eta^u-\eta^\ell)$ is nonzero,
some of this shift will survive a four-way chop as well. The
dominant contribution to $\delta \eta$ is likely from
$\delta\Delta$, rather than from $\delta g_F$. Assuming this
limit, the systematic error surviving is
\begin{equation}
  8\frac{\delta\Delta}{\Delta}\eta \delta {\mathcal E}_{\rm chop} \approx
  6\left(\frac{d_{\rm mf}{\mathcal E}_{\rm rot}}{{\rm E}_{\rm hf}}\right)^2
  \eta\delta{\mathcal E}_{\rm chop}.
\end{equation}

For a large but not inconceivable value for $\delta
\mathcal{E}_{\rm chop}$ of 100~$\mu$V/cm, and for other values as
in the Appendix, this works out to comfortably less than
100~$\mu$Hz, and is therefore not a problem. But this error would
scale as $\mathcal{E}_{\rm rot}^{-5}$, and thus could cause
trouble if for other reasons we chose to decrease
$\mathcal{E}_{\rm rot}$. The science signal is roughly independent
of ${\mathcal E}_{\rm rot}$, which should allow for the source of
error to be readily identified.

\subsection{Perpendicular ${\mathcal B}$-Fields}
\label{ss:perpB}

The quantization axis is essentially defined by ${\mathcal
E}_\mathrm{rot}$. The shift of the various levels $|a\rangle$,
$|b\rangle$, $|c\rangle$, $|d\rangle$ due to a component of the
magnetic field perpendicular to $\mathcal{E}_\mathrm{rot}$ is on
the order of
\begin{equation}
  \pm\frac{3}{4}\frac{(g_F \mu_B {\mathcal B}_{\bot})^2}{
    \gamma_{F} d_\mathrm{mf}\mathcal{E}_\mathrm{rot}}
\end{equation}
for the upper/lower states.  In the absence of rotation, the
lowest-order correction to $W^{u/\ell}({\mathcal
B}_\mathrm{rot})$ goes as
\begin{equation}
  -\frac{3}{2}\frac{ g_F^3 \mu_B^3 {\mathcal B}_\bot^2{\mathcal B}_{\rm rot}}
       {(\gamma_{F}d_{\rm mf}\mathcal{E}_{\rm rot})^2}.
\end{equation}
For reasonable experimental parameters, this will be a negligible
number.  The lowest-order correction to the state-chop
combination, $W^u({\mathcal B}_\mathrm{rot}) - W^{\ell}({\mathcal
B}_\mathrm{rot})$ is smaller still and goes as
\begin{equation}
  \frac{3}{2}\frac{g_F^3 \mu_B^3  {\mathcal B}_\bot^2{\mathcal B}_{\rm rot}}
       {\gamma_{F}d_{\rm mf}\mathcal{E}_{\rm rot}{\rm E}_{\rm hf}}.
\end{equation}
It is similar in form to the difference in $g$-factors caused by
the rotation of the field.

When we turn on rotation, there is an additional larger
contribution to $W^{u/\ell}({\mathcal B}_\mathrm{rot})$.   If we
assume (as a worst case) that ${\mathcal B}_{\bot}$ is purely
axial,  not azimuthal, then the lowest-order effect of ${\mathcal
B}_{\bot}$ is to tilt the quantization axis by angle given by
\begin{equation}
  \frac{\pm g_F \mu_B \mathcal{B}_{\bot}}
       {\gamma_{F}d_{\rm mf}\mathcal{E}_\mathrm{rot}
     \pm \mu_B g_F {\mathcal B}_{\rm rot}},
\end{equation}
with the +(-) in the numerator corresponding to the upper(lower)
states and the +(-) in the denominator corresponding to the
$\Omega$ = -1(+1) states.  This has the leading order effect on
$W^{u/\ell}$ of
\begin{equation}
  \frac{3\omega_{\rm rot} g_F^2 \mu_B^2 {\mathcal B}_\bot {\mathcal B}_{\rm rot}}
       {(\gamma_{F}d_{\rm mf}\mathcal{E}_{\rm rot})^2}
\end{equation}
even a rudimentary nulling of the Earth's magnetic field, say to
below 25 mG, will leave this term negligible, for parameters in
the Appendix.  Its contribution to the state chop, $W^u({\mathcal
B}_\mathrm{rot}) - W^\ell({\mathcal B}_\mathrm{rot})$, is still
smaller by $d_{\rm mf}{\mathcal E}_{\rm rot}/{\rm E}_{\rm hf}$
\begin{equation}
  \frac{\omega_{\rm rot} g_F^2 \mu_B^2
    {\mathcal B}_\bot {\mathcal B}_{\rm rot}}
       {\gamma_{F}d_{\rm mf}\mathcal{E}_{\rm rot}{\rm E}_{\rm hf}}
\end{equation}

\subsection{Stray Contributions to ${\mathcal B}_{||}$: Uniform or Time-Varying ${\mathcal B}$ Fields}
\label{ss:strayB}

In the previous section we have seen that the effects of
${\mathcal B}_\bot$ are small.  Spatial or shot-to-shot variation
in ${\mathcal B}_{||}$, on the other hand,  can limit coherence
time through its contribution to $W^{u/\ell}$. The biggest
contribution to ${\mathcal B}_{||}$ is of course the intentionally
applied rotating field ${\mathcal B}_\mathrm{rot}$.  Let's examine
the various other contributions to ${\mathcal B}_{||}$.

Static, uniform fields:  ${\mathcal B}$ fields of this nature are
relatively harmless.  ${\mathcal B}_{||}$ is defined relative to
the quantization axis $\hat{\mathcal{E}}_{\rm rot}$. The
time-average of ${\mathcal B}_{||}$ is $\langle {\mathcal B}_{||}
\rangle = \langle \vec{{\mathcal B}} \cdot \hat{\mathcal{E}}_{\rm
rot}\rangle$. Since $\hat{\mathcal{E}}_{\rm rot}$ sweeps out a
circle with angular velocity $\omega_\mathrm{rot}$, the
contribution to the time-averaged ${\mathcal B}_{||}$ from a
uniform, static magnetic field averages nearly to zero in a single
rotation of the bias electric field, and still more accurately
after a few cycles of axial and radial motion in the trap.  The
average electric field in the ion trap must be very close to zero,
or the ions would not remain trapped. In the case of certain
anharmonicities in the trapping potential, however, one can find
that the average value of $\hat{\mathcal{E}}_{\rm rot}$ is
nonzero, even if the average value of $\vec{\mathcal{E}}_{\rm
rot}$ is zero. For instance, an electrostatic potential term
proportional to $z^3$, along with a uniform axial magnetic field
${\mathcal B}_z$, will for an ion with nonzero axial secular
motion, yield a nonzero $\langle {\mathcal B}_{||} \rangle$. In
addition, nonzero ${\mathcal B}_z$ will interact with the tilt of
$\vec{\mathcal{E}}_{\rm rot}$ oscillating with an ion's axial
motion at $\omega_z$ to cause a frequency modulation similar to
the one discussed in Sec.~\ref{ss:axialosc}.  A uniform magnetic
field in the x-y plane will cause a frequency modulation at
$\omega_\mathrm{rot}$. If the modulation index for either of these
modulations approaches one, the modulation will begin to suppress
the contrast of spectroscopy performed at the carrier frequency.
For uniform magnetic fields with amplitude less than 10~mG
(achievable for instance by roughly nulling the earth's field with
Helmholtz coils), frequency modulation indices will be small, and,
barring pathologically large $z^3$ electrostatic terms, the mean
shifts from uniform, static ${\mathcal B}$ fields will be less
than 1 Hz and can be can be dealt with by means of an applied
${\mathcal B}_\mathrm{rot}^{\rm shim}$ as discussed in
Sec.~\ref{ss:fourchop}.

Time-varying magnetic fields with frequency near
$\omega_\mathrm{rot}$ can cause more trouble.  If the time between
the two Ramsey pulses used to interrogate the frequency is
$t_\mathrm{Ramsey}$, then the dangerous bandwidth is
$1/t_\mathrm{Ramsey}$, centered on $\omega_\mathrm{rot}$. We
discuss in order (i) thermally generated fields from the
electrodes, (ii) ambient magnetic field noise in laboratory, (iii)
magnetic fields associated with the application of
$\mathcal{E}_\mathrm{rot}$, oscillating coherently with
$\mathcal{E}_\mathrm{rot}$, (iv) shot to-shot variation in
magnitude of applied ${\mathcal B}_\mathrm{rot}$, and (v) spatial
inhomogeneities in ${\mathcal B}_\mathrm{rot}$.

(i)  Proposed EDM experiments on trapped atomic species such as
Cesium are vulnerable to magnetic field noise generated by
thermally excited currents in conductors located close to the
trapped species~\cite{MUN05}.  In our case, the effect is less
worrisome because, vis-a-vis the trapped atom experiments, our
bandwidth of vulnerability is centered at much higher frequency
fields, because our molecules are trapped considerably further
from the nearest conductors, and because the sensitivity of our
measurement of $d_e$ to magnetic field noise, which goes as $g_F
\mu_B/ \mathcal{E}_\mathrm{eff}$ is down by a factor of $10^4$.
The spectral density of thermal magnetic field noise (which is
calculated in reference~\cite{HPW99} in the simplified geometry of
a semi-infinite planar conductor) will surely be less than 1
pG/Hz$^{1/2}$ in our bandwidth of vulnerability. This effect is
negligible.

(ii) Like thermal magnetic noise, technical magnetic noise in our
lab arising for instance from various nearby equipment will not so
much decohere an individual measurement as generate shot-to-shot
irreproducibility between measurements.  What level of noise are
we sensitive to?  As we discuss in Sec.~\ref{s:conclusion} below,
the precision of a single trap load is unlikely to be better than
300 mHz, meaning magnetic field noise less than 0.2
$\mu$G/Hz$^{1/2}$ won't hurt us, for a 1 s interrogation time.
 Measurements made in our lab show that there are a number of
magnetic field ``tones" of very narrow bandwidth, associated with
harmonics of 60~Hz power and various power supplies.  As long as
we choose $\omega_\mathrm{rot}$ to not coincide with one of these
frequencies, in the range of 50~kHz to 300~kHz ambient magnetic
frequency noise in our lab has spectral density typically less
than 0.02 $\mu$G/Hz$^{1/2}$. For this reason, at least for the
first generation experiment, there will be no explicit effort to
shield ambient magnetic field other than to use Helmholtz coils to
roughly null the earth's dc field. The steel vacuum chamber will
in addition provide some shielding at 100 kHz.

(iii) In traditional eEDM experiments, one of the most difficult
unwanted effects to characterize and bring under control is
magnetic fields generated by leakage currents associated with the
high voltages on the electrodes that generate the principal
electric field.  In our case the bulk of the electric field
$\mathcal{E}_\mathrm{eff}$ is generated inside the molecule. The
laboratory electric fields are measured in V/cm, not kV/cm, and
leakage currents as traditionally conceived will not be a problem
for us.  On the other hand, the electric field does rotate
rapidly, and thus the electrode potentials must constantly
oscillate. Displacement currents in the trapping volume between
the electrodes, and real currents in the electrodes themselves and
in the wire leads leading to them, will generate magnetic fields
with spatial gradients and strengths that oscillate coherently
with $\mathcal{E}_\mathrm{rot}$ at the frequency
$\omega_\mathrm{rot}$.

The spatial structure of the oscillating magnetic fields will
depend on the geometry of the electrodes and in particular on the
layout of the wire leads that provide the current to charge
them.  In principle, shim coils can be constructed just outside
the trap electrodes and driven with various phases and amplitude
of current oscillating at $\omega_\mathrm{rot}$, all in order to
further control the shape of the magnetic field.  The one
immutable fact is the Maxwell equation, $\nabla \times
\vec{{\mathcal B}} = c^{-2}
\partial \vec{\mathcal{E}} / \partial t$.

The dominant time dependence of the electric field is from the
spatially uniform rotating field, and thus for a circular field
trajectory, the dominant contribution to the magnetic field
structure goes as
\begin{equation}
    \nabla \times \vec{{\mathcal B}} = k \hat{y}'
\end{equation}
with
\begin{equation}
\label{e:k}
    k  =  c^{-2} \mathcal{E}_\mathrm{rot} \omega_\mathrm{rot}
         = 350\ \textrm{nG/cm} \times \left(\frac{\omega_\mathrm{rot}}{2\pi \times
        100\ \textrm{kHz}}\right) \left(\frac{\mathcal{E}_\mathrm{rot}}{5\
        \textrm{V/cm}}\right),
\end{equation}
where $\hat{y}'$ is the direction in the x-y plane orthogonal to
the instantaneous electric field.

The curl determines only the spatial derivatives of ${\mathcal
B}$; ${\mathcal B}$ itself only depends on the boundary
conditions.  An idealized arrangement of current carrying leads
and shim coils could in principle force the ${\mathcal B}$ field
to be
\begin{equation}
    \vec{{\mathcal B}}^\mathrm{ideal} =  k x' \hat{z}.
\end{equation}
where $k$ is given by Eq.~\ref{e:k} and $x'$ is displacement in
the x-y plane along the direction of the instantaneous rotating
electric field.  These fields would be perpendicular to the
quantization axis provided by the electric field, and would have
negligible effect on the transitions of interest.

While realizing such an idealized displacement field would be very
difficult, there are relatively simple steps to take to minimize
the displacement fields. For instance, each rod-like electrode can
be charged up by two leads, one connected to each end of the rod,
with the leads running along respective paths symmetric in
reflection in the z=0 plane to a common oscillating voltage source
outside of the vacuum can, at z=0.  It is worth considering a
maximally bad electrode layout, to put a limit on worst-case
performance. Our electrodes will be spaced by about 10 cm and
mounted in such a way that their capacitance to each other or to
ground will be at worst 5 pF.  If the charging current is provided
entirely by a single lead connected to one end of the rod, the
peak current running down the rod near its center will be
80~$\mu$A, leading to a worst-case field magnitude at the trap
center of about 20~$\mu$G, and a contribution to $W^{u/\ell}$ of
perhaps 2.5 Hz. Spatial gradients of this effect, and shot-to-shot
irreproducibility of this effect will not contribute to
decoherence at the 0.1 Hz level. As for its contribution to
systematic error, this shift will survive the ${\mathcal B}$-chop,
but will be suppressed in the four-way chop by the factor $(\delta
g_F / g_F)$, perhaps a factor of a thousand. For still better
accuracy the shift should be nulled out of the ${\mathcal B}$ chop
by adjusting ${\mathcal B}_{\rm rot}^{\rm shim}$, as discussed in
Sec.~\ref{ss:fourchop}.

(iv)Given that the main effect of ${\mathcal B}_\mathrm{rot}$ is
to apply an offset frequency, $3g_F\mu_B {\mathcal
B}_\mathrm{rot}$ of perhaps 8 Hz, and given that (see
Sec.~\ref{s:conclusion}) the single-shot precision is unlikely to
be any better than 300 mHz, the shot-to-shot reproducibility of
${\mathcal B}_{\rm rot}$ need be no better than a part in 30, a
very modest requirement on stability. Decoherence then is not a
problem, but a potential source of systematic error arises if the
the $\mathcal{B}$ chop is not ``clean" that is if ${\mathcal
B}_\mathrm{rot}$ before the chop is not exactly equal to
$-{\mathcal B}_\mathrm{rot}$ after the chop. This sort of error
could arise for instance from certain offset errors in op-amps
generating the oscillating current. Experimentally, one adjusts
${\mathcal B}_\mathrm{rot}^{\rm shim}$ to cancel these offsets,
but even in the absence of that procedure, the four-way chop
cleans up these sorts of errors. For a rather egregious fractional
deviation from ${\mathcal B}$-chop cleanliness of, for instance,
1\%, and for $(\delta g_F / g_F) <0.001$, the systematic error
remaining after the four-way chop is $10^{-5}$ of the offset
frequency of perhaps 8 Hz. In HfF$^+$ this is a systematic error
on $d_e$ of $10^{-29}$ e cm.  For ThF$^+$ the error as referred to
$d_e$ is smaller still, and of course if we avail ourselves  of
${\mathcal B}_\mathrm{rot}^{\rm shim}$ so as to null the post
${\mathcal B}$-chop signal to $<$100~mHz, the systematic error on
$d_e$ will be less than $10^{-29}$ e cm for either species.

 (v) The largest single contribution to
decoherence (with the exception of spontaneous decay of the
${}^3\Delta_1$ line to a lower electronic state) will likely be
due to spatial inhomogeneity in the applied rotating bias field
${\mathcal B}_\mathrm{rot}$. That is to say, spatial
inhomogeneities in $\vec{{\mathcal B}}$ that rotate in the x-y
plane at frequency $\omega_\mathrm{rot}$.  First-order spatial
gradients in ${\mathcal B}_\mathrm{rot}$ are not important,
because ion secular motion in the trap will average away the
effects of these gradients leaving only the value of ${\mathcal
B}_\mathrm{rot}$ at the center of the trap. Second-order spatial
gradients on the other hand will lead to nonzero average frequency
shifts whose value will vary from ion to ion in a thermal sample
of ions, depending on conserved quantities of individual ion
motion like the axial secular energy E$_{\rm z}$ or radial secular
energy E$_\rho$, quantities with thermally averaged values of
$kT_z$ and $kT_\rho$, respectively, and with ion-to-ion variation
comparable to their mean values. The ${\mathcal B}_\mathrm{rot}$
will be generated by current-carrying rods which are of necessity
within the vacuum chamber because of the screening effects of a
metal vacuum chamber.  Unless particular care is taken in the
design of these rods, the second-order spatial gradients in
${\mathcal B}_\mathrm{rot}$ will scale as $1/X^2$, where $X$ is
the characteristic size (and spacing) of the current carrying
rods. The contribution to the inhomogeneity of the time-averaged
value of ${\mathcal B}_\mathrm{rot}$ experienced by a thermal
sample of ions orbiting in a cloud with r.m.s size $r$ is then of
order $(r^2/X^2) {\mathcal B}_\mathrm{rot}$, leading to an
ion-to-ion frequency variability of order $(r^2/X^2) 3 g_F \mu_B
{\mathcal B}_\mathrm{rot}$ For planned parameters of the
experiment, $(r^2/X^2)$ is of order 0.01. We have seen from
Sec.~\ref{ss:axialosc} above that the quantity $3 g_F \mu_B
{\mathcal B}_\mathrm{rot}$ must be at least about five times
larger than $\Delta$ in order to make the eigenstates in the
rotating fields be states of good $m_F$. Thus in the absence of
explicit apparatus design to null the second-order spatial
gradient in ${\mathcal B}_\mathrm{rot}$ (The rod-like electrodes
that bear the charge that generates $\mathcal{E}_\mathrm{rot}$ are
in the second-generation trap the same objects that carry the
current that generates ${\mathcal B}_\mathrm{rot}$ and thus their
shape is already subject to multiple design constraints) we may
have to live with a decoherence rate from this effect on the order
of $0.05 \Delta$, perhaps 0.5 s$^{-1}$, for the experimental
values given in the Appendix.

The inhomogeneity in ${\mathcal B}_\mathrm{rot}$ should reverse
quite cleanly with the ${\mathcal B}$ chop, and residual
imperfections there will be cleaned up with the four-way chop, and
thus the effects of the second-order gradients in ${\mathcal
B}_\mathrm{rot}$ are expected to be predominantly a source of
decoherence, rather than systematic error on measured $d_e$.

\subsection{Stray Contributions to ${\mathcal B}_{||}$: Static ${\mathcal B}$-Field Gradients}
\label{ss:Bgrad}

We now return to discussing static magnetic fields, now including
the effects of spatial gradients. With the characteristic size of
the ion cloud $r$ being smaller than the characteristic distance
$X$ from cloud center to source of magnetic field by a ratio of
0.1 or smaller, it makes sense to expand the field about the
uniform value at the trap center. The most general first-order
correction to a static magnetic field in the absence of local
sources can be characterized by five linearly independent
components as follows:
\begin{eqnarray}\nonumber
\label{e:bgrads}
  \vec{{\mathcal B}} &=& {\mathcal B}'_\mathrm{axgrad}(z \hat{z} -\frac{x}{2} \hat{x} -
  \frac{y}{2} \hat{y})\\
  \nonumber && + {\mathcal B}'_\mathrm{trans} (x \hat{x} - y \hat{y}) \\
  \nonumber && + {\mathcal B}'_1  (y \hat{x} + x \hat{y}) \\
  \nonumber && + {\mathcal B}'_2  (z \hat{x} + x \hat{z}) \\
  && + {\mathcal B}'_3   (y \hat{z} + z \hat{y})
\end{eqnarray}
By far the most important effect of these terms is the
``micromotion-axial gradient interaction." As discussed in
Sec.~\ref{ss:paultrap} above, the displacement of an ion's
circular micromotion $\vec{r}_{\rm rot}$ is exactly out of phase
with the rotation of its quantization axis $\hat{\mathcal{E}}$,
see Eq.~\ref{e:rotDisplacement}. Averaged over a cycle of
$\omega_\mathrm{rot}$, this will give rise to a nonzero average
contribution to ${\mathcal B}_{||}$ and cause a shift in
$W^{u/\ell}$ given by $3 g_F \mu_B {\mathcal B}'_\mathrm{axgrad}
r_\mathrm{rot} = 3 g_F \mu_B {\mathcal B}'_\mathrm{axgrad} e
{\mathcal E}_\mathrm{rot}/(M \omega_\mathrm{rot}^2)$. A guess for
a possible value of stray ${\mathcal B}'_\mathrm{axgrad}$ is 2
mG/cm, which for anticipated experimental parameters would lead to
a shift in $W^{u/\ell}$ of order 4 Hz, and this shift would
survive the ${\mathcal B}$ chop. As with the effect of
displacement currents, one expects the systematic effect of the
shift to be reduced after the four-way chop by $(\delta g_F /
g_F)$, but for maximum accuracy the effect should be shimmed out
of the ${\mathcal B}$ chop, either by adjusting the value of
${\mathcal B}_\mathrm{rot}^{\rm shim}$, or by applying (say with
anti-Helmholtz coils external to the vacuum chamber) a
compensating value of ${\mathcal B}'_\mathrm{axgrad}$.

A smaller effect arises from the interaction of the magnetic field
gradient with the component of the electric fields responsible for
providing ion confinement, which after averaging over cycles of
$\omega_{\rm rot}$ and $\omega_{\rm rf}$, always point inward,
giving rise to a net inward-pointing time average of
$\hat{\mathcal{E}}$. If we look at only the component of the
first-order magnetic field gradient that points towards or away
from the trap center
\begin{equation}
    \vec{{\mathcal B}}^\mathrm{central} =
    ({\mathcal B}^\prime_\mathrm{trans}-{\mathcal B}^\prime_\mathrm{axgrad}/2) x \hat{x} +
    (-{\mathcal B}^\prime_\mathrm{trans}-{\mathcal B}^\prime_\mathrm{axgrad}/2) y \hat{y} +
    {\mathcal B}^\prime_\mathrm{axgrad} z \hat{z}.
\end{equation}

The net contribution to  ${\mathcal B}_{||}$ comes from
integrating, along the rf and rotation micromotion trajectories,
over first a rotational cycle, and then an rf cycle, and then a
secular cycle in a given direction.  We assume that the trap is
sufficiently harmonic that there is no cross-dimensional mixing of
secular energy, that $\omega_{\rm x}$, $\omega_{\rm y}$, and
$\omega_{\rm z}$ are incommensurate and with principle axes as
defined in Eq.~\ref{e:bgrads}, and that hard-momentum-changing
collisions are rare enough so that, during the duration of a
spectroscopic measurement, there is no change in ${\rm E}_i$, the
sum of the kinetic and potential energy associated with an
individual ion's secular motion in the $i$th direction.  The
contribution to ${\mathcal B}_{||}$ is then,
\begin{equation}
  \delta{{\mathcal B}}_{||} =  -(
  ({\mathcal B}^\prime_{\rm trans}/2-{\mathcal B}^\prime_{\rm axgrad}/4) {\rm E}_x +
  (-{\mathcal B}^\prime_{\rm trans}/2-B^\prime_{\rm axgrad}/4) {\rm E}_y +
    {\mathcal B}^\prime_{\rm axgrad} {\rm E}_z)/(e\,{\mathcal E}_{\rm
    rot}).
\end{equation}
The contribution to ${\mathcal B}_{||}$ averaged over a thermal
sample of ions is given by the above expression with ${\rm E}_i$
replaced by $T_i$. Note that for $T_x = T_y = T_z$, several terms
cancel and the thermally averaged contribution to ${\mathcal
B}_{||}$ is just ${\mathcal B}^\prime_{\rm axgrad} {\rm
E}_z/(2e\mathcal{E}_\mathrm{rot})$. The decohering effect is
comparable because within a thermal sample, ${\rm E}_x$, ${\rm
E}_y$, and ${\rm E}_z$ will in general differ from one another for
a given ion, and between different ions.  For ${\mathcal
B}^\prime_\mathrm{axgrad}$ and ${\mathcal
B}^\prime_\mathrm{trans}$ each about 2 mG/cm, ion temperatures
about 15 K, the mean shift in W$^{\rm u/\ell}$ for typical
experimental parameters given in the Appendix might be 30~mHz,
with a comparable contribution to dephasing.

The three remaining terms in the first-order gradient, ${\mathcal
B}^\prime_1$, ${\mathcal B}^\prime_2$ and ${\mathcal B}^\prime_3$,
will contribute to a shift in ${\mathcal B}_{||}$ only when
combined with other (usually small) trap imperfections, for
instance the plane of rotation of $\mathcal{E}_\mathrm{rot}$ being
tilted with respect to the principal axes of the confining
potential. The net effects will be correspondingly smaller than
those from ${\mathcal B}^\prime_\mathrm{trans}$.

Just as with the second spatial derivative of ${\mathcal
B}_\mathrm{rot}$, the spatial derivative of ${\mathcal B}'_{\rm
axgrad}$, coupled to a thermal spread in the size of ion orbits,
can give rise to decoherence. Of course, ${\mathcal B}_{\rm
axgrad}$ is defined already as a first spatial derivative of a
magnetic field, thus the dephasing arises from a third derivative
of the field, and its rate should be down from the mean size of
the shift (roughly estimated above at 4 Hz) by a factor of order
$(r/X)^2$, or a factor of one hundred. Even spatially uniform
${\mathcal B}'_\mathrm{axgrad}$ could give rise to decoherence if
there is a spatial dependence in $r_\mathrm{rot}$. The fractional
change in $r_{\rm rot}$ is the same as the fractional change in
$\mathcal{E}_\mathrm{rot}$. As discussed in
Sec.~\ref{ss:paultrap}, this should be smaller than 0.5\% over the
typical size of the ion sample.

As a coda to this subsection, it is worth considering that
applying a very spatially uniform ${\mathcal B}_\mathrm{rot}$ may
be very challenging because of difficult-to-model eddy currents
induced in electrodes and light-gathering mirrors.  On the other
hand a purposely applied static ${\mathcal B}'_\mathrm{axgrad}$
would be perturbed only by the magnetic permeability of trap
construction materials, which can be minimized and modeled.  One
way or another we will need to bias away from the avoided crossing
discussed in Sec.~\ref{ss:bigangle}, but it may turn out that this
can be accomplished with greater spatial uniformity and thus with
a lower total decoherence rate by omitting the applied ${\mathcal
B}_{\rm rot}$ altogether, and providing the bias with a
deliberately applied ${\mathcal B}'_\mathrm{axgrad}$ field.  The
${\mathcal B}$ chop could be accomplished by chopping the sign of
${\mathcal B}^\prime_\mathrm{axgrad}$. The parity invariance
argument of Sec.~\ref{ss:bigangle} above can readily be modified
to describe a chop of ${\mathcal B}^\prime_\mathrm{axgrad}$ rather
than a chop in ${\mathcal B}_\mathrm{rot}$.

To sum up subsections IV.J and IV.K, we have looked at a range of
ways in which various contributions to ${\mathcal B}_{||}$ can
shift W$^u$ and W$^{\ell}$. Decoherence due to shot-to-shot
fluctuations or spatial inhomogeneity should not be a problem out
to beyond 1 s coherence times. Various effects can shift W$^u$ and
W$^{\ell}$ by as much as a few Hz, and this shift can survive a
$\mathcal{B}$ chop. With $\delta g_F / g_F$ on order of 10$^{-4}$,
and $\mathcal{E}_{\rm eff}$ estimated at 90~GV/cm in ThF$^+$,
after a four-way chop the remaining systematic error will be a few
$10^{-29}$ e cm, but this can be dramatically reduced by tuning
away the post-$\mathcal{B}$-chop signal with $\mathcal{B}^{\rm
shim}$. The most dangerous systematic error would be if
$\mathcal{B}_{\rm rot}$ were systematically different between
measurements on the upper and on the lower states.  Chopping
between upper and lower states will be determined by variations in
optical pumping, which should be well decoupled from the
mechanisms that generate $\mathcal{B}_{\rm rot}$.

\subsection{Relativistic (Ion-Motion-Induced) Fields}
\label{ss:relativistic}

The largest component of the velocity on the ions is that of the
micromotion induced by $\mathcal{E}_\mathrm{rot}$; for reasonable
experimental parameters it will be less than 1000 m/s.  In typical
lab-frame magnetic fields of a few mG, the motion will give rise,
through relativistic transformation, to electric fields of order
of a few $\mu$V/cm, which are irrelevant to our measurement.
Conversely, motion at 1000 m/s in typical lab-frame electric
fields of 10 V/cm generates a magnetic field of 0.1 $\mu$G. This
field will be rigorously perpendicular to the electric field, the
quantization axis, and thus represents only a negligible
modification to the generally unimportant ${\mathcal B}_{\bot}$.

\begin{figure}
\begin{center}
\includegraphics{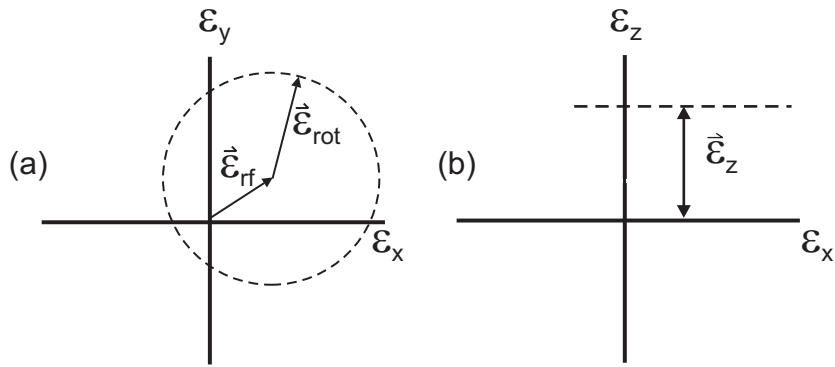}
\caption{Over one rotation of $\mathcal{E}_\mathrm{rot}$, both
$\mathcal{E}_\mathrm{rf}$ and $\mathcal{E}_z$ are quasistatic. The
total electric field is the sum of all three and its trajectory
over one cycle of $\mathcal{E}_\mathrm{rot}$ is plotted as the
dotted line projected onto (a) the x-y and (b) the x-z planes. The
electric field trajectory is a circle of radius
$\mathcal{E}_\mathrm{rot}$, parallel to and displaced from x-y
plane, a circle whose center is offset from the z-axis by
$\mathcal{E}_\mathrm{rf}$. In the limit
$\mid\mathcal{E}_\mathrm{rf}\mid$ $\ll$ $\mid{\mathcal
E}_\mathrm{rot}\mid$, the solid angle subtended from the origin by
this circle differs only slighlty from that subtended by a circle
with vanishing $\mathcal{E}_\mathrm{rf}$. The magnitudes of both
$\mathcal{E}_\mathrm{rf}$ and $\mathcal{E}_z$ relative to
$\mathcal{E}_\mathrm{rot}$ are very much exaggerated for clarity.
\label{f:Erot}}
\end{center}
\end{figure}

\subsection{Effect of RF Fields}

The effects of the rf electric fields providing Paul trap
confinement are best understood by putting them in the context of
a three-tier hierarchy of electric field magnitudes and
frequencies.

(i) $\mathcal{E}_{\rm rot}$, the nominally uniform, rotating
electric field, with field magnitude of perhaps 5 V/cm and
frequency $\omega_{\rm rot} = 2\pi \times 100$~kHz.

(ii)  $\mathcal{E}_{\rm rf}$, the Paul-trap fields, are highly
inhomogeneous, but at a typical displacement in the x-y plane of
perhaps 0.5 cm, the field strength might be 75 mV/cm, or two
orders of magnitude less than that of $\mathcal{E}_{\rm rot}$,
oscillating at a frequency, $\omega_{\rm rf} = 2\pi\times 15$~kHz
which is one order of magnitude less than $\omega_{\rm rot}$.  At
a fixed point in space, the rf fields average rigorously to zero
over time, but averaged instead along an ion's rf micromotion
trajectory, the rf fields contribute to

(iii) the inward-pointing trapping electric field, again very
inhomogeneous but with typical strength down from peak rf-field
values by factor of ($\omega_{\rm rf}$/$\omega_i$), another order
of magnitude, to perhaps 5 mV/cm. From the ion's perspective, the
direction of the trapping fields oscillate with the ion's secular
motions in the trap,  at frequencies $\omega_i$ of perhaps
$2\pi\times 1$~kHz, the slowest time scale by an order of
magnitude.

The effects of the strong, fast $\mathcal{E}_{\rm rot}$ have been
discussed extensively throughout Sec.~\ref{s:spectroscopy}, and
those of the weak, slow trapping fields were covered in
Sec.~\ref{ss:Bgrad} above. In this subsection we argue that the rf
electric fields, intermediate in both frequency and strength, are
the least significant of the three categories.

The effects of the rf fields averaged over the rf micromotion
trajectory are discussed in Sec.~\ref{ss:Bgrad}. The remaining
part averages to zero in one rf cycle, but is roughly frozen at a
single value over the duration of one cycle of $\omega_{\rm rot}$.
The dominant source of the rf fields' time-averaged contribution
to transitions W$^{u/l}$ is in very small corrections to Berry's
phase energy associated with the rotation of $\mathcal{E}_{\rm
rot}$. See Fig.~\ref{f:Erot}. The correction to the solid angle
arising from E$_{\rm rf}$ goes as (E$_z$/E$_{\rm rot}$)(E$_{\rm
rf}$/E$_{\rm rot}$)$^2$. If we include a factor of $\omega_{\rm
rot}$ to get a Berry's energy shift and evaluate for typical
experimental parameters, the magnitude of the resulting frequency
shift will be about 20 mHz, and will oscillate in sign with the
axial secular motion.  The magnitude of radial rf fields scales
linearly with the radial secular displacement. If secular
freqencies were commensurate, in particular if $\omega_z$ =
2$\omega_r$, then this 20 mHz shift could contribute to a
decoherence rate at the negligible level of a few tens of mHz. For
incommensurate ratios of $\omega_z$/$\omega_x$ or
$\omega_z$/$\omega_y$, the rf fields will be still less important.

\subsection{Systematic Errors Associated with Trap Asymmetries}

The symmetry argument of Sec.~\ref{ss:bigangle} was based on
parity invariance. This argument is only as good as reflection
symmetry of the electric and magnetic fields in the region of the
trapped ions. In this section we look, as an example, at the
consequences of a symmetry imperfection.

The electrodes used to generate $\mathcal{E}_{\rm rot}$ have been
numerically designed to make $\mathcal{E}_{\rm rot}$ as spatially
uniform as possible, but imperfections in design and construction
of the trap and imperfect drive electronics will lead to some
residual field nonuniformity. Suppose that the magnitude of the
$\mathcal{E}_{\rm rot}$ was consistently larger in the region of
the trap for which z$>$0, so that the value of $\mathcal{E}_{\rm
rot}$ over the z$>$0 half of an axial secular oscillation is about
0.3\% larger than that experienced over the z$<$0 half.  Thus the
frequency modulation of perhaps $\pm$~500 Hz, discussed in
Sec.~\ref{ss:axialosc} will no longer average to precisely zero
over an axial cycle but instead a net contribution of about 1.5 Hz
to W$^u$. Such a frequency shift would survive a $\mathcal{B}$
chop, and, following the protocol discussed in section IV.F, we
could very likely incorrectly identify this shift as arising from
the presence of a B$^{\rm stray}_{\rm rot}$, and apply a value of
B$^{\rm shim}$ to largely null the 1.5 Hz shift. After a complete
four-way chop, we would be left with a systematic error on the
order of ($\delta g_F/g_F$)$\times$1.5 Hz, or about 0.4 mHz.

For the value of $\mathcal{E}_{\rm eff}$ estimated for HfF$^+$, a
0.4 mHz error corresponds to a systematic error on d$_e$ of the
order of a few $10^{-29}$ e cm. For ThF$^+$, the error on d$_e$
would be about three times smaller.  We continue a more general
discussion on systematic errors in Sec.~\ref{ss:systerr} below.

\section{Collisions}
\label{s:collisions}

The overarching strategy of the trapped-ion approach to precision
spectroscopy is to accept low count rates in exchange for very
long coherence times.  In some previous precision measurement
experiments with trapped ions, the very best results have come
from taking this to the extreme limit of working with only one
ion~\cite{CHA04,MAT08,CHW09,MAR04,DUB05,TAM07,HOS05,OSK06}, or in
some cases a pair of ions~\cite{CHK10}, in the trap at any given
time. More often however, optimal precision is achieved working
with a small cloud of trapped ions. In this section we evaluate
various detrimental effects of ion-ion interactions.

\subsection{Mean-Field}

With no electrons present to neutralize overall charge, even a
relatively low density cloud of ions can have a significant
mean-field potential.  A spherically symmetric sample of
$N_\mathrm{ion}$ ions confined within a sphere of radius $r$ will
give rise to a mean-field potential
\begin{equation}
\label{e:meanfield1}
    \frac{U_\mathrm{mean-field}}{k_B} \approx 3~\mathrm{K} \times \left(\frac{N_\mathrm{ion}}{1000}\right)\left(\frac{r}{0.5\ \mathrm{cm}}\right)^{-1}.
\end{equation}
At values of the mean-field interaction energy comparable to or
larger than $k_B T$, there is a risk of instabilities, viscous
heating, and other undesirable effects; even in their absence,
systematic errors are more difficult to analyze in the strong
mean-field limit.  Ion-trap experiments have been performed at
much higher mean-field strengths, and indeed there have been
precision spectroscopy experiments done in systems for which the
interaction potential even between an individual pair of
nearest-neighbor ions is much larger than $k_B T$.  However, these
systems exhibit a high degree of spontaneous symmetry breaking
including crystallization~\cite{TBJ95}.

For the purpose of this paper, we assume the experiments will be
done in the low mean-field limit, say
\begin{equation}
\label{e:meanfield2}
    U_\mathrm{mean-field} \lesssim \frac{1}{3} k_B T.
\end{equation}
In this limit, mean-field effects are relatively benign, and can
be modeled as a modest decrease in the trap confining frequencies,
$\omega_i$, plus the addition of some anharmonic terms to the
potential.  Crucially for the arguments presented in
Sec.~\ref{ss:bigangle}, these additional modifications do not
break any of the reflection- or rotation-based symmetries of the
trapping fields. We note that
Eqs.~\ref{e:meanfield1}~and~\ref{e:meanfield2} combine to set
limits on various combinations of the ion number,
$N_\mathrm{ion}$, ion temperature, $T$, cloud radii, $r_i \propto
\sqrt{\frac{k_B T}{M\omega_i^2}}$, and mean ion density, $n
\propto \frac{N_\mathrm{ion}}{r_x r_y r_z}$.  This necessitates
making various compromises in selecting operating parameters.

\begin{figure}
\begin{center}
\includegraphics{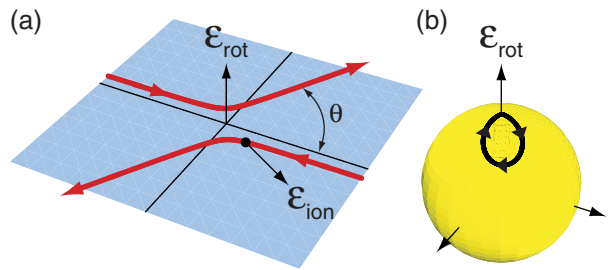}
\caption{Geometric phases accumulated during an ion-ion collision.
(a) A typical ion-ion collision trajectory (red), resultant
Rutherford scattering angle, $\theta$, and ion-ion interaction
electric field, $\mathcal{E}_\mathrm{ion}$, are shown in the
collision plane (blue).  For clarity, the collision plane has been
taken perpendicular to the instantaneous direction of
$\mathcal{E}_\mathrm{rot}$. (b) During an ion-ion collision the
molecular axis adiabatically follows the net electric field
vector, $\vec{\mathcal{E}}_\mathrm{rot} +
\vec{\mathcal{E}}_\mathrm{ion}$, and traces out the contour
(black) on the unit sphere (yellow).  The solid angle, $\Delta
\mathcal{A}(\theta)$, subtended by this contour gives rise to a
geometric phase accumulated by the eigenstates during the
collision.  This leads to decoherence of the spectroscopic
transition, see text.\label{f:collision}}
\end{center}
\end{figure}

In Sec.~\ref{ss:axialosc}, we saw that the axial component of the
electric field at the ion's location, $\mathcal{E}_z$, tilts the
rotating electric field and gives rise to an apparent shift of the
energy of our spectroscopic transition, linear in $\mathcal{E}_z$.
This energy shift integrated over time in turn gives rise to an
oscillatory phase shift, $\Delta \phi =
\frac{\omega_\mathrm{rot}}{\mathcal{E}_\mathrm{rot}} \int
\mathcal{E}_z dt$.  In a one-component ion cloud, the effects of
long-range, grazing-angle ion-ion collisions may be thought of as
simply a fluctuating component to the local electric field, and
the integrated effect of those fluctuations will make a random
contribution to the phase shift.  We present a simple argument to
show that the resulting rms spread in phase does not continue to
increase with time but reaches a steady-state asymptote.  This is
because $\mathcal{E}_z$ not only shifts the transition energy, it
also causes an axial force and corresponding acceleration, which,
like the shift in transition energy, is linear in $\mathcal{E}_z$.
Integrated over time, $\Delta p_z = e \int \mathcal{E}_z dt$, this
fluctuating force results in a fluctuating momentum.  But we know
that the combined effect of a trapping field and a large number of
random collisions will not cause the rms momentum to randomly walk
without bound but rather to be loosely bounded by a characteristic
thermal value, $\sqrt{\langle p_z^2 \rangle} \approx \sqrt{M k_B
T_z}$.  This is the nature of the thermal equilibration process --
once an ion has developed a super-thermal momentum, further
collisions are biased to reduce the momentum.  Since both the
phase excursion and the momentum excursion are linear in the
time-integrated axial electric field, we can estimate
$\sqrt{\langle \Delta \phi^2 \rangle} \approx
\frac{\omega_\mathrm{rot}}{\mathcal{E}_\mathrm{rot}} \sqrt{\langle
p_z^2 \rangle} \approx \sqrt{\frac{k_B T_z}{2 {\rm
E}_\mathrm{rot}}}$. Again, as discussed in Sec.~\ref{ss:axialosc},
if ${\rm E}_\mathrm{rot} \gtrsim 30 k_B T_z$, the phase
fluctuations for each ion's spectroscopic transition will be
bounded by a value less than one radian, so that there will be no
loss in spectroscopic contrast in a Ramsey-type experiment.

The argument in the paragraph above hinges on the assumption that
the electric field arising from the ion cloud's mean-field
distribution and from grazing-angle collisions is small in
magnitude compared to $\mathcal{E}_\mathrm{rot}$, so that the
shift in Berry energy is linear in the axial component of the
electric field.  For higher values of the ion temperature or
lower values of $\mathcal{E}_\mathrm{rot}$, a pair of colliding
ions can get so close to each other that the electric field is,
transiently, comparable to or larger than
$\mathcal{E}_\mathrm{rot}$.  We discuss the consequences in the
next subsection.

\subsection{Geometric Phases Accumulated During an Ion-Ion Collision}
\label{ss:ecollision}

As discussed in Sec.~\ref{ss:bigangle}, when a spin adiabatically
follows a time-varying quantization axis it acquires a geometric
(Berry's) phase.  For the eigenstates in
Fig.~\ref{f:hyperfine_stutz}(b), the geometric phase factor can be
written as $\exp{(\pm i m_F \mathcal{A})}$, where $\mathcal{A}$ is
the solid angle subtended by the contour on the unit sphere traced
out by the time-varying quantization axis. Thus, the relative
phase generated between the $|F=3/2,m_F= \pm 3/2\rangle$ states
used for spectroscopy is $\phi = 3 \mathcal{A}$. The concern of
this subsection is how ion-ion collisions cause uncontrolled
excursions of the quantization axis leading to random geometric
phase shifts and decoherence between spin states.  These
uncontrolled phase shifts will be written as $\Delta\phi = 3\Delta
\mathcal{A}$ to distinguish them from the calibrated geometric
phases in the experiment.

The instantaneous quantization axis for the molecular ion
eigensates is defined by the net electric field vector at the
location of the ion.  During a collision, this axis is defined by
the vector sum of the rotating electric field,
$\vec{\mathcal{E}}_\mathrm{rot}$, and the ion-ion interaction
electric field, $\vec{\mathcal{E}}_\mathrm{ion}$.  Both of these
are time-varying vectors, however typical ion-ion collisions have
a duration short compared to the rotation period of
$\mathcal{E}_\mathrm{rot}$ so for the purpose of this discussion
$\mathcal{E}_\mathrm{rot}$ will be taken as stationary.  Thus,
the problem is reduced to calculating the excursion of the
quantization axis under the time variation of
$\mathcal{E}_\mathrm{ion}$.  A typical ion-ion collision is shown
in Fig.~\ref{f:collision}(a) and the effect of this collision on
the quantization axis is shown in Fig.~\ref{f:collision}(b).

At the temperatures of our trapped ion samples, no two ions are
ever close enough for the details of the intermolecular potential
to matter.  Only monopole-monopole and monopole-dipole
interactions matter.  Further, the translational degree of freedom
may be treated as purely classical motion in a $1/r$ ion-ion
potential, with the initial condition of a given collisional event
characterized by an impact parameter and relative velocity.  The
outcome of the collision depends not only on the magnitudes of the
impact parameter and of the velocity, but also on their angles
with respect to the ambient electric bias field,
$\mathcal{E}_\mathrm{rot}$.  Each initial condition contributes a
particular amount to the variance in the phase between the
relevant internal states.  These contributions can be converted to
partial contributions to a decoherence rate, and a numerical
integral over a thermal distribution of collisional initial
conditions can yield the total decoherence rate.  We have pursued
this program to a greater or lesser extent with the decoherence
mechanisms discussed in this subsection and the one immediately
following, but the results are not especially illuminating and we
have used them primarily to confirm that the power-law expressions
discussed below represent only overestimates of the decoherence
rate, and that for experimental parameters of interest, the
decoherence rate will be conservatively less than 1~s$^{-1}$.

The main question is whether $T$ is high enough to include
significant phase space for collision trajectories for which the
peak value of $\mathcal{E}_\mathrm{ion} >
\mathcal{E}_\mathrm{rot}$ (which is to say, large enough to
transiently tip the direction of the total field by more than a
radian).  If so, then  a single collision can cause decoherence
and one can get a simple estimate of the cross-section for
decoherence simply from the size of the impact parameter that
leads to those events.  There is a significant probability for
collisions with $\mathcal{E}_\mathrm{ion} \gtrsim \mathcal{E}_{\rm
rot}$ when
\begin{equation}
    T \gg 18\ K \left(\frac{\mathcal{E}_\mathrm{rot}}{5\ \mathrm{V/cm}}\right)^{1/2},
\end{equation}
which leads to a decoherence rate
\begin{equation}
\label{e:hightdecoherencerate} \tau^{-1} \approx 0.47  \times
\left(\frac{n}{1000\ \mathrm{cm}^{-3}}\right)\left(\frac{T}{15\
\mathrm{K}}\right)^{1/2}\left(\frac{\mathcal{E}_\mathrm{rot}}{5\
\mathrm{V/cm}}\right)^{-1}\ \mathrm{s}^{-1}.
\end{equation}

If $T$ is instead so low that the Coulomb barrier suppresses
collisions that could lead to a sufficiently large value of
$\mathcal{A}$ and cause decoherence with a single collision, then
decoherence will arise only from the combined effects of many
collisions each causing small phase shifts that eventually random
walk the science transition into decoherence.  In this regime, the
decoherence rate falls off very fast at low temperatures.  For
\begin{equation}
    T \ll 18\ K \left(\frac{\mathcal{E}_\mathrm{rot}}{5\ \mathrm{V/cm}}\right)^{1/2},
\end{equation}
typical collisions have $\mathcal{E}_\mathrm{ion} \lesssim
\mathcal{E}_{\rm rot}$ and the decoherence rate is
\begin{equation}
\label{e:lowtdecoherencerate} \tau^{-1} \approx 0.13 \times
\left(\frac{n}{1000\ \mathrm{cm}^{-3}}\right)\left(\frac{T}{15\
\mathrm{K}}\right)^{13/2}\left(\frac{\mathcal{E}_\mathrm{rot}}{5\
\mathrm{V/cm}}\right)^{-4}\ \mathrm{s}^{-1}.
\end{equation}

\begin{figure}
\begin{center}
\includegraphics{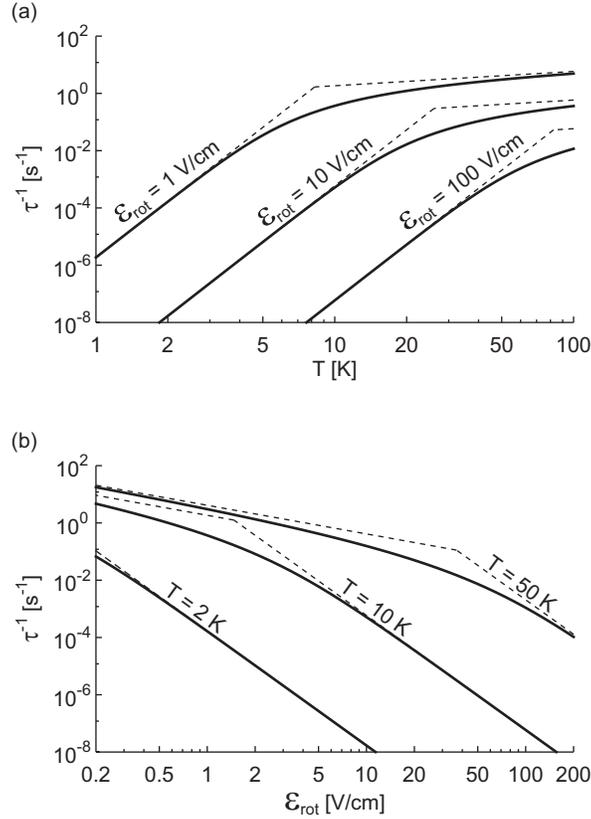}
\caption{Inverse coherence times, $\tau^{-1}$, due to geometric
phases accumulated during ion-ion collisions as a function of (a)
collision energy in temperature units and (b)
$\mathcal{E}_\mathrm{rot}$.  Dotted lines are approximations given
in Eqs.~\ref{e:hightdecoherencerate} and
\ref{e:lowtdecoherencerate}.  Solid lines are more involved
estimates based on integrals over collision parameters, but are
still based on approximations so as to be conservative.  The ion
density was taken to be $n=1000$~cm$^{-3}$.\label{f:taucoll}}
\end{center}
\end{figure}

Both Eqs.~\ref{e:hightdecoherencerate} and
\ref{e:lowtdecoherencerate} represent conservative estimates of
the decoherence rate, and for an intermediate range of
temperature, the decoherence rate will be less than whichever
estimate gives the smaller value (Fig.~\ref{f:taucoll}).

\subsection{$m$-Level Changing Collisions}
\label{ss:mcollision}

A second source of decoherence can arise from ion-ion collisions
that induce transitions between internal levels of a molecule.
The dominant inelastic channel will be transitions between $m_F$
levels induced by a sufficiently sudden tilt in the quantization
axis defined by the instantaneous local electric field.  There
are two conditions for such a transition to occur: (i) the
direction of the total field must change by nearly a radian or
more, so that there is significant amplitude for, e.g., an $m_F =
+3/2$ level in the unperturbed electric field to suddenly have
non-negligible projection on an $m_F = +1/2$ level in the
collision-perturbed field, and (ii) the time rate of change of
the electric field direction must be comparable to or larger than
the energy splitting between an $m_F = 3/2$ level and its nearest
$m_F = 1/2$ level in the field $\mathcal{E}_\mathrm{rot}$.

Note that the first requirement is the same as the requirement
for picking up an appreciable single collision Berry's phase.
However, not all collisions that satisfy the first requirement
will satisfy the second requirement.  In particular, if the
relative velocity in a collision is too low, then the time rate
of change of the electric field direction will be too slow to
satisfy the second requirement.  Thus, given that the first
requirement is satisfied, then the second requirement will
\textit{not} be satisfied when
\begin{equation}
    T <  5\ \mathrm{K} \times \frac{\mathcal{E}_\mathrm{rot}}{5\ \mathrm{V/cm}},
\end{equation}
In this limit, the second requirement is more stringent than the
first requirement, which means that the rate of $m$-level
changing collisions will be smaller than the rate of
single-collision Berry's phase-induced decoherence.  In the
opposite limit, we expect the second requirement will be met
whenever the first requirement is met, and thus we would expect
that the two channels of decoherence, $m$-level changing and
single-collision Berry's phase, will be comparable in magnitude.

Looking at particular collision trajectories in more detail, we
see that there are trajectories that can cause an $m$-level change
but for which there is no contribution to Berry's phase because
the electric field traces out a trajectory with no solid angle
(for instance, if the classical impact parameter $\vec{b}$ is
parallel to $\vec{\mathcal{E}}_\mathrm{rot}$).  We also note that
our formulation of the requirement of sweep rate for $m$-level
changing collisions neglects the fact that
$\mathcal{E}_\mathrm{ion}$ will not only change the direction of
the total electric field ($\vec{\mathcal{E}}_\mathrm{ion} +
\vec{\mathcal{E}}_\mathrm{rot}$) but also in general will change
its magnitude.  For most impact parameters, the magnitude of the
total electric field will increase, thus suppressing nonadiabatic
effects.  However, a narrow range of impact parameters exists
where the magnitude of the total electric field decreases, thus
enhancing nonadiabatic effects.  However, the above scaling laws
account for the majority of collisions.

In the end, we are less interested in the actual rates than we
are in putting conservative limits on decoherence rates.  For
instance, in calculating the curves in Fig.~\ref{f:taucoll}, we
pessimistically took a worst-case geometry,
$\mathcal{E}_\mathrm{rot}\ \bot\ \mathcal{E}_\mathrm{ion}$, which
gives an upper limit on the size of the effect.  Thus we estimate
that:

\begin{itemize}
  \item For $T <  5\ \mathrm{K} \times \frac{\mathcal{E}_\mathrm{rot}}{5\ \mathrm{V/cm}}$ the \textit{total} collisional decoherence, including both $m$-level-changing and Berry's-inducing effects, will be less than or equal to the value given by solid curves in Fig.~\ref{f:taucoll}, while
  \item For $T >  5\ \mathrm{K} \times \frac{\mathcal{E}_\mathrm{rot}}{5\ \mathrm{V/cm}}$, the \textit{total} collisional decoherence will be no greater than twice as large as the rate indicated by those curves.
\end{itemize}

\section{Conclusions: Precision and Accuracy}
\label{s:conclusion}

Recall from Sec.~\ref{s:intro} the three components to the
sensitivity figure-of-merit:

\subsection{Coherence Time}

Conclusions of Sec.~\ref{s:spectroscopy} and~\ref{s:collisions}:
Taking into account only collisional decoherence, and all the
questions associated with being in rotating fields and in trapping
fields, we would anticipate a coherence time longer than one
second. Black-body thermal excitation of the J=1 rotational level
will also be well over one second. Vibrational black-body
excitation for the v=0 state is estimated at 6 s for HfF$^+$ in a
300 K environment. Thus the dominant limitation to coherence will
likely be the radiative lifetime of the $^3\Delta_1$ state,
estimated~\cite{PMT09} at 390 ms for HfF$^+$, and still longer for
ThF$^+$, for which the $^3\Delta_1$ state is predicted to be still
lower in energy. The largest uncertainty in the lifetime
calculation is the uncertainty in the $^3\Delta_1 \rightarrow$
$^1\Sigma$ decay energy, calculated to be 1600 cm$^{-1}$ in
HfF$^+$.

\subsection{$\mathcal{E}_{\rm eff}$}

$\mathcal{E}_{\rm eff}$ in HfF$^+$ is calculated by Meyer and
coworkers to be 30~GV/cm~\cite{MB08}, and by Titov \textit{et al.}
to be 24~GV/cm~\cite{PMI07}. For ThF$^+$, Meyer calculates
90~GV/cm~\cite{MB08}.  The uncertainties in these numbers are hard
to assess, but they are very likely accurate to better than a
factor of two and, if ongoing spectroscopic studies provide
experimental values of hyperfine and fine structure that confirm
the ab initio values predicted by the St.\ Petersburg group, our
confidence in the precision of calculated $\mathcal{E}_{\rm eff}$
will be much higher.

\subsection{Count Rate and Summary of Expected Precision}

We are producing HfF$^+$ ions by photoionization in a relatively
narrow range of quantum states, and can estimate yield per quantum
level within the desired trapping volume at perhaps 100 ions per
shot, but we have just begun to characterize the efficiency of the
process and very little optimization has been done. Our design
efficiency for reading out spin states of trapped ions via
laser-induced fluorescence (LIF) is 4\%, but that has not been
verified yet. With a large uncertainty, then, we may detect about
one ion per shot with four shots per second. Overall, precision in one hour could be
about 10 mHz. For ten hours of data, we anticipate (very roughly)
a raw precision at $5 \times 10^{-28}$ e~cm in HfF$^+$, and $1.5
\times 10^{-28}$ e~cm in ThF$^+$. We are investigating several more efficient alternatives to LIF for spin readout, including in particular
resonantly enhanced photodissociation or second photoionization. Even if we detect as
many as four ions in a shot, single shot precision will be no better than 300 mHz, which sets a relaxed requirement for suppressing experimental shot-to-shot noise.

\subsection{Accuracy, Systematic Error}
\label{ss:systerr}

We have not completed a systematic study of the consequences of
all possible violations of reflection symmetry in the trapping
fields, but work in this direction is ongoing.

For now, we make the following three observations:

i) For the field asymmetries we have analyzed to date, realistic
estimates for the magnitude in as-constructed field imperfections
lead to systematic errors on the order of a few 10$^{-29}$ e cm or
less. While this is not yet as accurate as our ultimate ambitions,
it would represent roughly a factor of thirty improvement on the
existing best experimental limit.

ii) Asymmetries analyzed to date lead to systematic errors whose
signs reverse when the direction of rotation $\omega_{\rm rot}$
reverses. If we combine measurements made with clockwise and
counterclockwise field rotation, the errors vanish.  Ideally, we'd
like to design sufficient accuracy into the experiment so that the
chop in field rotation is not needed to achieve desired accuracy,
but as a practical matter we will of course run the experiment
both ways, averaging the results to get ultimate accuracy, and
differencing them to diagnose experimental flaws.

iii) Auxiliary measurements are envisioned to characterize and
shim out flaws in the as-constructed trap. For instance, we plan
to be able to shim the equilibrium position of the ion cloud up
and down along the trap axis, and  at each location measure the
energy difference E$_b$-E$_d$. Unlike E$_a$-E$_b$, E$_b$-E$_d$ is
highly electric-field sensitive. The result will be a precise
measurement of any spatial gradient in $\mathcal{E}_{\rm rot}$.

iv) All systematic errors we have analyzed to date have strong
dependencies on quantities such as $\omega_{\rm rot}$,
$\mathcal{B}_{\rm rot}$, $\mathcal{E}_{\rm rot}$, and on the
ion-cloud temperature and density, and the trap confining
frequencies. A true signal from a nonzero value of d$_e$ will be
largely independent of all those quantities. We anticipate making
a number of auxiliary measurements with the experimental
parameters tuned far away from their optimal values to
deliberately exaggerate the size of systematic errors and allow us
thus to characterize their dependencies in less integration time
than that required for ultimate sensitivity. Even so, and as is
often the case in precision measurement experiments, sensitivity
and accuracy are coupled. To the extent we can measure d$_e$ to
high precision at many combinations of experimental parameters, we
will better be able to detect and reject false signals.

We believe the experiment as we have described it should have the capability to improve the limit on the electron's electric dipole moment to 10$^{-29}$ e cm.  As of this writing, the largest contribution to the uncertainty in our ultimate capability has to do with unknown efficiencies of state preparation and read out.  More specialized publications from our group addressing progress in these areas are forthcoming.

\subsection{Summary}

Until now, molecular ions have not been viable candidates for symmetry violation searches largely due to the fact that applying electromagnetic fields to manipulate the internal states of the molecule would also violently perturb the translational motion of the ions.  In this work, we have proposed a technique to overcome this obstacle -- namely applying an electric field that rotates at radio frequencies.  The specifics of performing high-resolution electron spin resonance spectroscopy under these conditions were analyzed.  In particular, we have shown that a significant advance towards detecting the permanent electric dipole moment of the electron can be made by probing the valence electrons in a ground or metastable ${^3}\Delta_1$ level of an ensemble of trapped diatomic molecular ions.

\textit{Note added in proof:} Since the submission of this work, a new experimental limit on the electric dipole moment of the electron has been achieved using YbF molecules: $|d_e| < 10.5 \times 10^{-28}$~e~cm~\cite{Hudson2011}.

\section*{Appendix: Typical Experimental Parameter Values}
\label{s:app}

%\appendix{Appendix: Typical Experimental Parameter Values}
%\label{s:app}

\noindent $d_e{\mathcal E}_{\rm eff} = 2\pi \times 0.36$~mHz,
transition energy between $m_F = +3/2$ and $m_F = -3/2$ states in
ThF$^+$ if $d_{e}$ = 1.7 x 10$^{-29}$ e cm.

\noindent $d_\mathrm{mf}$ = +1.50 a.u.\ $\approx  2\pi \times
2$~MHz/(V/cm), electric dipole moment of HfF$^+$ in the molecular
rest frame.

\noindent $\mathcal{E}_\mathrm{rot}$ =  5 V/cm, rotating electric
field.

\noindent $\omega_\mathrm{rot} = 2\pi \times 100$~kHz, frequency
of rotating electric field.

\noindent ${\rm E}_{\rm rot} = 1800$ K, typical kinetic energy in
rotational micromotion.

\noindent $r_{\rm rot}$ = 0.6 mm, radius of circular micromotion.

\noindent $d_\mathrm{mf}\mathcal{E}_\mathrm{rot} =   2\pi \times
10$~MHz.

\noindent (3/2)$\gamma_{F=3/2}  d_\mathrm{mf}
\mathcal{E}_\mathrm{rot} = 2\pi \times 5$~MHz, Stark shift of $m_F
= \pm 3/2$ states of $^3\Delta_1$ levels in rotating electric
field.

\noindent $\omega_\mathrm{ef} = 2\pi \times 10$~kHz,
$\Lambda$-doublet splitting between opposite parity $^3\Delta_1$
J=1 states.

\noindent $g_{F=3/2}  =  0.03$, magnetic g-factor in $^3\Delta_1$
$m_F = \pm 3/2$ states.

\noindent $B_\mathrm{rot}$ =  70 $\mu$G, rotating magnetic field.

\noindent $3g_F \mu_B  \mathcal{B}_\mathrm{rot}  = 2\pi \times
8$~Hz, Zeeman splitting between $m_F = +3/2$ and $m_F = -3/2$
states due to $\mathcal{B}_\mathrm{rot}$.

\noindent $\delta g_{F=3/2}$/$g_{F=3/2}$ $\approx$ 3 $\times$
10$^{-4}$, fractional difference of magnetic g-factor for upper
and lower levels, for parameters shown in the Appendix.

\noindent $\Delta$ $\approx$ $2\pi \times 2$~Hz, splitting at the
avoided crossing between $m_F$ = +3/2 and $m_F$ = -3/2 levels, for
parameters shown in the Appendix.

\noindent $\mathcal{B}_\bot  = 25$ mG, anticipated scale of
transverse magnetic field.

\noindent r = 0.5 cm, characteristic rms radius of trapped ion
cloud.

\noindent T = 15 K, characteristic temperature  of trapped cloud.

\noindent $\noindent \omega_i = 2\pi \times 1$~kHz, typical trap
confining frequency.

\noindent $\mathcal{E}_z = 5$ mV/cm, typical axial electric field
applied for confinement.

\noindent $\mathcal{E}_{\rm rf} = 75$ mV/cm, typical Paul trap
electric field strength, at typical cloud radius.

\noindent $<\mathcal{E}_{\rm rf}>$ = 5 mV/cm, typical radial
confining electric field, averaged over one Paul cycle.

\noindent $\omega_{\rm rf} = 2\pi \times 15$~kHz, typical ``rf
freq" for Paul trap.

\noindent ${\rm E}_{\rm rf} = 15$ K, typical kinetic energy in
Paul micromotion.

\noindent ${\rm E}_\mathrm{hf} = 2\pi \times 45$~MHz, hyperfine
splitting between F = 1/2 and F = 3/2 states of $^3\Delta_1$ J=1
level.

\section*{Acknowledgments}
We gratefully acknowledge many useful discussions.
        On the topic of molecular and atomic spectroscopy, we benefited
from discussions with Peter Bernath, Carl Wieman, Tom Gallagher,
Ulrich Hechtfische, and Jim Lawler.
    On molecular structure, Andrei Derevianko, Svetlana Kotochigova,
Richard Saykally, Laura Gagliardi, and especially the St.
Petersburg group, Anatoly Titov, Mikhail Kozlov, and Alexander
Petrov.
    On molecular dynamics, we had useful discussions with Carl
Lineberger, David Nesbitt, and especially Bob Field.
    On EDM measurements, Pat Sandars, Dave DeMille, and Neil
Shafer-Ray.
    We thank Gianfraco DiLonardo for the loan of a hollow cathode lamp
and Tobin Munsat for the loan of a high-current power supply.
    We acknowledge useful contributions in our lab from Herbert
Looser, Tyler Yahn, Tyler Coffey, Matt Grau, and Will Ames.
    We thank Jun Ye and members of his group for sharing their
innovative ideas in comb-based spectroscopy, and Konrad Lehnert
for enlightening us about sensitive microwave detection.
    This work was supported by NIST, NSF, and funds from a Marsico Chair of Excellence.

% Bibliography
%\bibliography{RevisedJMolSpecteEDM110715}

\begin{thebibliography}{139}
\bibitem[{uni()}]{units}
\bibinfo{note}{Energies will be quoted as angular frequencies, in
  radians/second, or temperatures, in Kelvin, where the factors of $\hbar$ and
  $k_B$ are omitted from equations, respectively.}
\bibitem[{gfa()}]{gfactors}
\bibinfo{note}{This is a good approximation at the $10^{-3}$ level.}
\end{thebibliography}

\end{document}